\documentclass[prx,twocolumn,citeautoscript,amsmath,superscriptaddress]{revtex4-1}

\usepackage{graphicx}
\usepackage{newtxtext}
\usepackage{epsfig}
\usepackage{amsmath,bbm,amssymb,bm,times}
\usepackage[colorlinks,linkcolor=blue,citecolor=blue]{hyperref}

\begin{document}
\title{Topological synchronization of coupled nonlinear oscillators}
\author{Kazuki Sone}
\email{sone@noneq.t.u-tokyo.ac.jp}
\affiliation{Department of Applied Physics, The University of Tokyo, 7-3-1 Hongo, Bunkyo-ku, Tokyo 113-8656, Japan}
\author{Yuto Ashida}
\affiliation{Department of Applied Physics, The University of Tokyo, 7-3-1 Hongo, Bunkyo-ku, Tokyo 113-8656, Japan}
\affiliation{Department of Physics, The University of Tokyo, 7-3-1 Hongo, Bunkyo-ku, Tokyo 113-0033, Japan}
\affiliation{Institute for Physics of Intelligence, The University of Tokyo, 7-3-1 Hongo, Tokyo 113-0033, Japan}
\author{Takahiro Sagawa}
\affiliation{Department of Applied Physics, The University of Tokyo, 7-3-1 Hongo, Bunkyo-ku, Tokyo 113-8656, Japan}
\affiliation{Quantum-Phase Electronics Center (QPEC), The University of Tokyo, 7-3-1 Hongo, Bunkyo-ku, Tokyo 113-8656, Japan}

\begin{abstract} 
Synchronization of coupled oscillators is a ubiquitous phenomenon found throughout nature. Its robust realization is crucial to our understanding of various nonlinear systems, ranging from biological functions to electrical engineering. On another front, in condensed matter physics, topology is utilized to realize robust properties like topological edge modes, as demonstrated by celebrated topological insulators. Here, we integrate these two research avenues and propose a nonlinear topological phenomenon, namely topological synchronization, where only the edge oscillators synchronize while the bulk ones exhibit chaotic dynamics. We analyze concrete prototypical models to demonstrate the presence of positive Lyapunov exponents and Lyapunov vectors localized along the edge. As a unique characteristic of topology in nonlinear systems, we find that unconventional extra topological boundary modes appear at emerging effective boundaries. Furthermore, our proposal shows promise for spatially controlling synchronization, such as on-demand pattern designing and defect detection. The topological synchronization can ubiquitously appear in topological nonlinear oscillators and thus can provide a guiding principle to realize synchronization in a robust, geometrical, and flexible way.
\end{abstract}

\maketitle

\section{Introduction\label{section1}}
Nonlinear oscillators ubiquitously appear in a variety of fields from biology \cite{Buck1988,Luo1991} to engineering \cite{Kozyreff2000,Ludwig2013,Blaabjerg2006,Wiesenfeld1996}. They often exhibit frequency-synchronization \cite{Acebron2005}, where (even inhomogeneous) interacting oscillators vibrate at the same frequency. Synchronization plays a crucial role in classical nonlinear systems, e.g., circadian rhythm \cite{Dibner2010}, and even in quantum systems \cite{Lee2013,Laskar2020}. Controlling synchronization and desynchronization \cite{Sieber2014,Isele2016,Zhang2020a} of nonlinear oscillators is essential to fulfill their functions. However, nonlinear oscillators are often irregularly affected by their uncontrollable circumstances, and therefore it is desirable to investigate universal principles to robustly control and design nonlinear oscillators against such disorder.

On another front, topology represents the property of matter unchanged under continuous deformations and thus can provide a guiding principle to realize robust systems. Numerous studies in condensed matter physics have focused on such utility of topology, where a remarkable example is a topological insulator \cite{Kane2005,Hasan2010,Qi2011} exhibiting a metallic surface and an insulating bulk. Topological insulators exhibit topologically nontrivial bulk associated with edge-localized modes robust against disorder, as a consequence of a principle called bulk-edge correspondence \cite{Hasan2010,Qi2011}. Such edge modes exhibit gapless dispersion relations and backscattering-free edge current. Because bulk topology is intrinsically related to Hamiltonians that govern linear dynamics, topology in physics has been studied mostly in linear systems. In recent studies, it has been revealed that topological edge modes can affect the nonlinear dynamics \cite{Smirnova2020,Ota2020,Kotwal2021} in, e.g., photonics \cite{Zhang2020b} and mechanical lattices \cite{Chen2014}, where topological edge soliton \cite{Leykam2016} and bulk-localized topological modes \cite{Lumer2013} can emerge. Despite these recent advancements, topology in nonlinear systems is still largely unexplored even at the conceptual level.

Given these fundamental problems, we here propose a topological mechanism of robust edge-localized frequency-synchronization, where only nonlinear oscillators at the edge of the system synchronize, whereas dynamics of the bulk oscillators is chaotic. We demonstrate the emergence of such a synchronized state by numerical calculations of nonlinear oscillators with linear couplings corresponding to a topological Hamiltonian. We term the proposed state as a topological synchronized state (TSS). 

By calculating the Lyapunov exponents \cite{Oseledets1968} and vectors \cite{Takeuchi2009,Ginelli2013}, we confirm the chaotic behavior of bulk oscillators and find that topological edge modes are represented by the edge-localized Lyapunov vectors. While such coexistence of synchronization and desynchronization is apparently reminiscent of chimera states \cite{Kuramoto2002,Abrams2004,Panaggio2015,Wolfrum2011,Hohlein2019,Majhi2019,Tinsley2012,Martens2013}, we emphasize that the mechanism of the proposed TSS is distinct from conventional chimera states because of its topological origin. Moreover, nonlinearity leads to another unique topological phenomenon: unconventional extra boundary modes appear at the emergent effective boundary. Such extra boundary modes can increase the number of synchronized oscillators in the TSS. 

\begin{figure*}[t]
  \includegraphics[width=160mm,bb=0 0 920 540,clip]{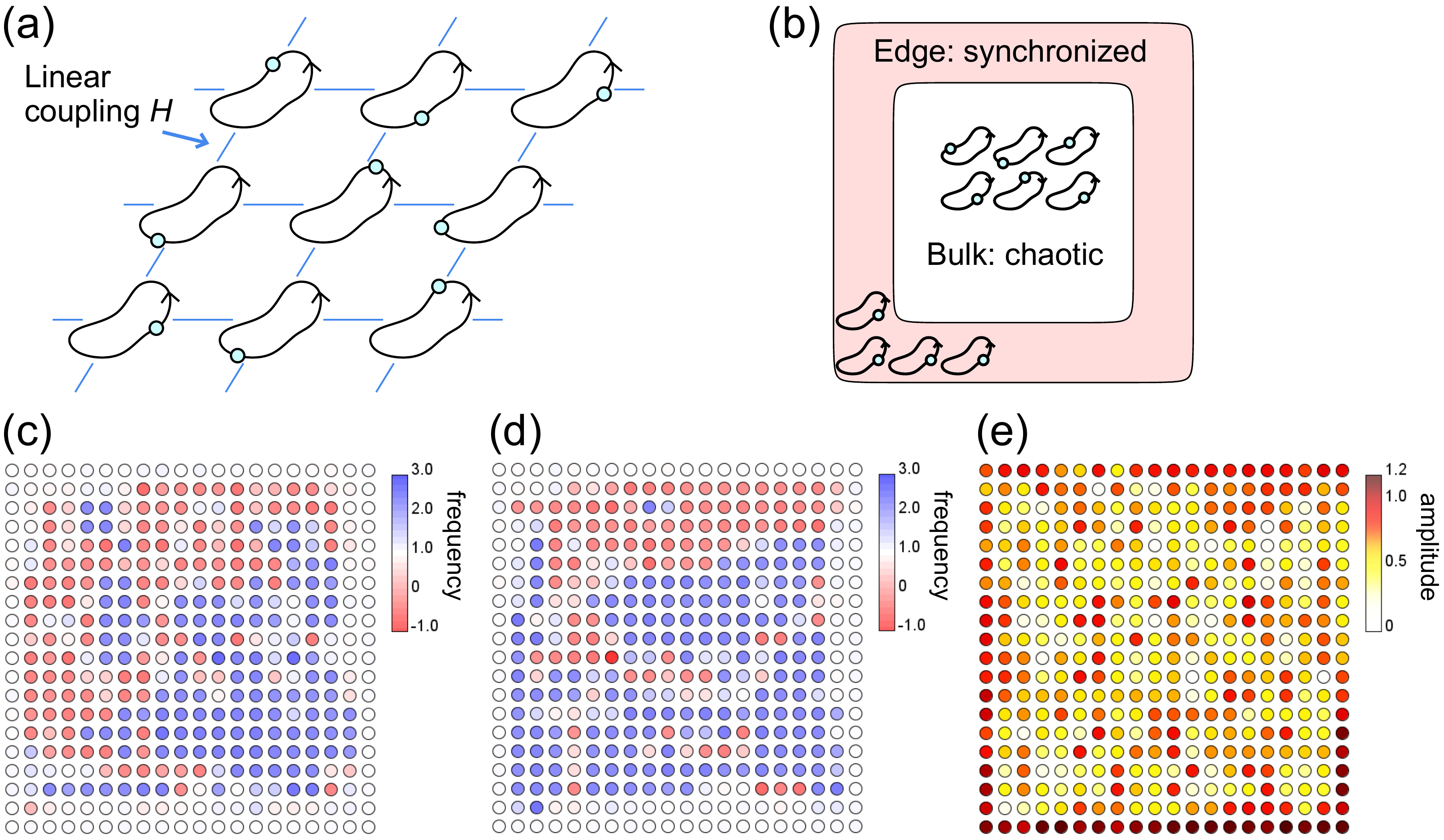}
\caption{\label{fig1} Dynamical formation of topological synchronized states (TSSs). (a) Schematic of the considered nonlinear system. Nonlinear oscillators are placed on a square lattice. Each oscillator is linearly coupled to other oscillators at the nearest neighbor sites. The linear coupling corresponds to the Hamiltonian of a topological insulator laser, which exhibits amplified edge oscillations. (b) TSS observed under the open boundary condition. The nonlinear oscillators near the boundary synchronize, while those away from the boundary exhibit chaotic dynamics instead of being synchronized. (c),(d) Frequency of the first component of oscillators at each site. We numerically calculate the dynamics of the first model of the TSS (Eqs.~\eqref{model1}, \eqref{hamiltonian_TIL_wavenumber}). Panels (c) and (d) are the snapshots at the time $t=100$ and $t=200$ for each. Empty sites represent the oscillators oscillating around their natural frequencies. We confirm that the frequencies of the edge oscillators are almost homogeneous and constant, which indicates their frequency-synchronization. Meanwhile, the oscillators inside exhibit inhomogeneous frequencies. One can also confirm that their frequencies vary over time, thus indicating the desynchronization and instability of the bulk oscillators. The parameters used are $u=-1$, $b=0.5$, $\alpha=0.5$, $\beta=1$, $\omega_0=1$, and $\Delta\omega=0.2$. (e) Amplitudes of the oscillators. Here, we numerically simulate the dynamics of the model as in panels (c) and (d). We check that the amplitudes of most of the edge oscillators are larger than those of the bulk oscillators. The same parameters as panels (c) and (d) are used in the calculation.
}
\end{figure*}

As applications of the TSS, we propose to arrange the synchronized oscillators in an on-demand pattern, and to detect defective structures by observing the lasing patterns around the defects. We also propose a concrete electrical-circuit realization of the TSS. While we focus on several types of non-Hermitian Hamiltonians \cite{Shen2018,Gong2018,Yao2018,Kunst2018,Zhao2019,Fruchart2021} of lasing edge modes \cite{Harari2018,Song2020,Sone2020} to realize the TSS, we expect that the TSS can ubiquitously emerge in nonlinear oscillators with topological linear couplings. In addition, the robustness of the topological edge modes guarantees the stability of the TSS against detuning parameters and thus provides a robust and universal mechanism to control synchronization.

\section{Concept of topological synchronization\label{section2}}
Before moving to detailed results of numerical analyses, we illustrate the general concept of TSS proposed in this article. Here, we consider coupled nonlinear oscillators that self-vibrate even if they are decoupled. Such oscillators maintain their self-oscillations at the balance of the injection and dissipation of energy and exhibit nonlinear features that cannot arise in linear harmonic oscillators, such as the robustness of the amplitude. Dynamics of nonlinear oscillators can be described as a limit cycle; one of the simplest models for this is the (complex) Stuart-Landau oscillators \cite{Stuart1960} described as $\dot{Z} = (i\omega+\alpha-\beta |Z|^2) Z$, where $Z$ is the complex-valued state variable. In the following, we focus on the case that $\beta$ is real. This model exhibits self-excited oscillations with frequency $\omega$ and amplitude $\sqrt{\alpha/\beta}$ and is the normal form of a nonlinear oscillator around the parameters where it begins self-oscillations.

{\renewcommand{\arraystretch}{1.0}
\begin{table*}[t]
\caption{Summary of the properties of our models.}
\begin{center}
%\begin{tabular}{|p{0.18\linewidth}p{0.18\linewidth}p{0.18\linewidth}p{0.18\linewidth}p{0.18\linewidth}|}
\begin{tabular}{|l@{\hspace{2.0mm}}||l@{\hspace{2.0mm}}|l@{\hspace{2.0mm}}|l@{\hspace{2.0mm}}|l@{\hspace{2.0mm}}|}
%\begin{tabularx}{0.9\linewidth}{|XXXXX|}
\hline
Model & TSS & Protection & Coupling & Nonlinearity-induced \rule[0pt]{0pt}{10pt} \\
{} & {} & mechanism & {} & boundary modes \\[1mm]\hline \hline
Model using exceptional & Yes & Exceptional & Non-Hermitian & Yes \rule[0pt]{0pt}{10pt} \\
edge modes (Eqs.~\eqref{hamiltonian_TIL_wavenumber}, \eqref{hamiltonian_TIL}) & {} & edge modes & {} & {} \\[1mm]\hline
Model using & Yes & Conventional & Non-Hermitian & No \rule[0pt]{0pt}{10pt} \\
conventional bulk & {} & bulk topology & {} & {} \\
topology (Eq.~\eqref{hamiltonian_conventionaltopo_TIL}) & {} & {} & {} & {} \\[1mm]\hline
Model using Hermitian & Yes (only 1st and & Exceptional & Hermitian & No \rule[0pt]{0pt}{10pt} \\
linear couplings & 4th components & edge modes & (Non-Hermiticity from & {} \\
(Eqs.~\eqref{Hermitian_coupling}, \eqref{Hermitian_coupling_real}) & synchronize) & {} & nonidentical oscillators) & {} \\[1mm]\hline
\end{tabular}
\end{center}
\label{table1}
\end{table*}
}

TSS is defined as a coexistence of synchronized edge oscillators and desynchronized bulk oscillators due to the nontrivial topology of the system. Such synchronization of the edge oscillators can appear in coupled nonlinear oscillators where the linear coupling is described by a Hamiltonian that is topological in the wavenumber space. We can construct concrete models of the TSS by utilizing linearly coupled Stuart-Landau oscillators arranged on a square lattice (see Fig.~\ref{fig1}(a)). In the present work, we assume that four oscillators exist at each site and they interact with ones at the same site as well as the nearest neighbor sites (four or more oscillators are necessary to realize the desirable properties of the linear coupling, such as its lasing edge modes discussed in the following). The dynamics of our model is described by the following set of equations,
\begin{eqnarray}
 \frac{d}{dt} Z_j(\mathbf{x}) &=&  (i\omega_j(\mathbf{x})+\alpha-\beta |Z_j(\mathbf{x})|^2)Z_j(\mathbf{x})\nonumber\\
 &{}& -i \sum_{k,\mathbf{x}'} H_{jk}(\mathbf{x},\mathbf{x}') Z_k(\mathbf{x}'), \label{model1}
\end{eqnarray}
where $\mathbf{x}=(x,y)$, $\mathbf{x}'=(x',y')$ represent the location of the site in the square lattice, and $j,k=1,\cdots,4$ distinguish four oscillators at each site. $H_{jk}(\mathbf{x},\mathbf{x}')$ represents the linear coupling between the oscillators at the same site or the nearest neighbor sites. We assume $Z_k(\mathbf{x}') = 0$ for $\mathbf{x}'$ corresponding to the outside of the system, which realizes open boundaries. We note that this choice of the boundary condition is just for the simplicity of the models, and TSS is independent of the detail of boundary conditions on the condition that both edges are disconnected. We adopt a Hamiltonian of topological lasing modes \cite{Harari2018,Song2020,Sone2020} as the matrix $H_{jk}(\mathbf{x},\mathbf{x}')$ describing the linear coupling, and obtain TSS under the open boundary condition. To investigate the robustness of the TSS against disorders, we introduce the inhomogeneity into the system such that natural frequencies of each oscillator, $\omega_j(\mathbf{x})$, are uniformly distributed around the mean value $\omega_0$ with the distribution width being $\Delta\omega$.

In the following sections, we analyze three models of TSS. These models are constructed in the same spirits as described above, while the linear couplings are different. In our first model (Sec.~\ref{section3}), we introduce the Hamiltonian featuring the exceptional edge modes \cite{Sone2020} to describe the linear coupling. Exceptional edge modes are protected in an unconventional mechanism unique to non-Hermitian systems and enable us to realize TSS in a simple system. To show that we can also realize TSS by using conventional topological edge modes, we construct another model in Sec.~\ref{section4}. In Sec.~\ref{section5}, we propose another model utilizing Hermitian linear couplings by modifying our first model. The properties of our three models are summarized in Table \ref{table1}. In Sec.~\ref{section6}, we propose applications of TSS to on-demand pattern designing and defect detection with their numerical demonstration. We also present a schematic of the possible realization of TSS using an electrical circuit. Section \ref{section7} summarizes the main results and discusses several open problems.

\begin{figure}[b]
  \includegraphics[width=70mm,bb=0 0 685 600,clip]{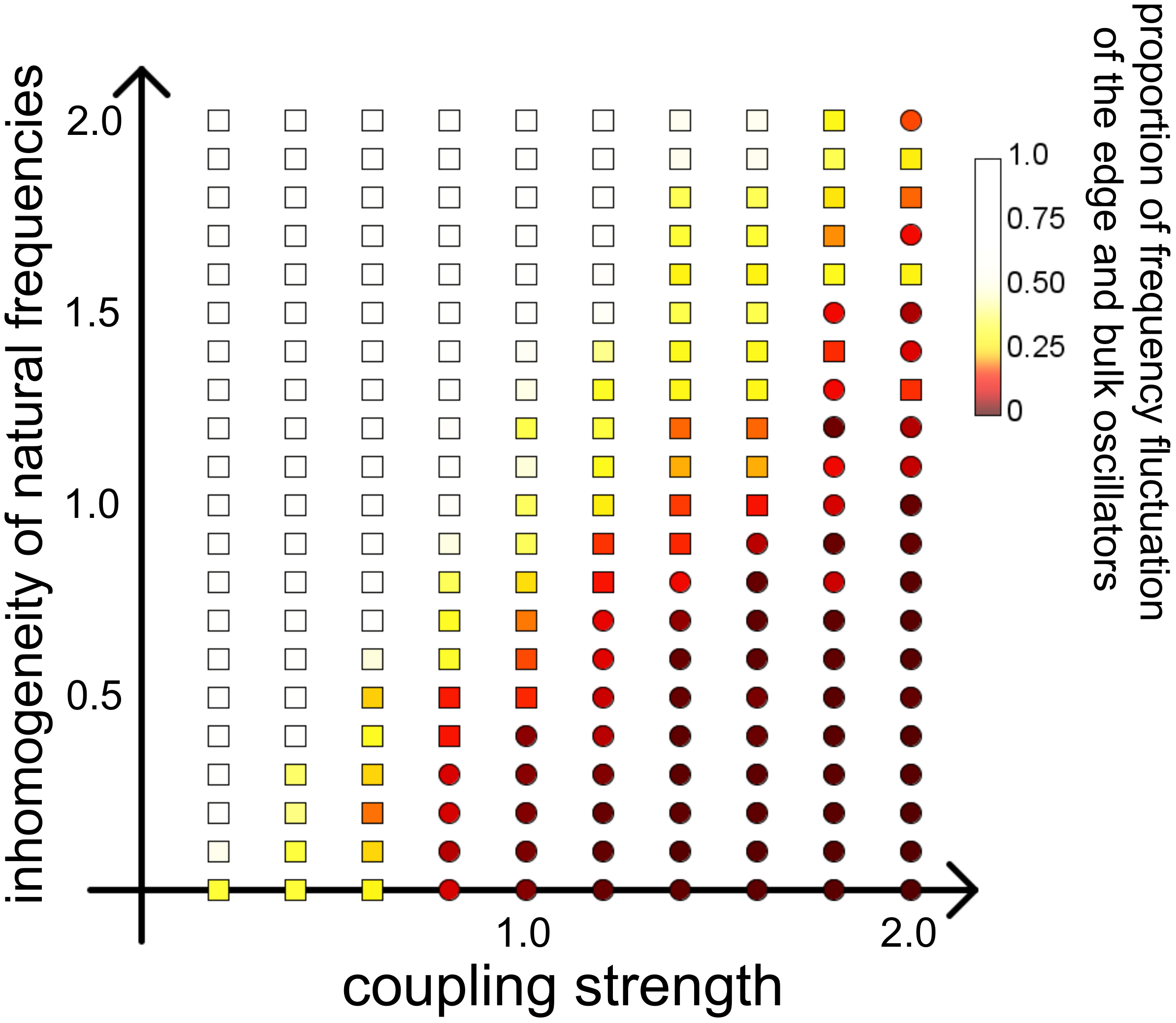}
\caption{\label{fig2} Phase diagram of the model of the TSS. The ratio of frequency fluctuations of the edge oscillators to those of the bulk ones is depicted at each parameter. We calculate the dynamics of the model of the TSS (Eqs.~\eqref{model1}, \eqref{hamiltonian_TIL_wavenumber}) under different strengths of the linear coupling and the inhomogeneity of the natural frequencies. The red region (right bottom) shows parameter regimes where the TSS emerges. The circles and squares represent the classification of data of frequency fluctuations of the edge and bulk oscillators obtained by a numerical clustering method. The circles correspond to the parameters for which the presence of the TSS is identified. The phase boundary obtained here corresponds to the parameter points at which the bandgap of the effective Hamiltonian closes and thus the bulk band can change its topology.
}
\end{figure}

\section{Model of topological synchronization utilizing exceptional edge modes\label{section3}}
\subsection{Model and numerical calculation of its dynamics}
To realize the TSS in a simple model, we first adopt the Hamiltonian of exceptional edge modes \cite{Sone2020} as  $H_{jk}(\mathbf{x},\mathbf{x}')$, which utilizes the robustness of singularities called exceptional points, topological gapless structures unique to non-Hermitian systems. The exceptional edge modes are protected by the topology of the dispersion relations around their branchpoint structures \cite{Shen2018} and exhibit lasing behavior \cite{Harari2018,Song2020} amplifying edge oscillations. In nonlinear oscillators, such edge modes can lead to qualitatively different synchronization behavior between edge and bulk oscillators without judicious designing of the system.

In more detail, we construct the linear coupling starting from the Hamiltonian of a Chern insulator with two internal degrees of freedom, $H_{\rm QWZ}=(u+\cos k_x + \cos k_y) \sigma_z + \sin k_x \sigma_x + \sin k_y \sigma_y$, called the Qi-Wu-Zhang (QWZ) model \cite{Qi2006}. This corresponds to the lattice version of the Dirac Hamiltonian, which is a prototype of topological insulators and exhibits edge-localized modes associated with its topological invariants (Chern numbers). Then, we combine the Chern insulator and its time-reversal counterpart $H^{\ast}_{\rm QWZ}$ by the non-Hermitian coupling $i b \sigma_x \otimes \sigma_x$ and obtain the linear coupling that exhibits amplification of edge modes protected by exceptional points. Such Hamiltonian used in our prototypical model is expressed in the wavenumber space as
\begin{eqnarray}
 H(\mathbf{k}) &=& (u+\cos k_x + \cos k_y) I_2 \otimes \sigma_z + \sin k_y I_2 \otimes \sigma_y \nonumber\\
 &{}& + \sin k_x \sigma_z \otimes \sigma_x + ib \sigma_x \otimes \sigma_x \label{hamiltonian_TIL_wavenumber}
\end{eqnarray}
where $I_2$ is the $2\times2$ identity matrix and $\sigma_{x,y,z}$ are the Pauli matrices 
\begin{eqnarray}
 \sigma_x = \left(
  \begin{array}{cc}
   0 & 1 \\
   1 & 0 
  \end{array}
  \right),\ \sigma_y = \left(
  \begin{array}{cc}
   0 & -i \\
   i & 0 
  \end{array}
  \right),\ \sigma_z = \left(
  \begin{array}{cc}
   1 & 0 \\
   0 & -1 
  \end{array}
  \right).
\label{sigma}
\end{eqnarray}
We can obtain the following real-space description by the inverse Fourier transformation:
\begin{eqnarray}
 &{}& H_{jk}(\mathbf{x},\mathbf{x}') = \nonumber\\ 
 &{}& \left(u\delta_{\mathbf{x},\mathbf{x}'}+\frac{\delta_{\mathbf{x}+\mathbf{e}_{x},\mathbf{x}'}+\delta_{\mathbf{x}-\mathbf{e}_{x},\mathbf{x}'}+\delta_{\mathbf{x}+\mathbf{e}_{y},\mathbf{x}'}+\delta_{\mathbf{x}-\mathbf{e}_{y},\mathbf{x}'}}{2} \right) \nonumber\\
 &{}& \times (I_2 \otimes \sigma_z)_{jk} + \frac{\delta_{\mathbf{x}+\mathbf{e}_{y},\mathbf{x}'}-\delta_{\mathbf{x}-\mathbf{e}_{y},\mathbf{x}'}}{2i} (I_2 \otimes \sigma_y)_{jk}\nonumber\\ 
 &{}& + \frac{\delta_{\mathbf{x}+\mathbf{e}_{x},\mathbf{x}'}-\delta_{\mathbf{x}-\mathbf{e}_{x},\mathbf{x}'}}{2i} (\sigma_z \otimes \sigma_x)_{jk} + ib \delta_{\mathbf{x},\mathbf{x}'} (\sigma_x \otimes \sigma_x)_{jk}, \nonumber\\ \label{hamiltonian_TIL}
\end{eqnarray}
where $\delta_{\mathbf{x},\mathbf{x}'}$ represents the Kronecker delta and $\mathbf{e}_{x,y}$ are the lattice vectors in the $x$ or $y$ direction. $i b \sigma_x \otimes \sigma_x$ represents the non-Hermitian coupling between the QWZ model and its time-reversal counterpart. We note that there are coupling terms with imaginary coefficients, and these can be realized by doubling the internal degrees of freedom or time-delayed coupling as seen in electrical circuits \cite{Lee2018}.

We numerically calculate the dynamics of our model under the open boundary condition and confirm the emergence of the TSS, i.e., the synchronization (desynchronization) of the oscillators at the edge (in the bulk). Figures \ref{fig1}(c),(d) show the frequency distribution obtained from the numerical simulations. One can see that the edge oscillators exhibit constant and homogeneous frequencies, indicating their frequency-synchronization, while the bulk ones oscillate at time- and space-varying frequencies. It is noteworthy that the fluctuations of natural frequencies are irrelevant to the TSS (see Appendix \ref{appx_C} for the related numerical calculation). The existence of TSS under the inhomogeneous natural frequencies indicates the robustness of TSS guaranteed by its topological nature. 

We carefully examine the robustness of TSS by calculating the dynamics of our model at different strengths of the linear coupling and inhomogeneity of the natural frequencies. Figure \ref{fig2} shows the phase diagram determined on the basis of frequency fluctuations of the edge and bulk oscillators, which signifies the presence/absence of the TSS. If the inhomogeneity of the natural frequencies is large and/or the linear coupling is weak, we obtain a fully desynchronized state. We note that the parameter where the TSS disappears corresponds to the point at which the dispersion relation of the Hamiltonian becomes gapless due to strong disorders and thus can change its topology. Therefore, the result in Fig.~\ref{fig2} supports the argument that TSS is topologically protected. We can also analyze the robustness of TSS by constructing a one-dimensional model imitating the dynamics of the synchronized edge oscillators in our model (see Appendix \ref{appx_D} and Eqs.~\eqref{1d-chain}, \eqref{1d-chainH}).

Here, we emphasize that the TSS is defined as the coexistence of synchronized edge oscillators and desynchronized bulk oscillators with a topological origin. If we set $\alpha$ negative in our model \eqref{model1}, we can damp the bulk oscillations (see Appendix \ref{appx_B}), while the edge ones still exhibit synchronized oscillations (note that a similar synchronized state is observed in previous research \cite{Wachtler2020}). However, such synchronization is not a TSS defined here because the bulk oscillators do not exhibit (chaotic) self-oscillations. Using another Hamiltonian as linear couplings, we can also realize cluster synchronization where the edge oscillators and bulk ones oscillate at different frequencies (see Appendix \ref{appx_B}), which is also out of the range of the TSS.

\begin{figure*}[t]
  \includegraphics[width=160mm,bb=0 0 1250 345,clip]{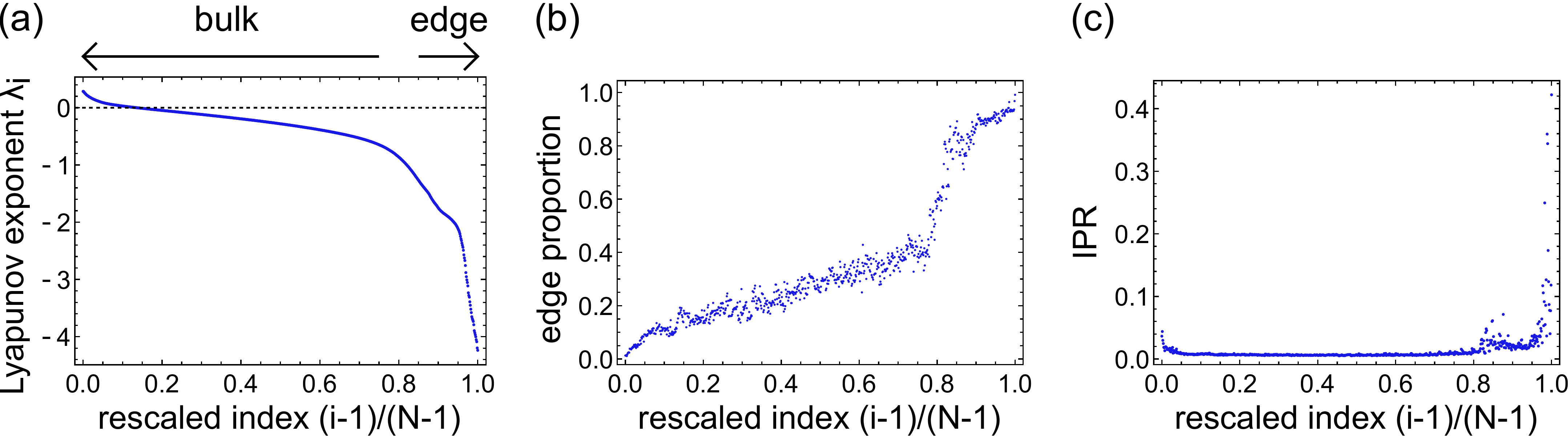}
\caption{\label{fig3} Lyapunov spectrum and the indices of the localization of the Lyapunov vectors. (a) Lyapunov exponents calculated from the first model of the TSS (cf. Eqs.~\eqref{model1}, \eqref{hamiltonian_TIL_wavenumber} and Fig.~\ref{fig1}). They are arranged in descending order. The index of the Lyapunov exponent is rescaled for the maximum to be unity. The dashed horizontal line corresponds to zero Lyapunov exponents. Some of the Lyapunov exponents are positive, i.e., placed above the dashed line, which indicates the chaotic behavior of the oscillators. The parameters used are $u=-1$, $b=0.5$, $\alpha=0.5$, $\beta=1$, $\omega_0=0.2$, and $\Delta\omega=0.2$. (b) Proportions of the edge amplitudes $P_{\rm edge}$ of the Lyapunov vectors in the TSS. We set the relative index of the Lyapunov vector to be the same as panel (a). The edge proportions rise steeply around the relative index $0.8$. Therefore, the Lyapunov vectors of the small indices spread in the bulk, while those of the large indices are localized at the edge of the system. This increase accompanies the decrease of the Lyapunov exponents in panel (a), which indicates that the edge localized Lyapunov vectors are dissipative modes. (c) Inverse participation ratios (IPRs) of the Lyapunov vectors. We set the relative index of the Lyapunov vector to be the same as in panel (a). The IPRs show a small rise around the relative index $0.8$, corresponding to the edge localization of the Lyapunov vectors. The IPRs also increase steeply at a larger relative index, which indicates that a few Lyapunov vectors are strongly localized at a few edge sites.
}
\end{figure*}

\subsection{Chaos in bulk oscillators and the edge-localized Lyapunov vectors}
We next show that the bulk oscillators exhibit chaotic dynamics. Chaos is characterized by the extreme sensitivity to initial conditions, i.e., exponential growth of the initial difference between trajectories. The infinitesimal rate of such exponential divergence is called the Lyapunov exponent \cite{Oseledets1968}, and its positivity gives a defining feature of chaos. We numerically calculate the Lyapunov exponents of the first model. Figure \ref{fig3}(a) shows the obtained ones and the existence of positive Lyapunov exponents indicating the chaos. We note that Stuart-Landau oscillators can exhibit two types of chaos, namely, amplitude chaos and phase chaos \cite{Shraiman1992}. In this model, we can see amplitude-chaotic behaviors, e.g., phase slips via zero amplitude (see Appendix \ref{appx_F}).

Lyapunov exponents also tell us the effective degree of freedom of chaotic attractor, Lyapunov dimension defined as $D_{\rm L}=\sum_{i\leq M} \lambda_i / | \lambda_{M+1} | + M$, where $\lambda_i$ represents Lyapunov exponents arranged in descending order, and $M$ is the smallest integer that satisfies $\sum_{i \leq M+1} \lambda_i < 0$. Previous studies \cite{Wolfrum2011,Hohlein2019} have revealed that the Lyapunov dimension corresponds to the number of the desynchronized oscillators in the chimera state. In our first model, we obtain the Lyapunov dimension $D_{\rm L}=254.568$, which is almost $90$ percent of the total degree of freedom of oscillators except for the first and second oscillators from the edge, $D_{\rm bulk}=288$. Our Lyapunov analysis thus clearly indicates that the bulk oscillators are desynchronized and exhibit chaotic dynamics.

Lyapunov vectors can be used to know the local geometric information of chaotic attractors \cite{Ginelli2013}. Specifically, perturbation parallel to a Lyapunov vector is amplified or attenuated in a forward and backward process expressed as $|\delta \vec{Z}(\pm t)| \sim |\delta \vec{Z}(0)|e^{\pm \lambda_i t}$, where $\lambda_i$ is the associated Lyapunov exponent. In our first model, we reveal that the Lyapunov vectors corresponding to the small Lyapunov exponents are localized at the edge of the system, thus realizing topological edge modes in nonlinear systems. To clearly show such localization to the edge, we define the following index of the proportion of amplitude of the edge oscillators,
\begin{equation}
 P_{\rm edge} = \sum_{i\in {\rm edge}} |v_i|^2, \label{edge_proportion}
\end{equation}
where $v_i$ is the $i$th component of the Lyapunov vector and we sum the squares of the components corresponding to the edge oscillators. This index takes a large value when a Lyapunov vector is localized furthermost to the edge. Figure \ref{fig3}(b) shows the proportions of the edge oscillators of the Lyapunov vectors in our first model. One can see the steep increase in the index, which indicates that the first about $80$ percent of Lyapunov vectors are extended to the bulk, while the others are localized to the edge. We demonstrate such localization and delocalization of the Lyapunov vectors by plotting them in the real space (see Appendix \ref{appx_G}). It is noteworthy that the Lyapunov exponents decrease around the index where the edge proportions increase. Thus, we expect that perturbation to the edge oscillators is attenuated, which implies stability of their synchronization.

We also evaluate the degree of localization in the entire Lyapunov vectors by calculating inverse participation ratios (IPRs) defined as
\begin{equation}
 {\rm IPR} = \sum_{i} |v_i|^4. \label{ipr}
\end{equation}
The IPR becomes large when a small number of components of a Lyapunov vector exhibit large amplitudes, that is, it is localized at only a few sites. Figure \ref{fig3}(c) shows the IPR of the Lyapunov vectors obtained in our first model. There is a slight increase in the IPR around the relative index of the Lyapunov vectors $0.8$ as a result of the localization to the edge. The smallness of the increase in the IPR indicates that most of the edge-localized Lyapunov vectors spread at many edge sites, which is reminiscent of conventional topological edge modes in linear systems. We can find another region where the IPR steeply increases. Those large IPRs imply the strong localization of the Lyapunov vectors to a few edge sites. We note that the Lyapunov exponents also decrease in this region, indicating the strong damping of the perturbation corresponding to these localized Lyapunov vectors. While the proportions of the edge oscillators in Eq.~\eqref{edge_proportion} reveal the existence of the edge-localized Lyapunov vectors, IPRs allow us to classify such edge-localized Lyapunov vectors into two groups, the extended ones corresponding to conventional edge modes and the strongly localized ones induced by nonlinearity.

\begin{figure*}[t]
  \includegraphics[width=160mm,bb=0 0 1125 610,clip]{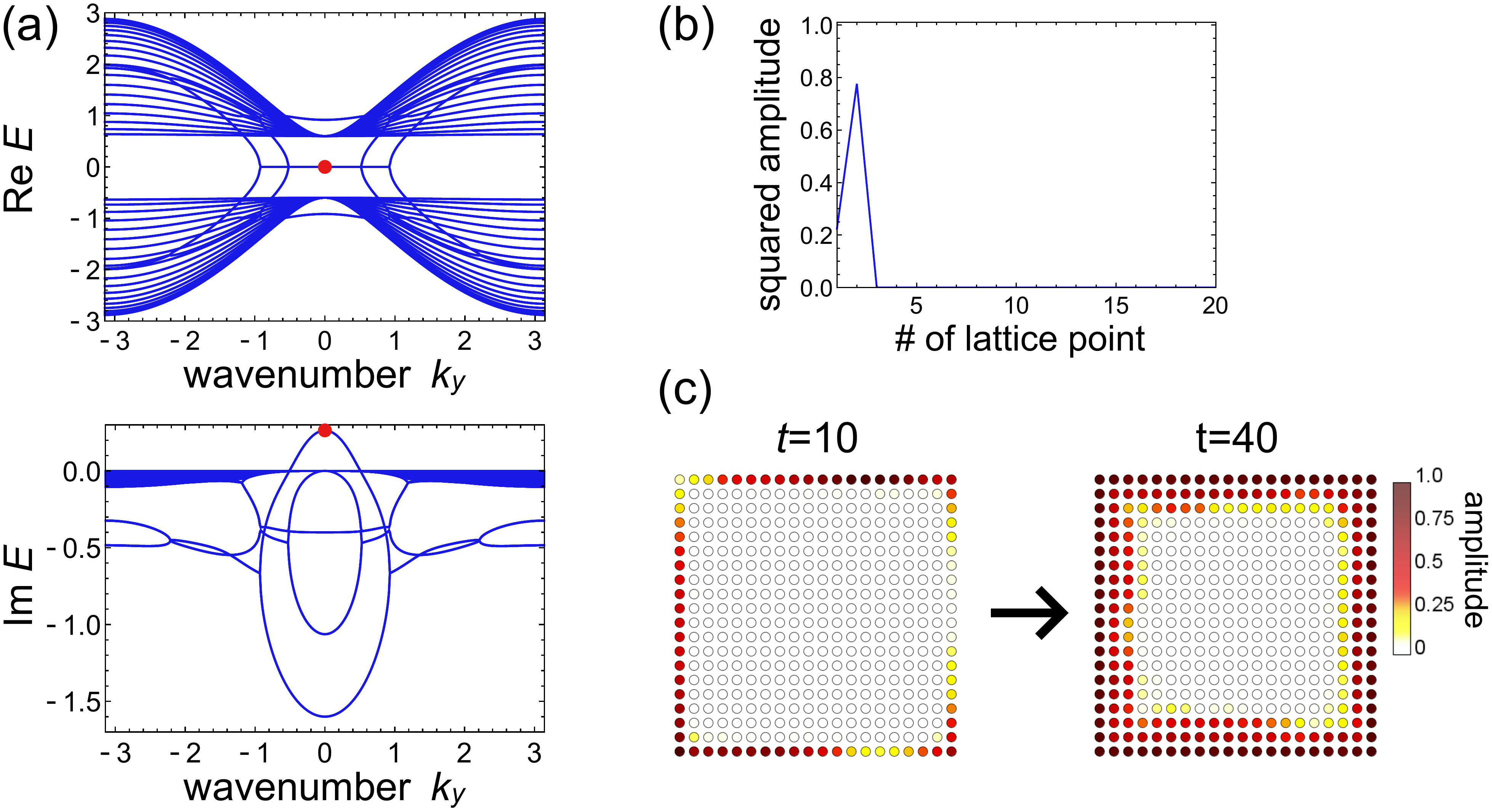}
\caption{\label{fig4} Dispersion relations and the dynamics demonstrating a spontaneous nonlinearity-induced boundary. (a) Edge dispersions of the Hamiltonian $\tilde{H}$ of topological insulator laser utilized in our first model under the existence of edge-localized on-site loss. We impose the open (periodic) boundary condition in the $x$ ($y$) direction. The edge-localized on-site loss corresponds to the effect of nonlinear terms in the case that the edge modes of the original Hamiltonian $H$ are amplified. We find the extra number of gapless modes. Those extra topological modes appear due to the effective boundary created by the nonlinearity-induced edge-localized loss. The parameters used are $u=-1$, $b=0.8$, $\alpha=0$, $\beta=1$, $\omega_0=0$, and $\Delta\omega=0$. (b) Density distribution of one of the extra topological modes corresponding to the red point in panel (a). One can see that this eigenstate exhibits the largest amplitude at the second site, which implies that it is localized at the effective boundary emerging between the first and second sites. (c) Dynamical formation of the extra topological boundary modes. We set the coefficient of the nonlinear term much smaller than that of the linear coupling. The snapshots at the time $t=10$ ($t=40$) are shown in the left (right) figure. When the edge oscillators are not fully amplified (corresponding to $t=10$), the edge-localized loss is not large enough to create an effective boundary. After the edge oscillators are fully amplified, the inside oscillators begin to largely oscillate, which implies the appearance of the extra topological boundary modes.
}
\end{figure*}

The edge-localized Lyapunov vectors and their negative Lyapunov exponents are related to the edge modes of the effective Hamiltonian obtained via the linearization of the equation. We note that if there are no nonlinear terms (i.e., $\beta=0$ in our model), Lyapunov exponents and vectors are identical to the imaginary parts of the eigenvalues and eigenvectors of the effective Hamiltonian, respectively. It is noteworthy that non-Hermiticity is essential to realize nonzero imaginary parts of eigenvalues, and nonlinear oscillators ubiquitously exhibit such dissipative and injective effects. In nonlinear systems, we can obtain the effective Hamiltonian by linearizing the equation around the state at each moment. The linear equation governed by the sequence of such effective Hamiltonians describes the time evolution of the difference between the perturbed trajectory and the original one. We numerically check that the effective Hamiltonian of our first model exhibits edge modes with negative imaginary parts of the eigenvalues (see Appendix \ref{appx_E}), which indicates the (short-term) stability of the synchronization of the edge oscillators. We note that nonlinear terms lead to a random on-site loss in the effective Hamiltonian. However, topological modes are robust against such disorders and thus still appear from the present effective Hamiltonian. 

We also note that some previous studies \cite{Hata2014,Hata2017} discuss the role of the localization of eigenvectors of the linearized equation in the localized patterns. However, the origins of the localization are different between such localized patterns and the TSS. The localized patterns in the previous research rely on the real-space configuration of the oscillator networks and are explained from the perturbation analysis, while the TSS utilizes its nontrivial topology in the wavenumber space. It is also noteworthy that the TSS cannot be observed in topologically trivial systems (cf. Appendix \ref{appx_I}), which indicates the crucial role of the wavenumber topology in the TSS.

\begin{figure*}[t]
  \includegraphics[width=160mm,bb=0 0 920 275,clip]{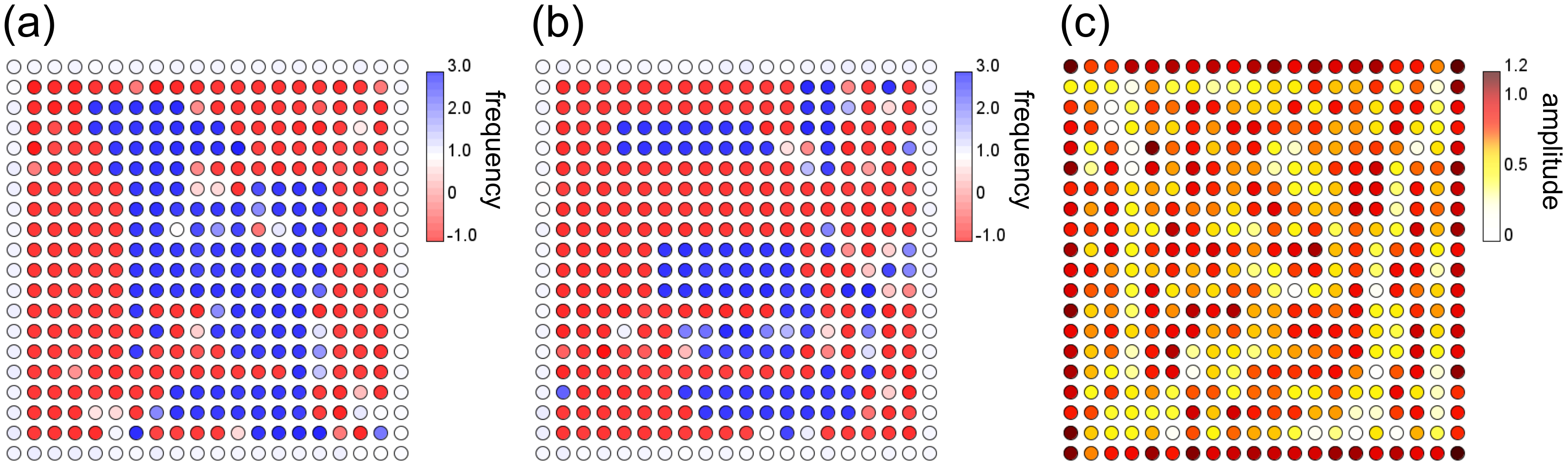}
\caption{\label{fig5} TSS induced by conventional topological edge modes. (a),(b) Frequency of the first component of oscillators at each site. We numerically calculate the dynamics of the model of the TSS \eqref{model1} with different topological linear couplings \eqref{hamiltonian_conventionaltopo_TIL} from those in Fig.~\ref{fig1}, where the edge modes are protected by conventional bulk topology. Panels (a) and (b) are the snapshots at the time $t=1000$ and $t=2000$ for each. Empty sites represent the oscillators oscillating around their natural frequencies. We confirm the coexistence of the frequency-synchronized edge oscillators and the desynchronized bulk oscillators. The parameters used are $u=-1$, $u'=0.02$, $a=2$, $b=0.5$, $\alpha=0.5$, $\beta=1$, $\omega_0=1$, and $\Delta\omega=0.2$. (c) Amplitudes of the oscillators obtained from the numerical simulation of the dynamics of the model considered in panels (a) and (b). We check that the amplitudes of most of the edge oscillators are larger than those of the bulk oscillators. The same parameters as panels (a) and (b) are used in the calculation.
}
\end{figure*}

\subsection{Emergence of extra topological modes by nonlinearity-induced boundary}
We find that nonlinearity can induce unconventional extra edge modes, which should be the origin of the synchronization of the second oscillators from the edge as discussed in the previous section. Namely, nonlinear systems exhibiting self-excited vibration, including our model, can generate effective boundaries and extra topological modes regardless of the initial conditions. Previous studies \cite{Lumer2013,Tuloup2020} have discussed that one can create effective boundaries by externally preparing a localized initial state, whereas effective boundaries in our first model spontaneously appear by utilizing the lasing edge modes. 

To show the existence of such extra boundary modes, we rewrite the equation of our model \eqref{model1} as $ \dot{Z}_j(\mathbf{x}) = -i\sum_k \tilde{H}_{jk}(\mathbf{Z};\mathbf{x},\mathbf{x}') Z_k(\mathbf{x}')$, where $\tilde{H}_{jk}(\mathbf{Z};\mathbf{x},\mathbf{x}') =H_{jk}(\mathbf{x},\mathbf{x}') +(-\omega_j(\mathbf{x})+i\alpha-i\beta |Z_j(\mathbf{x})|^2)\delta_{jk}\delta_{\mathbf{x},\mathbf{x}'}$ represents the state-dependent Hamiltonian which governs the time evolution at each moment. This state-dependent Hamiltonian includes an on-site loss at edge sites, because lasing edge modes of the original Hamiltonian $H$ of our first model amplify the edge oscillators as shown in Fig.~\ref{fig1}(e). We note that recent research \cite{Zhao2019} has shown that on-site loss creates an effective boundary, and topological boundary modes can appear at the boundary between the regions with gain and loss. While the previous study has focused on linear dynamics with an externally introduced boundary, in our model the on-site loss is an emergent feature in the sense that it is spontaneously induced by the nonlinearity.

To explicitly demonstrate the presence of such extra boundary modes, we numerically diagonalize the state-dependent Hamiltonian $\tilde{H}$ with edge-localized loss. Here, we consider the on-site loss that should be induced by lasing edge modes of the original Hamiltonian $H$ used in Fig.~\ref{fig1}. Figure \ref{fig4}(a) shows the dispersion relation of this state-dependent Hamiltonian. We obtain eight gapless modes, which are the twice number of gapless modes compared to the original Hamiltonian without on-site loss (see Appendix \ref{appx_H} for the dispersion relation of the original Hamiltonian). We find positive imaginary parts of eigenvalues and the localization to the second site of the corresponding eigenvectors, which leads to the amplification of the second oscillators from the edge.

We also directly confirm the emergence of the extra boundary modes from the numerical simulation. We set the coefficient of nonlinear terms to be small compared to those of linear coupling $H$. Figure \ref{fig4}(c) presents the snapshots of the relative amplitude of each site obtained from the simulation. In the beginning, only the outermost oscillators have large amplitudes, while after a sufficiently long time the inner ones also begin to oscillate with large amplitudes. This behavior represents the emergence of the extra boundary modes after the nonlinearity-induced loss grows enough to balance with the linear coupling and create an effective boundary.

The presence of the extra boundary modes above can alter the number of the synchronized oscillators in the TSS. Edge-localized on-site loss also exists in the effective Hamiltonian obtained from the linearization of the equation around the state at each moment. Therefore, the number of topological modes can increase in such an effective Hamiltonian via the same mechanism as in the state-dependent Hamiltonian. We confirm the existence of the extra boundary modes and negativity of the imaginary parts of their eigenvalues (see Appendix \ref{appx_E}), which leads to the increase in the number of dissipative Lyapunov vectors.

\begin{figure*}[t]
\includegraphics[width=160mm,bb=0 0 1250 345,clip]{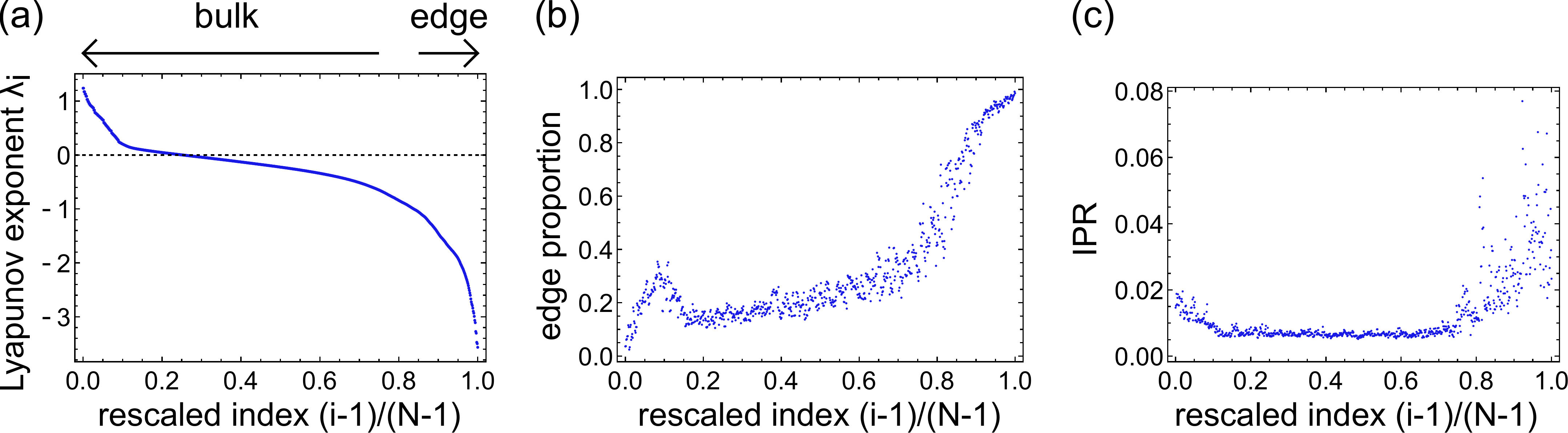}
\caption{\label{fig6} Lyapunov analysis of the model of TSS utilizing lasing edge modes protected by conventional bulk topology. (a) Lyapunov exponents of the model with the linear couplings exhibiting edge modes protected by conventional bulk topology \eqref{hamiltonian_conventionaltopo_TIL}. The index of each Lyapunov exponent is rescaled for the maximum to be unity. We obtain positive Lyapunov exponents (above the dashed line) as in Fig.~\ref{fig3}, which indicates the chaos of the bulk oscillators. The parameters used are $u=-1$, $u'=0.02$, $a=2$, $b=0.5$, $\alpha=0.5$, $\beta=1$, $\omega_0=0.2$, and $\Delta\omega=0.2$. (b) Proportions of the edge amplitudes $P_{\rm edge}$ \eqref{edge_proportion} of the Lyapunov vectors. We set the relative index of the Lyapunov vector to be the same as panel (a). The edge proportions rise steeply around the relative index $0.8$, which indicates that the Lyapunov vectors of the small indices spread in the bulk, while those of the large indices are localized at the edge of the system. (c) Inverse participation ratios (IPRs) of the Lyapunov vectors. We set the relative index of the Lyapunov vector to be the same as in panel (a). The IPRs increase around the relative index $0.8$, corresponding to the edge localization of the Lyapunov vectors. Since all the IPRs are less than $0.08$, the strongly localized modes are absent unlike Fig.~\ref{fig3}. 
}
\end{figure*}

\section{TSS utilizing conventional bulk topology\label{section4}}
\subsection{Model and its dynamics}

While we utilize the Hamiltonian featuring exceptional edge modes in our first model (Eq.~\eqref{hamiltonian_TIL_wavenumber} and Fig.~\ref{fig1}), we can also realize TSS by using a Hamiltonian of topological edge modes protected by conventional bulk topology. To demonstrate this, we consider another non-Hermitian Hamiltonian exhibiting lasing edge modes, which is described in the real space as
\begin{eqnarray}
 &{}& H_{jk}(\mathbf{x},\mathbf{x}') = \nonumber\\
 &{}& \left(u\delta_{\mathbf{x},\mathbf{x}'}+\frac{\delta_{\mathbf{x}+\mathbf{e}_{x},\mathbf{x}'}+\delta_{\mathbf{x}-\mathbf{e}_{x},\mathbf{x}'}+\delta_{\mathbf{x}+\mathbf{e}_{y},\mathbf{x}'}+\delta_{\mathbf{x}-\mathbf{e}_{y},\mathbf{x}'}}{2} \right) \nonumber\\
 &{}& \times\left[ \frac{a+1}{2}(I_2 \otimes \sigma_z)_{jk} + \frac{a-1}{2}(\sigma_z \otimes \sigma_z)_{jk} \right] \nonumber\\
 &{}& + \frac{\delta_{\mathbf{x}+\mathbf{e}_{y},\mathbf{x}'}-\delta_{\mathbf{x}-\mathbf{e}_{y},\mathbf{x}'}}{2i} \nonumber\\
 &{}& \times \left[ \frac{a+1}{2}(I_2 \otimes \sigma_y)_{jk} + \frac{a-1}{2}(\sigma_z \otimes \sigma_y)_{jk} \right] \nonumber\\
 &{}& + \frac{\delta_{\mathbf{x}+\mathbf{e}_{x},\mathbf{x}'}-\delta_{\mathbf{x}-\mathbf{e}_{x},\mathbf{x}'}}{2i} \nonumber\\
 &{}& \times \left[ \frac{a+1}{2}(I_2 \otimes \sigma_x)_{jk} + \frac{a-1}{2}(\sigma_z \otimes \sigma_x)_{jk} \right] \nonumber\\
 &{}& + ib \delta_{\mathbf{x},\mathbf{x}'} (\sigma_x \otimes I_2)_{jk} + iu' \delta_{\mathbf{x},\mathbf{x}'} (\sigma_z \otimes I_2)_{jk} \label{hamiltonian_conventionaltopo_TIL},
\end{eqnarray}
where $a$, $b$, $u$, and $u'$ are real parameters, and $\sigma_i,\delta_{\mathbf{x},\mathbf{x}'}$ represent the $i$th component of the Pauli matrices and the Kronecker delta. As in our first model, this Hamiltonian is constructed from two layers of the QWZ model (a Chern insulator model), $H_{\rm QWZ}$ and $aH_{\rm QWZ}$ combined by the non-Hermitian coupling $ib\sigma_x \otimes I_2$. Thus, we can rewrite the Hamiltonian as 
\begin{eqnarray}
H = \left(
  \begin{array}{cc}
   aH_{\rm QWZ} + iu' I_2 & ib I_2 \\
   ib I_2 & H_{\rm QWZ} - iu' I_2
  \end{array}
  \right),
\end{eqnarray}
where $\pm iu' I_2$ represent the additional gain and loss. The modification of the hopping amplitude (parametrized by $a$) and non-Hermitian couplings enable the amplification of only the edge modes derived from the nontrivial topology of $H_{\rm QWZ}$.

Figures \ref{fig5}(a),(b) show the frequency distributions of nonlinear oscillators with such topological linear couplings. We can confirm the emergence of the TSS, i.e., the edge oscillators are frequency-synchronized while the bulk ones are chaotic. We can also confirm the amplification of the edge oscillators in Fig.~\ref{fig5}(c) as in Fig.~\ref{fig1}(e).

\subsection{Lyapunov analysis}
\noindent
We also calculate the Lyapunov exponents and vectors to confirm the chaos of the bulk oscillators and the existence of the edge-localized Lyapunov vectors (see Appendix \ref{appx_A} for the detailed numerical method). Figure \ref{fig6}(a) shows the Lyapunov exponents of the model considered here. There are positive Lyapunov exponents which indicate the chaos of the bulk oscillators. 

Figure \ref{fig6}(b) shows the proportions of the edge amplitudes \eqref{edge_proportion} of the Lyapunov exponents. We can confirm the existence of the edge-localized Lyapunov vectors as in the model using exceptional edge modes \cite{Sone2020} (cf. Eq.~\eqref{hamiltonian_TIL_wavenumber} and Fig.~\ref{fig3}(b)). However, the behavior of the inverse participation ratios (IPRs) \eqref{ipr} shown in Fig.~\ref{fig6}(c) is different from that in Fig.~\ref{fig3}(c), because there are no steep increases in the IPRs. Thus, the strongly localized Lyapunov vectors disappear from the model considered here. This is because the nonlinear effects are not large enough to generate such strongly localized modes. We also note that nonlinearity-induced boundary modes are absent in this model as can be seen in Fig.~\ref{fig5}. This may also be due to the smallness of nonlinearity.

\begin{figure}[t]
\includegraphics[width=86mm,bb=0 0 685 520,clip]{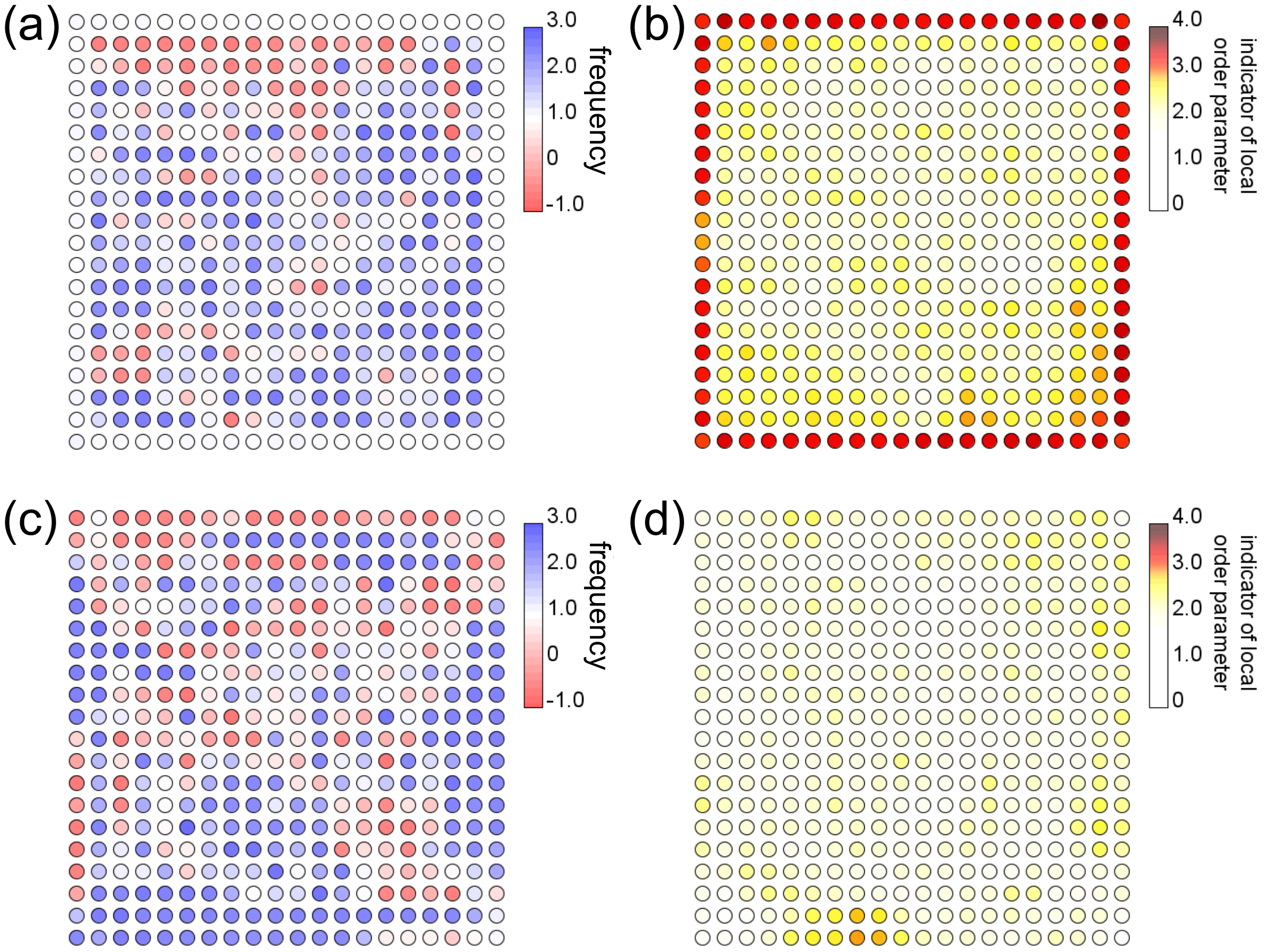}
\caption{\label{fig7} Frequency and amplitude distributions of the model utilizing Hermitian linear couplings. We numerically simulate the model utilizing Hermitian linear couplings \eqref{Hermitian_coupling}. The frequency ((a) and (c)) and the indicator of the local order parameter $M(x,y)$ \eqref{orderparameter_component} ((b) and (d)) at each site are shown. Panels (a) and (b) ((c) and (d)) present the frequencies and the indicators of the local order parameters of the first (second) component of oscillators at time $t=100$. The edge oscillators at the first components of the lattice points exhibit the homogeneous and constant frequency and the large indicators of the local order parameters, which implies their frequency-synchronization. In contrast, the second components of oscillators are desynchronized. The parameters used are $u=-1$, $b=0.5$, $\alpha=1$, $\beta=1$, $\omega_0=1$, and $\Delta\omega=0.2$.
}
\end{figure}

\section{TSS utilizing Hermitian linear couplings\label{section5}}
\subsection{Model and its dynamics}

\begin{figure*}[t]
\includegraphics[width=160mm,bb=0 0 1300 350,clip]{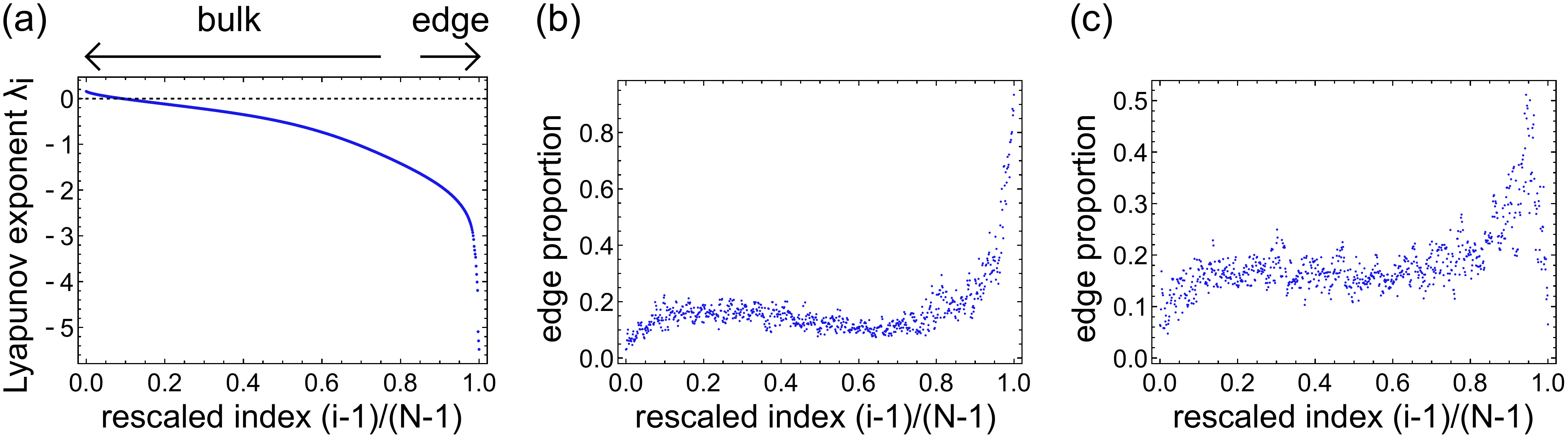}
\caption{\label{fig8} Lyapunov analysis of the model utilizing Hermitian linear couplings. (a) Lyapunov exponents of the model with the Hermitian linear couplings \eqref{Hermitian_coupling} plotted in descending order. The index of each Lyapunov exponent is rescaled for the maximum to be unity. Positive Lyapunov exponents (above the dashed line) exist as in Fig.~\ref{fig3}, indicating the chaos of the bulk oscillators. The parameters used are $u=-1$, $b=0.5$, $\alpha=1$, $\beta=1$, $\omega_0=0.2$, and $\Delta\omega=0.2$. (b),(c) Proportions of the edge amplitudes $P_{\rm edge}$ \eqref{edge_proportion} of the Lyapunov vectors. We plot those of the first and fourth components in panel (b) and those of the second and third components in panel (c). We set the relative index of the Lyapunov vector to be the same as panel (a). The edge proportions of the first and fourth components rise steeply around the relative index $0.95$, which indicates that the Lyapunov vectors of the large indices are localized at the first and fourth components of the edge sites. On the other hand, the edge proportion of the second and third components decreases in that region around the relative index $0.95$. Therefore, the second and third components of the edge oscillators exhibit chaotic behavior.
}
\end{figure*}

We also find that the TSS can emerge in nonidentical nonlinear oscillators with linear couplings described by a Hermitian Hamiltonian. To construct a model of such TSS, we consider the unitary transformation of the Hamiltonian considered in our first model \eqref{hamiltonian_TIL_wavenumber} and obtain 
\begin{eqnarray}
 H(\mathbf{k}) &=& (u+\cos k_x + \cos k_y) \sigma_x \otimes \sigma_z + \sin k_y I_2 \otimes \sigma_y \nonumber\\
 &{}& + \sin k_x I_2 \otimes \sigma_x + i b \sigma_z \otimes \sigma_z, \label{Hermitian_coupling}
\end{eqnarray}
where $I_2$ is the $2\times2$ identity matrix and $\sigma_{x,y,z}$ are the Pauli matrices. This Hamiltonian is described in the real space as
\begin{eqnarray}
 &{}& H_{jk}(\mathbf{x},\mathbf{x}') = \nonumber\\
 &{}& \left(u\delta_{\mathbf{x},\mathbf{x}'}+\frac{\delta_{\mathbf{x}+\mathbf{e}_{x},\mathbf{x}'}+\delta_{\mathbf{x}-\mathbf{e}_{x},\mathbf{x}'}+\delta_{\mathbf{x}+\mathbf{e}_{y},\mathbf{x}'}+\delta_{\mathbf{x}-\mathbf{e}_{y},\mathbf{x}'}}{2} \right) \nonumber\\
 &{}& \times (\sigma_x \otimes \sigma_z)_{jk} + \frac{\delta_{\mathbf{x}+\mathbf{e}_{y},\mathbf{x}'}-\delta_{\mathbf{x}-\mathbf{e}_{y},\mathbf{x}'}}{2i} (I_2 \otimes \sigma_y)_{jk} \nonumber\\
 &{}& + \frac{\delta_{\mathbf{x}+\mathbf{e}_{x},\mathbf{x}'}-\delta_{\mathbf{x}-\mathbf{e}_{x},\mathbf{x}'}}{2i} (I_2 \otimes \sigma_x)_{jk} + ib \delta_{\mathbf{x},\mathbf{x}'} (\sigma_z \otimes \sigma_z)_{jk}, \nonumber\\ \label{Hermitian_coupling_real}
\end{eqnarray}
where $\delta_{\mathbf{x},\mathbf{x}'}$ is the Kronecker delta and $\mathbf{e}_{x,y}$ are the lattice vectors in the $x$ or $y$ direction. Here, we perform the unitary transformation of the Hamiltonian in our first model \eqref{hamiltonian_TIL_wavenumber} to convert the non-Hermitian term $ib \delta_{\mathbf{x},\mathbf{x}'} (\sigma_z \otimes \sigma_z)_{jk}$ into a diagonal one. Such diagonal non-Hermitian terms can be introduced as the modulation of the parameter $\alpha$ in the Stuart-Landau equation (cf. Eq.~\eqref{model1}). Thus, we can consider Stuart-Landau oscillators with this linear coupling as nonidentical oscillators whose linear coupling is Hermitian. 

We numerically calculate the dynamics of such coupled nonidentical oscillators. Figure \ref{fig7}(a) shows the frequency distribution of oscillators at the first components of lattice points. We confirm the TSS, i.e., the edge oscillators synchronize while the bulk ones exhibit chaos. Unlike the other models of TSS, however, the second components of the oscillators exhibit chaotic motion even at the edge of the system (Fig.~\ref{fig7}(c)). Such synchronization and desynchronization of the edge oscillators can be confirmed from the indicator of the local order parameter restricted to the first and fourth (second and third) components
\begin{widetext}
\begin{equation}
 M(x,y;t) = \frac{1}{2T} \int_{t-T}^{t} dt' \sum_{k=1,4\ (\text{or}\ k=2,3)} |Z_k(x,y)+Z_k(x+1,y)+Z_k(x-1,y)+Z_k(x,y+1)+Z_k(x,y-1)|, \label{orderparameter_component}
\end{equation}
\end{widetext}
as shown in Fig.~\ref{fig7}(b),(d). The first components of edge oscillators exhibit large indicators of the local order parameter indicating their synchronization, while those of the second components have small values.

\subsection{Lyapunov analysis}
We also calculate the Lyapunov exponents and vectors to confirm the chaos of the bulk oscillators as in Figs.~\ref{fig3}, \ref{fig6}. Figure \ref{fig8}(a) shows the existence of positive Lyapunov exponents, which implies the chaos in this system. We compare the proportion of amplitude of the edge oscillators of the Lyapunov vectors \eqref{edge_proportion} between the first and fourth components and the second and third ones, as shown in Fig.~\ref{fig8}(b),(c). The Lyapunov vectors corresponding to small Lyapunov exponents are localized at the first and fourth components of the edge sites. On the other hand, the amplitudes of Lyapunov exponents at the second and third components are small at the large indices. Thus, the first and fourth components of edge oscillators are synchronized, while the other oscillators exhibit chaos.

\begin{figure}[t]
\includegraphics[width=86mm,bb=0 0 815 625,clip]{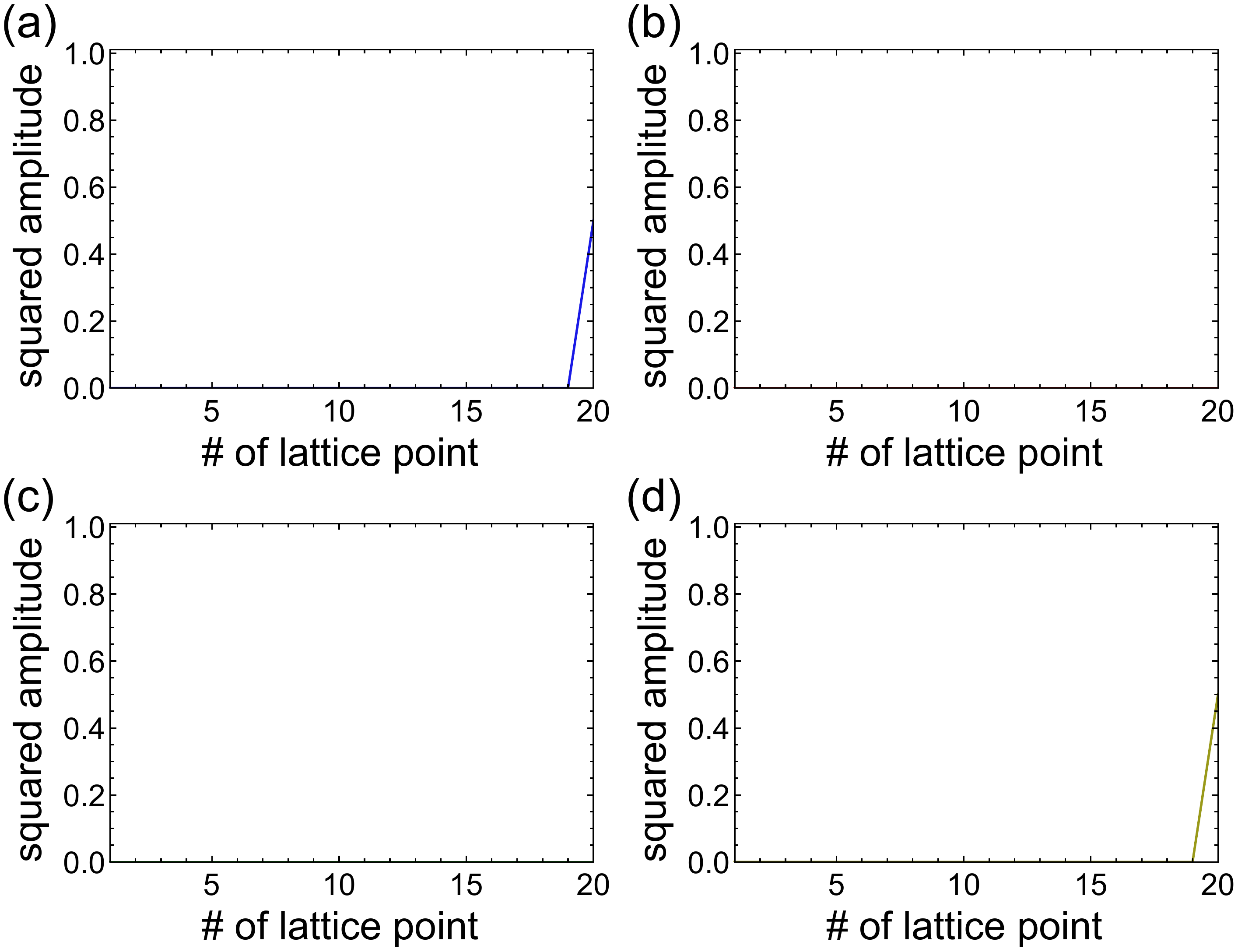}
\caption{\label{fig9} Amplitude distribution of the lasing edge modes. The squared amplitude of the lasing edge mode of the linear couplings \eqref{Hermitian_coupling} is shown. The wavenumber is $k=0$ and the eigenvalue is $E=ib$. Panels (a), (b), (c), and (d) show the amplitude distributions of the first, second, third, and fourth components for each. The amplitudes of lasing edge modes are localized at the first and fourth components. The parameters used are $u=-1$ and $b=0.5$.
}
\end{figure}

Desynchronization of the second and third components of the edge oscillators is explained from the amplitude distribution of the lasing edge modes shown in Fig.~\ref{fig9}. The lasing edge modes with the maximum imaginary part of the eigenvalue exhibit zero amplitude at the second and third components of the lattices and edge-localized amplitudes at the first and fourth components. Therefore, the lasing edge modes obtained from the linear coupling \eqref{Hermitian_coupling} only amplify the first and fourth components of the edge oscillators. Since such amplification leads to the different behavior between the edge and bulk oscillators (see also Appendix \ref{appx_H}), only the first and fourth components of the edge oscillators can synchronize, while the second and third components exhibit chaotic oscillations.

\section{Applications of TSS\label{section6}}
Damped bulk oscillators in the TSS will create an effective edge, along which the oscillators can be newly synchronized. We here propose several applications of such emerging synchronized oscillators, e.g., on-demand synchronization pattern designing and defect detection.

First, by placing damped oscillators into the bulk, one can arrange synchronized oscillators in an arbitrary pattern. We demonstrate such flexible pattern designing in Fig.~\ref{fig10} (see Appendix \ref{appx_A} for the numerical method to introduce damped oscillators). In Fig.~\ref{fig10}(b), we define the indicator of the local order parameter as
\begin{widetext}
\begin{equation}
 M(x,y;t) = \frac{1}{4T} \int_{t-T}^{t} dt' \sum_{k=1}^4 |Z_k(x,y)+Z_k(x+1,y)+Z_k(x-1,y)+Z_k(x,y+1)+Z_k(x,y-1)|, \label{orderparameter}
\end{equation}
\end{widetext}
which becomes large when the phase of the oscillator matches those of the nearest neighbors (we note that this includes information of all the components unlike that defined in Eq.~\eqref{orderparameter_component}). The oscillators around damped ones exhibit the homogeneous frequencies and the large values of the indicator $M$, thus being synchronized. We note that the synchronization pattern disappears in a topologically trivial system (see Appendix \ref{appx_I}), demonstrating the crucial role played by the topology of the linear coupling.

\begin{figure}[t]
  \includegraphics[width=86mm,bb=0 0 685 635,clip]{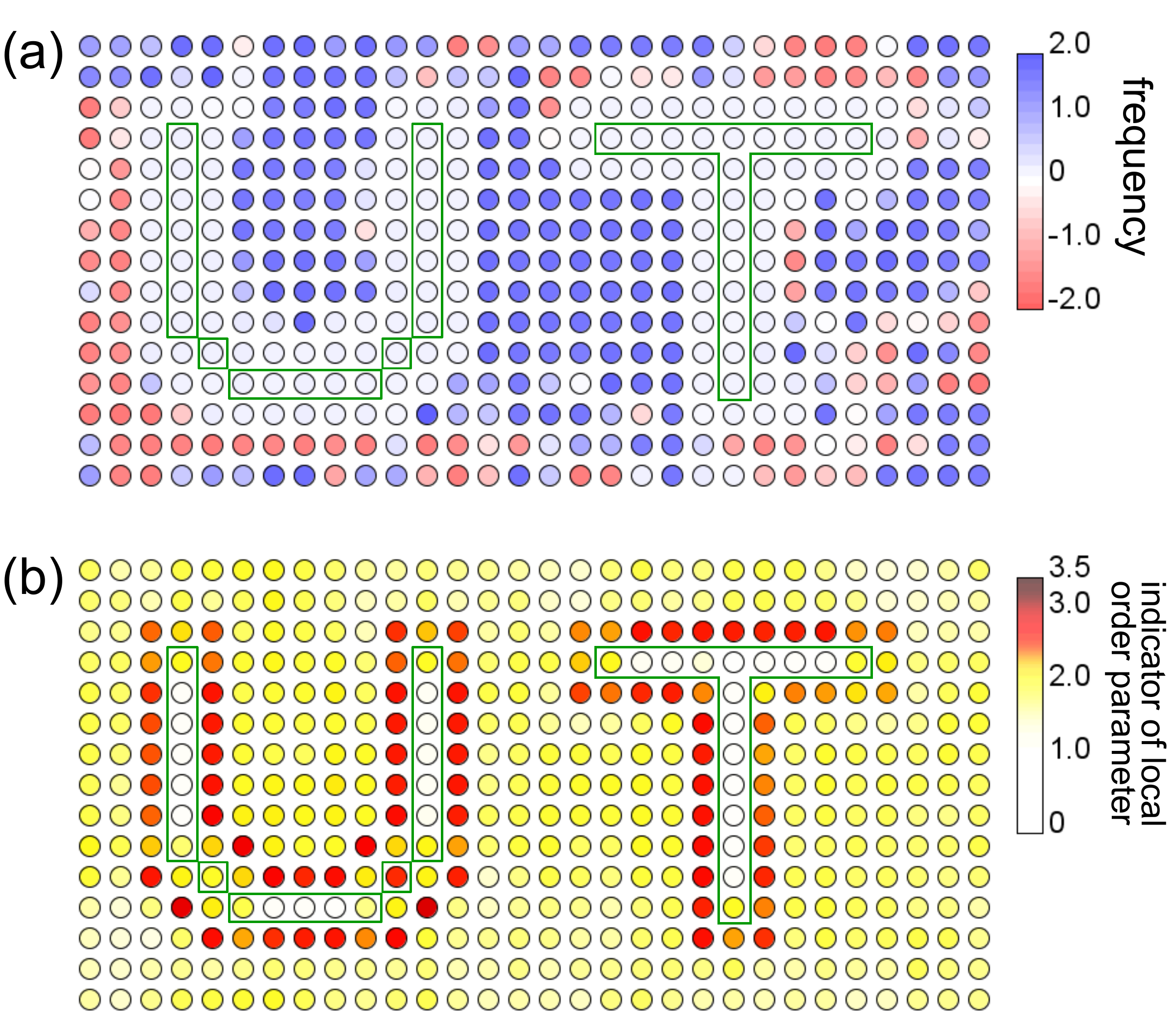}
\caption{\label{fig10} On-demand pattern designing of the synchronized oscillators. We numerically demonstrate that the on-demand pattern designing of the synchronized oscillators is possible by utilizing the TSS. To arrange the synchronized oscillators in the shape of characters, UT, we damp the oscillators encircled by the green boxes and numerically calculate the dynamics of the first model of topological synchronization (cf. Eqs.~\eqref{model1}, \eqref{hamiltonian_TIL_wavenumber} and Fig.~\ref{fig1}) under the periodic boundary condition. The parameters used are $u=-1$, $b=0.3$, $\alpha=0.5$, $\alpha_d=-100$, $\beta=1$, $\omega_0=0.1$, and $\Delta\omega=0.2$. In panel (a), the frequencies of the oscillators are shown. The oscillators around the damped ones exhibit a homogeneous frequency, indicating the frequency-synchronization of those oscillators. We obtain the desired shape of the collection of the synchronized oscillators. In panel (b), the indicator of the local order parameter $M(x,y)$ in Eq.~\eqref{orderparameter} at each site is shown. We also confirm the appearance of the desired synchronization pattern from this result.
}
\end{figure}

\begin{figure*}
  \includegraphics[width=120mm,bb=0 0 940 735,clip]{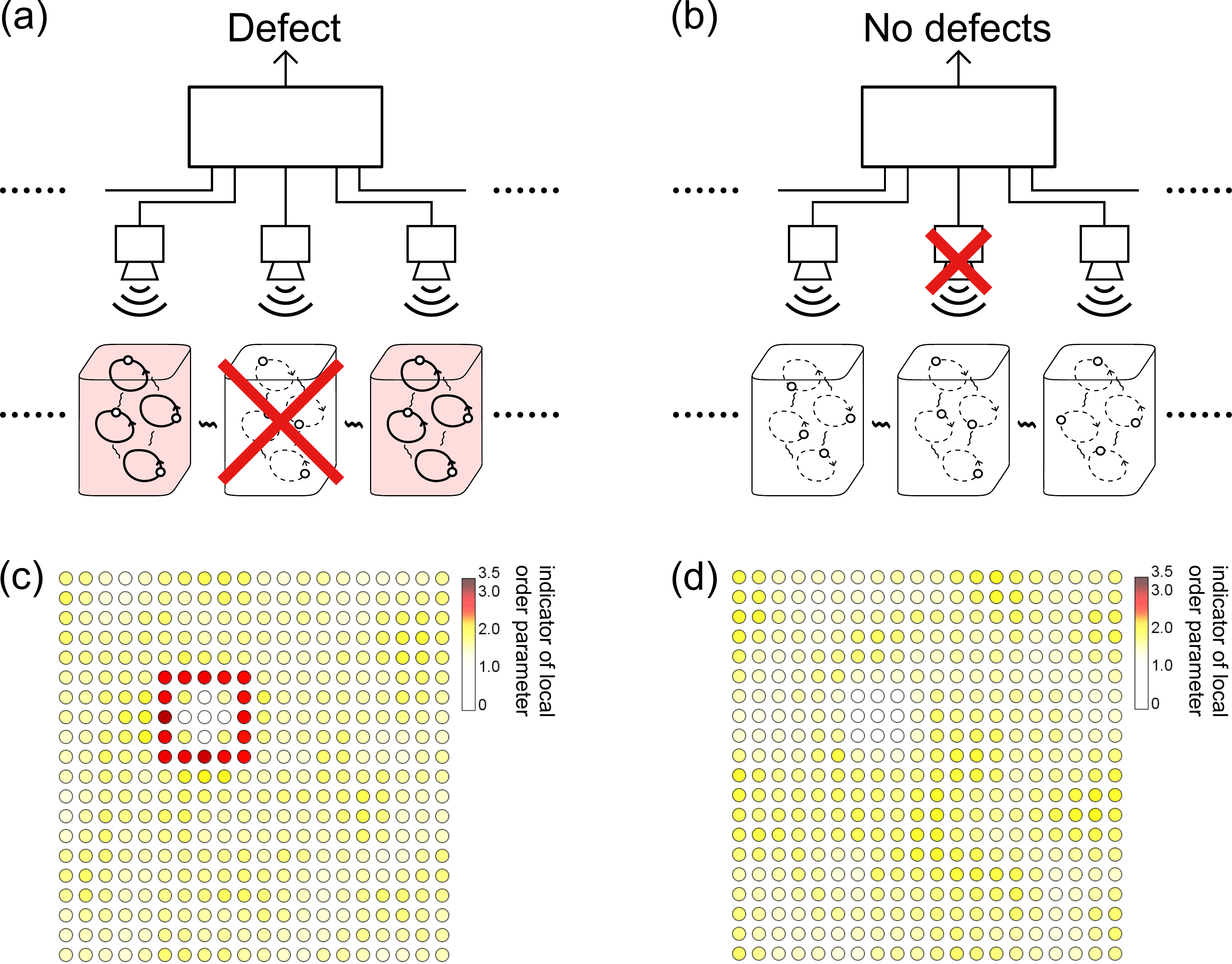}
\caption{\label{fig11} Applications of the TSS to the defect detection. (a),(b) Schematics of detection mechanism. We assume that the dynamics of nonlinear oscillators are detected by sensors. Information of the oscillators' state from the sensors is processed to judge the synchronization of oscillators and the existence of defects. If oscillators are broken and stop their self-oscillation, the oscillators around them are synchronized as shown in panel (a). By observing such synchronization, we can detect the breakdown of oscillators. When sensors are disordered and send no signals, the oscillators remain desynchronized as shown in panel (b). Therefore, this kind of breakdown of sensors is distinguishable from the breakdown of oscillators. (c) The indicator of the local order parameter $M(x,y)$ in Eq.~\eqref{orderparameter} at each site under the existence of the broken oscillators. Large values of the indicator show the synchronization of the oscillators at the corresponding site. One can judge that the oscillators surrounded by synchronized ones are broken. The parameters used are $u=-1$, $b=0.5$, $\alpha=0.5$, $\alpha_d=-100$, $\beta=1$, $\omega_0=0.1$, and $\Delta\omega=0.2$. (d) The indicator $M(x,y)$ at each site under the existence of the broken sensors. The indices are small everywhere, which shows that there are no synchronized oscillators, and thus no oscillators are broken. The parameters used are the same as panel (c).
}
\end{figure*}

If some of the oscillators suddenly stop their self-oscillations, the oscillators around them begin to synchronize. Thus, by observing the appearance of a synchronized cluster, we can continuously monitor the formation of the broken oscillators, i.e., defects. To demonstrate such possibility of defect detection using the TSS, we numerically calculate the dynamics of our first model and damp some oscillators in the middle of the simulation. Figure \ref{fig11}(c) shows the spatial distribution of the indicator $M(x,y)$ \eqref{orderparameter} in the steady-state regime. We find the synchronized oscillators that precisely lie along the edges of the defects. In contrast, when the sensors that detect the state of oscillators get broken and are unable to send signals to the central computing system  (see Fig.~\ref{fig11}(b)), our method obtains no enhanced signals indicating synchronization. Therefore, we can affirm the existence of broken oscillators from the large value of the indicator $M(x,y)$ \eqref{orderparameter} even under the nonnegligible possibility of this kind of breakdown of sensors. We numerically check the absence of the signal of synchronization in the case of sensors' disorder (see Fig.~\ref{fig11}(d)).

\begin{figure}
  \includegraphics[width=86mm,bb=0 0 530 470,clip]{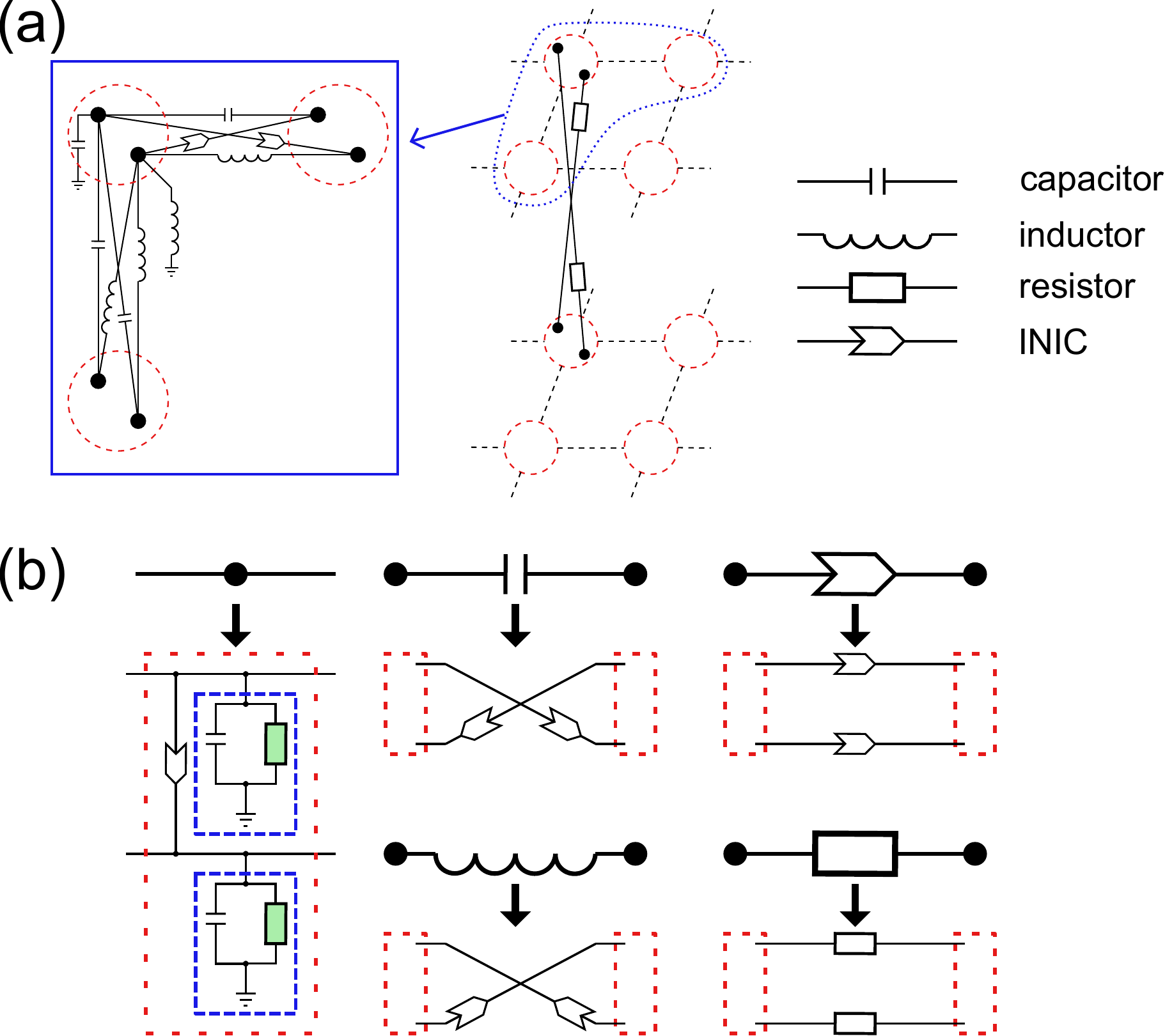}
\caption{\label{fig12} Proposed electrical circuit to realize topological synchronization. (a) Linear system described by the effective Hamiltonian of topological insulator laser constructed by an electrical circuit using capacitors, inductors, resistors, and negative impedance converters with current inversion (INICs). We consider two layers of square lattices, and each lattice point (a red dashed circle) has two sites (black dots). The dynamics of the voltages at the sites follows the Hamiltonian of topological insulator laser in Eq.~\eqref{hamiltonian_TIL_wavenumber}. The left inset represents the detail of the circuit in each layer (corresponding to the enlarged view of the sites encircled by the blue dot curve in the middle panel). The middle panel shows how to connect the sites of the different layers. The legend shows the correspondence of the figures and the circuit elements. (b) Substitution rules to construct the electrical circuit of TSS. We can construct the model of the TSS by replacing the circuit elements in panel (a). We substitute each site (a black dot) into two van-del-Pol circuits (blue dashed squares) coupled by an INIC. Green-filled boxes represent the nonlinear resistors. The other panels represent the substitution rules of capacitors, inductors, resistors, and INICs. We only use resistors and INICs to avoid frequency dependence of impedance. Red dashed squares represent the coupled van der Pol circuits in the left panel.
}
\end{figure}

Meanwhile, a system similar to our model can be realized by using an electrical circuit, which is shown in Fig.~\ref{fig12} as a schematic. Using nonlinear resistors, we propose to use van der Pol circuits \cite{Pol1926,Johnson2014} that simulate nonlinear oscillators. We also utilize negative impedance converters with current inversion \cite{Hofmann2019} to omit the frequency dependence. The dynamics of voltages in this electrical circuit imitates our first model, while the dynamics of van der Pol circuits is different from the Stuart-Landau oscillators (see Appendix \ref{appx_J} for the derivation of circuit equations). We note that one can realize the TSS without fine-tuning of the circuit constant of each element due to the topological protection of the TSS (see Appendix \ref{appx_D}). The electrical-circuit implementation of topological synchronization can have a potential application in information processing, as is discussed in the previous studies of topological electrical circuits \cite{Kotwal2021,Ezawa2020}. Our model should also be relevant to broad ranges of other physical systems. Especially, we expect that it is possible to realize the TSS in photonic systems, where topology \cite{Leykam2016,Lumer2013,Harari2018}, non-Hermiticity \cite{Zhao2019,El-Ganainy2007,Xu2016}, and nonlinearity \cite{Kozyreff2000,Ludwig2013} can be intertwined.

\section{Summary and Discussions\label{section7}}
We proposed a robust mechanism to control synchronization, namely, topological synchronization where edge oscillators synchronize while bulk ones exhibit chaotic dynamics. We termed such coexistence state of chaotic bulk motions and synchronized edge oscillators as the topological synchronized state (TSS) and analyzed the localized behaviors of its models. The edge modes of the topological linear couplings between oscillators are represented by the Lyapunov vectors localized at the edge of the system. The mechanism relies on topology in the wavenumber space and thus is different from the conventional mechanism  of chimera states; the latter can be explained from the self-consistency analysis and spatial variation of local order parameters \cite{Kuramoto2002,Abrams2004}. We also demonstrated that nonlinearity induces emerging effective boundaries, which leads to extra topological modes unique to nonlinear systems. As examples of applications, we showed that one can realize arbitrary patterns of synchronized oscillators by using topological synchronization. We also revealed that chaotic nonlinear systems can be used to detect disordered oscillators. Our model can be realized by utilizing an electrical circuit. Our proposal provides a general method to robustly and geometrically control the nonlinear oscillators by combining nonlinearity and topology.

While in our numerical calculations, we focused on $10\times 10$ and $20\times20$ lattices, TSS is independent of the system size, as long as the system is large enough to apply the band theory of bulk systems. This is because we here consider topology in the wavenumber space, and such analyses in the wavenumber space describe the bulk properties of infinite systems. Therefore, nontrivial topology and associated phenomena do not alter if we consider the larger lattice than analyzed in our calculations. We also used the Dirichlet-type open boundaries, while TSS appears under the other open boundary conditions (e.g. the Neumann boundary conditions), which may play an important role in some typical nonlinear systems such as fluids. We can understand this from the conventional argument in topological physics \cite{Hasan2010}. Such open boundaries connect topological systems and the vacuum, which have different topological invariants. However, we can change topological invariants only by closing the band gaps. Since the system and the vacuum has gapped bulk, such gap closing is possible only at the boundary. Therefore, boundary-localized gapless modes must appear in this situation. TSS shares the same origin with conventional topological edge modes and thus can appear under any type of open boundary conditions.

It is noteworthy that the TSS should ubiquitously appear in nonlinear systems with a topological effective Hamiltonian. In this paper, we confirmed the emergence of TSS in homogeneous Stuart-Landau oscillators whose linear couplings are described by Hamiltonians of two types of lasing edge modes protected by different topological mechanisms. We also discussed the TSS utilizing a Hermitian topological linear coupling and inhomogeneity of oscillators (cf. Table \ref{table1}). Furthermore, there is no need to judiciously adjust the parameters as discussed in Appendix \ref{appx_D}, which should be of practical advantage to realize the TSS in a variety of systems. Despite the broadness of models considered in this work, the TSS in identical oscillators induced by a Hermitian topological Hamiltonian still remains an intriguing problem. This may be possible by utilizing nonlinearity-induced non-Hermitian effects and realizing positive imaginary parts only in bulk modes of the effective Hamiltonian obtained by linearization of the equation (cf. Appendix \ref{appx_E}). Furthermore, we anticipate that other nonlinear oscillators than the Stuart-Landau model (e.g., the Fitzhugh-Nagumo model \cite{Fitzhugh1961,Nagumo1962}) can exhibit the TSS, because topological edge modes should be stable against the change of the on-site loss induced by the nonlinear terms.

\begin{acknowledgments}
We thank Eiji Saitoh and Shin-ichi Sasa for valuable discussions. K.S. is supported by World-leading Innovative Graduate Study Program for Materials Research, Industry, and Technology (MERIT-WINGS) of the University of Tokyo. K.S. is also supported by JSPS KAKENHI Grant Number JP21J20199. Y.A. is supported by JSPS KAKENHI Grant Number JP19K23424. T.S. is supported by JSPS KAKENHI Grant Numbers JP16H02211 and JP19H05796 and JST, CREST Grant No. JPMJCR20C1, Japan. T.S. is also supported by Institute of AI and Beyond of the University of Tokyo.
\end{acknowledgments}

\appendix
\section{Details of numerical calculations}\label{appx_A}
\subsection{Numerical simulation of the model of topological synchronization \label{appx_A1}} We simulate the dynamics of our first model (Eqs.~\eqref{model1}, \eqref{hamiltonian_TIL_wavenumber}) by using the fourth-order Runge-Kutta method. To calculate the time evolution of the complex-valued state variables, we deal with real and imaginary parts of those variables separately as $Z_i(x,y)=X_i(x,y)+iY_i(x,y)$. Real and imaginary coefficients are substituted as $aZ\rightarrow a(X,Y)^T $ and $ibZ\rightarrow -ib\sigma_y (X,Y)^T$, where $a$, $b$ are real, and $\sigma_{y}$ is the Pauli matrix. We arrange the $20\times20$ sites which have four complex Stuart-Landau oscillators for each. We impose the open boundary condition both in the $x$ and $y$ direction. We set the natural frequency of each oscillator $\omega_0+\Omega$ with $\Omega$ being a random value from uniform distributions ranging $[-\Delta\omega,\Delta\omega]$. The numerical calculation starts from the random initial condition, where real and imaginary parts of the initial state variables are the random value from uniform distributions ranging $[-0.1,0.1]$. In Fig.~\ref{fig1}, we set the time step $dt = 0.005$ and use the parameters $u=-1$, $b=0.5$, $\alpha=0.5$, $\beta=1$, $\omega_0=1$, and $\Delta\omega=0.2$. In Figs.~\ref{fig1}(c),(d), we demonstrate the frequency distribution. We define the phase of each oscillator as ${\rm Im} \log Z_i(x,y)$. We calculate and plot the time-averaged variation of the phase, setting the time window as $T_o=10$. In Figs.~\ref{fig1}(c),(d), we only plot the frequency of the first component of oscillators at each site. We also plot the amplitude of the first component of oscillators $|Z_1(x,y)|$ in Fig.~\ref{fig1}(e).

In Fig.~\ref{fig5}, we show the dynamics of nonlinear oscillators in the same way as in Fig.~\ref{fig1}. The parameters used are $u=-1$, $u'=0.02$, $a=2$, $b=0.5$, $\alpha=0.5$, $\beta=1$, $\omega_0=1$, and $\Delta\omega=0.2$. We set the time step $dt = 0.005$. We plot the time-averaged variation of the phase, i.e., the frequency of each oscillator at the first component of the lattice point in Figs.~\ref{fig5}(a),(b). We also plot the amplitude of the first component of oscillators $|Z_1(x,y)|$ in Fig.~\ref{fig5}(c).

In Fig.~\ref{fig7}, we also calculate the dynamics of nonlinear oscillators in the same way as in Fig.~\ref{fig1}. The parameters used are $u=-1$, $b=0.5$, $\alpha=1$, $\beta=1$, $\omega_0=1$, and $\Delta\omega=0.2$. We set the time step $dt = 0.005$. We plot the time-averaged variation of the phase, i.e., the frequency of each oscillator at the first (second) component of the lattice point in Fig.~\ref{fig7}(a) (Fig.~\ref{fig7}(c)). We also plot the indicator of the local order parameter \eqref{orderparameter} of the first and fourth (second and third) components in Fig.~\ref{fig7}(b) (Fig.~\ref{fig7}(d)).

\subsection{Calculations on the phase diagram} To obtain Fig.~\ref{fig2}, we numerically calculate the dynamics of our first model (Eqs.~\eqref{model1}, \eqref{hamiltonian_TIL}) at different strengths of the linear coupling and inhomogeneity of the natural frequencies. We introduce the parameter of the strength of the linear coupling $c$ by rewriting the equation of our model \eqref{model1} as
\begin{eqnarray}
  \frac{d}{dt} Z_j(\mathbf{x}) &=&  (i\omega_j(\mathbf{x})+\alpha-\beta |Z_j(\mathbf{x})|^2)Z_j(\mathbf{x}) \nonumber\\
  &{}& -ic \sum_{k,\mathbf{x}'} H_{jk}(\mathbf{x},\mathbf{x}') Z_k(\mathbf{x}'). \label{model_with_c}
\end{eqnarray}
We change the parameter $c$ from $0.2$ to $2$ at intervals of $0.2$. We also change the inhomogeneity of the natural frequencies $\Delta \omega$ from $0$ to $2$ at intervals of $0.1$. We fix the other parameters as $u=-1$, $b=0.5$, $\alpha=0.5$, $\beta=1$, and $\omega_0=1$. At each parameter, we calculate the dynamics and frequencies of oscillators in a $20\times20$ square lattice by using the fourth-order Runge-Kutta method as in Fig.~\ref{fig1}. Then, at time $t=100$, we calculate the standard deviations of the frequencies of the outermost oscillators and the bulk ones whose $x$ and $y$ coordinates satisfy $6\leq x \leq 15$ and $6\leq y \leq 15$. We plot the standard deviation of the frequencies of the outermost oscillators divided by that of the bulk ones in Fig.~\ref{fig2}. 

We also perform the numerical clustering of two-dimensional data composed of the pair of the standard deviations of the frequencies of the edge and bulk oscillators. We utilize the $k$-means method to conduct such clustering. We set the number of clusters as $k=2$. We determine the symbol used at each point in Fig.~\ref{fig2} based on which cluster the corresponding two-dimensional data point belongs to.

\subsection{Numerical methods of Lyapunov exponents and Lyapunov vectors} We calculate the Lyapunov exponents and vectors of our first model (Eqs.~\eqref{model1}, \eqref{hamiltonian_TIL}) by utilizing the Shimada-Nagashima algorithm \cite{Shimada1979} and Ginelli's algorithm \cite{Ginelli2007}. In those calculations, we arrange $10\times10$ sites and impose the open boundary condition. We deal with real and imaginary parts of the state variables separately as in the numerical calculation of our model (see Appendix \ref{appx_A1}). We first calculate the dynamics of our first model from the random initial condition as in Fig.~\ref{fig1}. Here, we set the time step as $\Delta t=0.1$. The obtained trajectory is used to determine the infinitesimal rate of change of the difference between the perturbed and original trajectories. We allow an initialization period of $T_{\rm ini}=1000$ to stabilize the dynamics into a chaotic attractor. We repeatedly perform the QR decomposition based on the matrices derived from the linearization of the equation and the obtained trajectory from $t=T_{\rm ini}$ to $t=T_{\rm ini}+20000$. The averages of the logarithmic diagonal entries of the $R$ matrices converge to Lyapunov exponents. To omit the effect of the initial condition, we use the $R$ matrices obtained from the time $t=T_{\rm ini}+10000$ to $t=T_{\rm ini}+20000$ to calculate the Lyapunov exponents. We also utilize those $R$ matrices to calculate the Lyapunov vectors. We iteratively multiply the inverse of obtained $R$ matrices to a randomly-obtained upper-triangle matrix. Finally, the product of the obtained upper-triangle matrix and the $Q$ matrix obtained in the calculation of the Lyapunov exponents represents the set of the Lyapunov vectors. 

We also utilize the Shimada-Nagashima algorithm \cite{Shimada1979} and Ginelli's algorithm \cite{Ginelli2007} to calculate the Lyapunov exponents and vectors of the other models (Eqs.~\eqref{hamiltonian_conventionaltopo_TIL}, \eqref{Hermitian_coupling_real}). The parameters of models are set to be equal to those utilized in the calculation of dynamics (Figs.~\ref{fig6}, \ref{fig8}). Furthermore, in the model utilizing the Hermitian linear coupling, we set the initialization period as $T_{\rm ini}=100000$. The other parameters used in the calculations of Lyapunov exponents and vectors are the same as those in the calculations on our first model.

\subsection{Numerical calculations on the emergent nonlinearity-induced boundary} To show the emergence of the extra topological boundary modes by the emergent nonlinearity-induced boundary, we diagonalize the state-dependent Hamiltonian $\tilde{H}(\mathbf{Z})$ under the open (periodic) boundary condition in the $x$ ($y$) direction. We perform the inverse Fourier transformation of the Hamiltonian in Eq.~\eqref{hamiltonian_TIL_wavenumber} used in our first model only in the $x$ direction. We divide the real and imaginary parts of the state variables as in the simulation of the dynamics of our model and consider the twice size of the matrix compared to the Hamiltonian in Eq.~\eqref{hamiltonian_TIL_wavenumber}. We consider the $1\times20$ supercell structure. To determine the strength of the nonlinear loss terms (i.e., $-\beta|z|^2z$), we first diagonalize the Hamiltonian without nonlinear loss terms. We obtain four edge modes $v^i_j(x)$ ($i=1,\cdots,4$ is the index of the edge modes and $j=1,\cdots,4$ represents the component of oscillators at each site) with maximum imaginary parts of eigenvalues ${\rm Im}\,E_0$ at the wavenumber $k=0$. Then, we introduce the on-site loss $-ia(|v^1_j(x)|^2+|v^2_j(x)|^2+|v^3_j(x)|^2+|v^4_j(x)|^2)Z_j(x)$, where the coefficient is set to be $a={\rm Im}\,E_0$ to balance gain and loss applied to the edge modes $v^i_j(x)$. Thus, this on-site loss corresponds to the case that the edge modes in the Hamiltonian without the nonlinearity-induced on-site loss are fully amplified. Finally, we diagonalize the Hamiltonian with this nonlinear on-site loss term at each wavenumber and obtain the dispersion relation in Fig.~\ref{fig4}(a) and the eigenvector of the edge mode in Fig.~\ref{fig4}(b).

We also calculate the dynamics of the extra boundary modes. As in the calculation of the dynamics in Fig.~\ref{fig1}, we use the fourth-order Runge-Kutta method and consider the $20\times20$ square lattice with the open boundary to simulate the dynamics of nonlinear oscillators with linear couplings described by the Hamiltonian in Eq.~\eqref{hamiltonian_TIL}. By using the parameters $\alpha=5\times10^{-5}$ and $\beta=1\times10^{-4}$, we realize the situation that the nonlinearity has a negligible effect on the dynamics at the initial stage of the simulation. We consider the random initial condition, where real and imaginary parts of the initial state variables are the random value from uniform distributions ranging $[-0.001,0.001]$. We set the time step $dt=0.005$. The other parameters used are $u=-1$, $b=0.8$, $\omega_0=0.1$, and $\Delta\omega=0.2$.

\subsection{Numerical demonstrations of the on-demand pattern designing} To demonstrate the possibility of arranging synchronized oscillators in a desired pattern, we calculate the dynamics of our first model (Eqs.~\eqref{model1}, \eqref{hamiltonian_TIL}) under the existence of damped oscillators (see Fig.~\ref{fig10}). We implement damped oscillators by setting $\alpha=-100$ at the corresponding sites. We arrange $15\times30$ sites and impose the periodic boundary condition both in the $x$ and $y$ direction. As in the numerical simulation of the dynamics of our model, we utilize the fourth-order Runge-Kutta method. We start the calculation from the random initial condition, where real and imaginary parts of the initial state variables are the random value from uniform distributions ranging $[-0.1,0.1]$. We set the time step $dt = 0.005$ and use the parameters $u=-1$, $b=0.3$, $\alpha=0.5$, $\beta=1$, $\omega_0=0.1$, and $\Delta\omega=0.2$. We calculate the frequency of each oscillator from the time-averaged variation of the phase, setting the time window as $T_o=20$. We also set the time window to calculate the indicators of local order parameters as $T=100$.

\subsection{Numerical demonstrations of the defect detection} In the demonstration of the defect detection by using the topological synchronized state (TSS) (cf. Fig.~\ref{fig11}), we calculate the dynamics of our first model (Eqs.~\eqref{model1}, \eqref{hamiltonian_TIL}) under the periodic boundary condition by the fourth-order Runge-Kutta method. To implement the broken oscillators in Fig.~\ref{fig11}(c), we change the parameters $\alpha$ at the corresponding sites from $0.5$ to $-100$ at the time $t=200$. On the other hand, to simulate the disorders of sensors, we assume $Z_i(x,y)=0$ only in the calculation of the indicator of the local order parameter \eqref{orderparameter}. We consider the random initial condition, where real and imaginary parts of the initial state variables are the random value from uniform distributions ranging $[-0.1,0.1]$. We set the time step $dt = 0.005$ and use the parameters $u=-1$, $b=0.5$, $\alpha=0.5$, $\beta=1$, $\omega_0=1$, and $\Delta\omega=0.2$. We calculate the frequency of each oscillator from the time-averaged variation of the phase, setting the time window as $T_o=5$. We also set the time window to calculate the indicators of local order parameters as $T=100$.

\section{Synchronization not categorized in the TSS}\label{appx_B}
\subsection{Cluster synchronization in the case of another effective Hamiltonian with topological lasing modes}\label{appx_B1}
\begin{figure}
\includegraphics[width=70mm,bb=0 0 410 610,clip]{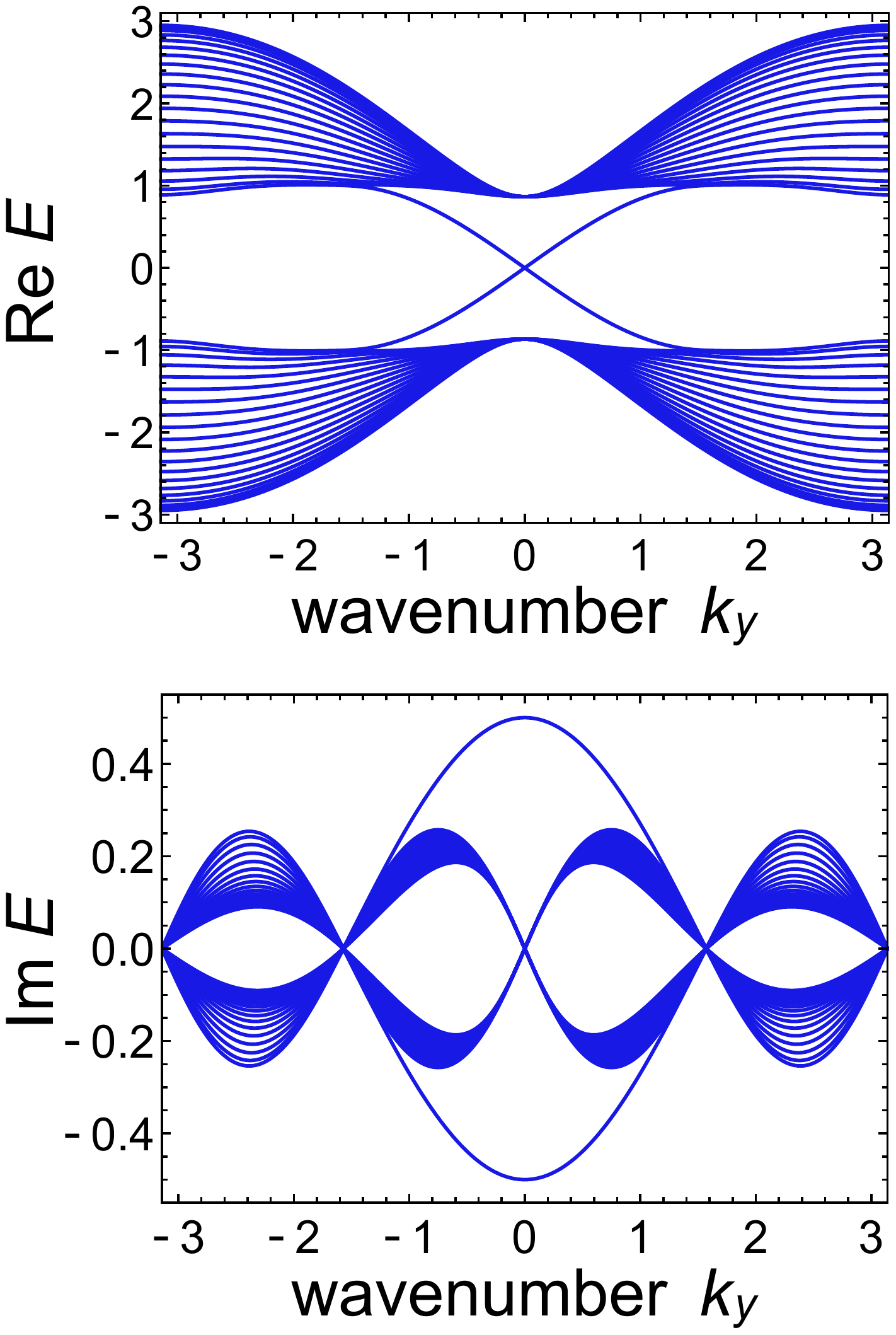}
\caption{\label{fig13} Dispersion relation of the lasing modes localized at the right edge. We calculate the dispersion relation of the Hamiltonian of topological lasing modes \eqref{lasing_QWZ}. We consider the open (periodic) boundary condition in the $x$ ($y$) direction and arrange $20$ sites in the $x$ direction. We obtain a gapless edge band with positive imaginary parts of eigenvalues. This gapless band corresponds to lasing edge modes localized at the right side. The parameters used are $u=-1$ and $\gamma=0.5$.
}
\end{figure}
\begin{figure}
\includegraphics[width=86mm,bb=0 0 610 280,clip]{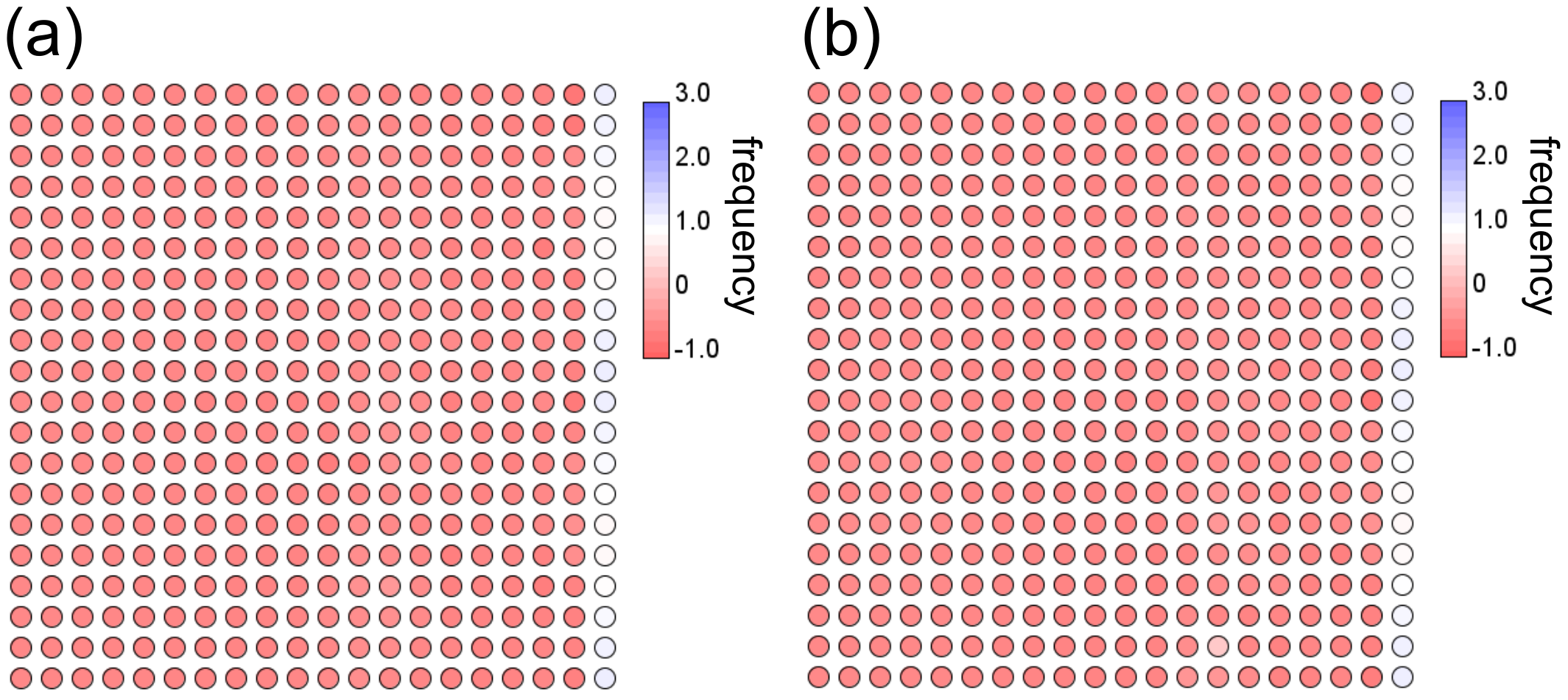}
\caption{\label{fig14} Cluster synchronization from a topological Hamiltonian. We calculate the dynamics of the Stuart-Landau oscillators with linear coupling described by a Hamiltonian of topological lasing modes that is different from the Hamiltonians considered in the main text \eqref{lasing_QWZ}. Panels (a) and (b) show the snapshots of the frequency distributions of the first component of oscillators at the time $t=1000$ and $t = 2000$ for each. The bulk and left-side oscillators exhibit homogeneous and constant frequency, which is different from that of the right-side oscillators. Therefore, the bulk oscillators form a synchronized cluster, indicating the absence of the TSS. The parameters used are $u=-1$, $\gamma=0.5$, $\alpha=0.5$, $\beta=1$, $\omega_0=1$, and $\Delta\omega=0.2$.
}
\end{figure}

The TSS needs the instability of bulk oscillators to desynchronize them. On the other hand, we numerically check that even if the linear coupling shows topological lasing edge modes, there is a possibility that bulk oscillators stably synchronize in some models, which indicates the absence of the TSS in such models. Here we consider the following Hamiltonian of topological lasing modes:
\begin{eqnarray}
 H(\mathbf{k}) &=& (u+\cos k_x + \cos k_y) \sigma_z \nonumber\\
 &{}&+ (\sin k_y+i\gamma \cos k_y) \sigma_y + \sin k_x \sigma_x, \label{lasing_QWZ}
\end{eqnarray}
where $\sigma_{x,y,z}$ are the Pauli matrices, and $I_2$ is the $2\times2$ identity matrix. This Hamiltonian exhibits lasing modes localized only at the right side of the system. Figure \ref{fig13} shows the band structure calculated under the open (periodic) boundary condition in the $x$ ($y$) direction, which shows the existence of lasing edge modes. We arrange two Stuart-Landau oscillators (cf. Eq.~\eqref{model1}) at each site of a $20\times20$ square lattice and combine them by the linear coupling described by the Hamiltonian \eqref{lasing_QWZ}. We assume the open (periodic) boundary condition in the $x$ ($y$) direction. By calculating the dynamics of this model as in Fig.~\ref{fig1}, we find that the bulk and left-side oscillators also synchronize. Figure \ref{fig14} shows the frequency of the oscillator at each site. One can confirm that the bulk and left-side oscillators exhibit a homogeneous frequency that is different from that of the right-side oscillators. Thus, the bulk and left-side oscillators form another synchronized cluster, which indicates the absence of the TSS. We note that similar cluster synchronization is studied in one- and two-dimensional nonlinear topological systems \cite{Kotwal2021}. 

\subsection{Coexistence of synchronized edge oscillators and damped bulk ones}\label{appx_B2}
\begin{figure}[t]
\includegraphics[width=86mm,bb=0 0 610 275,clip]{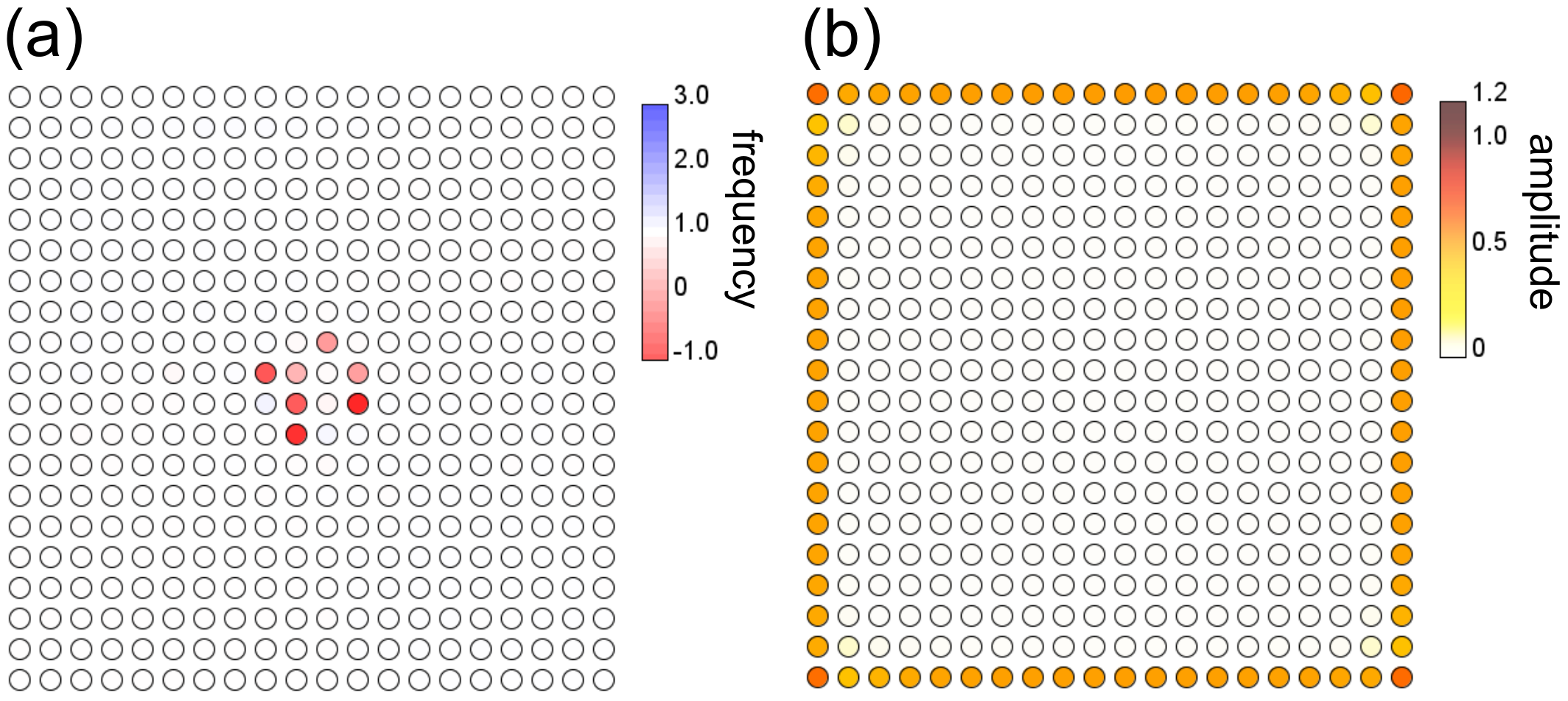}
\caption{\label{fig15} Coexistence of synchronized edge oscillators and damped bulk oscillators. (a) Frequency of the first component of oscillators at each site in a damping parameter region. We numerically simulate the first model of the TSS (Eqs.~\eqref{model1}, \eqref{hamiltonian_TIL_wavenumber}) at the parameters $u=-1$, $b=0.5$, $\alpha=-0.2$, $\beta=1$, $\omega_0=1$, and $\Delta\omega=0$. The figure shows the snapshot at time $t=100$. The edge oscillators exhibit homogeneous frequencies and thus synchronize. (b) Amplitudes of the first components of oscillators at each site at time $t=100$. We can confirm that the bulk oscillators exhibit almost zero amplitudes. The parameters used are the same as in panel (a).
}
\end{figure}

By tuning the parameters, we show that our model (Eqs.~\eqref{model1}, \eqref{hamiltonian_TIL_wavenumber}) can exhibit a different coexistence state of synchronization where the bulk oscillators are damped while the edge ones are synchronized. Here, we set the parameter $\alpha$ in the Stuart-Landau equation (cf. Eq.~\eqref{model1}) to be negative and larger than $-|b|$. The negativity of $\alpha$ implies the damping of self-oscillations of the bulk oscillators, while the lasing edge modes of linear couplings still lead to synchronized oscillations at the edge of the sample in this parameter region. Figure \ref{fig15} shows the frequency and amplitude distributions obtained from the numerical calculations of our model at $\alpha=-0.2$. We can confirm that the bulk oscillators exhibit almost zero amplitudes, while the edge ones synchronize. It is noteworthy that the bulk oscillators slightly oscillate at frequencies close to that of the edge ones due to their interaction. We define the TSS as the coexistence of the synchronized edge oscillators and the desynchronized bulk oscillators, while the synchronized state observed in this calculation exhibits bulk oscillators without self-oscillations and thus is not categorized in the TSS.

\section{TSS without fluctuations of the natural frequencies}\label{appx_C}
\begin{figure}[t]
\includegraphics[width=86mm,bb=0 0 610 280,clip]{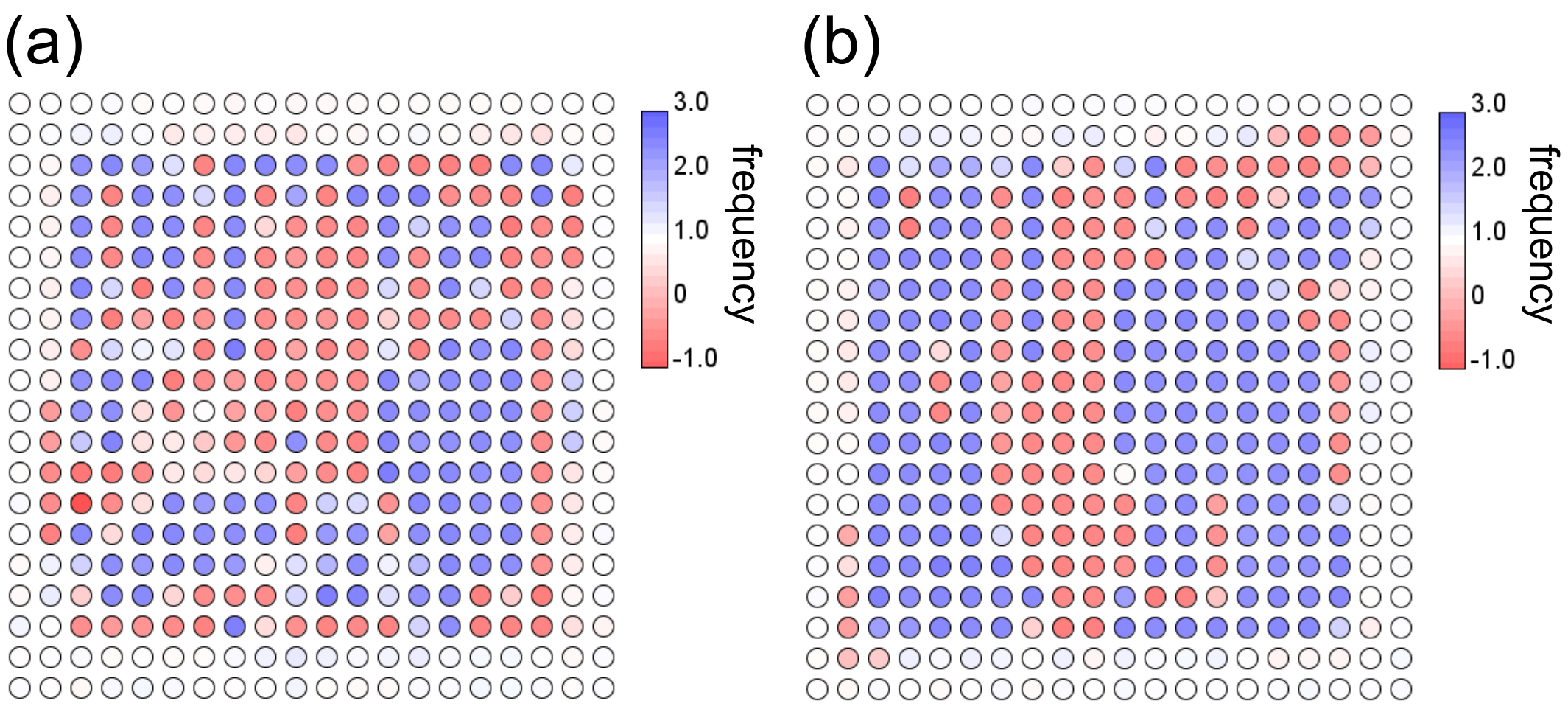}
\caption{\label{fig16} TSS without fluctuations of the natural frequencies. We simulate the first model of the TSS (Eqs.~\eqref{model1}, \eqref{hamiltonian_TIL_wavenumber}) under the condition of the homogeneous natural frequencies. Panels (a) and (b) show the snapshots of the frequency distributions of the first component of oscillators at the time $t=100$ and $t=200$ for each. As in the case of the fluctuating natural frequencies (Fig.~\ref{fig1}(c),(d)), the edge oscillators exhibit homogeneous and constant frequencies, while the bulk ones vibrate at space- and time-varying frequencies. Therefore, the edge oscillators synchronize, while the bulk ones desynchronize, which indicates the emergence of the TSS. The parameters used are $u=-1$, $b=0.5$, $\alpha=0.5$, $\beta=1$, $\omega_0=1$, and $\Delta\omega=0$.
}
\end{figure}

Here we show that the fluctuations of the natural frequencies are unnecessary to realize the TSS as in conventional chimera states \cite{Abrams2004,Panaggio2015}. We calculate the dynamics of the first model of the TSS (Eqs.~\eqref{model1}, \eqref{hamiltonian_TIL_wavenumber}) without fluctuations of natural frequencies, i.e., $\Delta\omega=0$. Figure \ref{fig16} presents the frequency distribution obtained under the homogeneous natural frequency. We can confirm that the edge oscillators vibrate at a homogeneous and constant frequency, while the bulk oscillators exhibit inhomogeneous and unstable frequencies. Therefore, the TSS appears even under the homogeneous natural frequency. We set the parameters as $u=-1$, $b=0.5$, $\alpha=0.5$, $\beta=1$, and $\omega_0=1$ and consider the random initial condition as in Fig.~\ref{fig1}.

\begin{figure}[t]
\includegraphics[width=60mm,bb=0 0 420 610,clip]{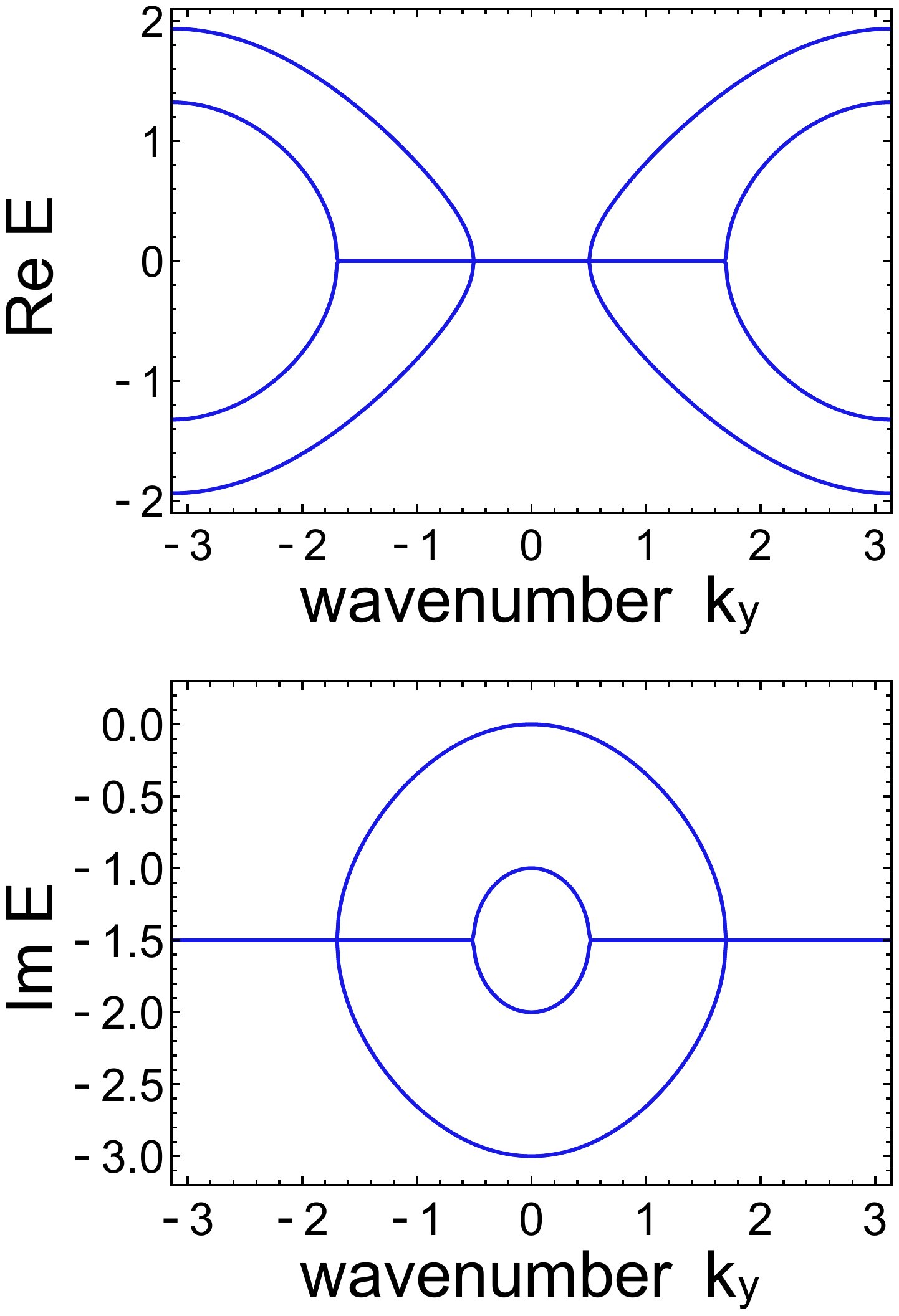}
\caption{\label{fig17} Dispersion relation of the linearized equation of one-dimensional chain model. We numerically calculate the dispersion relation of the coefficient matrix obtained from the linearization of the one-dimensional chain model around the stationary solution (Eqs.~\eqref{1d-chainH_k}, \eqref{1d-chain-SS}). The upper (lower) panel shows the real (imaginary) part of the eigenvalues. All the eigenvalues have nonpositive imaginary parts, which implies the linear stability of the stationary solution. The parameters used are $u=-1$, $b=0.5$, $\alpha=0.5$, $\beta=1$, $\omega_0=0$, and $\Delta\omega=0$.
}
\end{figure}

\section{Detailed analyses on the stability of the TSS}\label{appx_D}
\subsection{Stability of the TSS via a one-dimensional model}\label{appx_D1}

Here we discuss the stability of the TSS via the one-dimensional chain of Stuart-Landau oscillators which effectively describes the edge oscillators in our first model (Eqs.~\eqref{model1}, \eqref{hamiltonian_TIL_wavenumber}). Since the edge oscillators in our model exhibit larger amplitudes than those of the bulk ones (cf. Fig.~\ref{fig1}(e)), the interaction between the edge and bulk oscillators is smaller than that between the edge ones. Therefore, we can assume that the linear couplings between the edge and bulk oscillators only have a perturbative effect on the dynamics of the synchronized edge oscillators, and thus the model without the hopping terms in the $x$ direction,
\begin{eqnarray}
 \frac{d}{dt} Z_j(y) &=&  (i\omega_j(y)+\alpha-\beta |Z_j(y)|^2)Z_j(y) \nonumber\\
 &{}& -i \sum_{k,y'} H^{\rm 1d}_{jk}(y,y') Z_k(y') + \xi_j(y), \label{1d-chain} \\
 H^{\rm 1d}_{jk}(y,y') &=& \left( u\delta_{y,y'}+\frac{\delta_{y+1,y'}+\delta_{y-1,y'}}{2} \right) (I_2 \otimes \sigma_z)_{jk} \nonumber\\
 &{}&+\frac{\delta_{y+1,y'}-\delta_{y-1,y'}}{2i} (I_2 \otimes \sigma_y)_{jk} \nonumber\\
 &{}& +ib\delta_{y,y'} (\sigma_x \otimes \sigma_x)_{jk} \label{1d-chainH}
\end{eqnarray}
can describe the edge oscillators in our first model, where $\delta_{y,y'}$ is the Kronecker delta and $\sigma_i$ and $I_2$ are the $i$th component of the Pauli matrices and the $2\times2$ identity matrix respectively. We introduce $\xi_j(y)$ to describe the perturbative effect from the bulk oscillators, while we show that the TSS is robust against such perturbation and inhomogeneity of parameters. We note that the chaotic dynamics in the bulk vanishes the long-term correlation of interaction between the edge and bulk oscillators, and thus $\xi_j(y)$ can be considered as noise.

First, we obtain the stationary synchronized solution of Eq.~\eqref{1d-chain} under the absence of the interaction from bulk oscillators and the homogeneous natural frequencies. We conduct the Fourier transformation of the Hamiltonian used in the one-dimensional model \eqref{1d-chainH} and obtain
\begin{equation}
 H^{\rm 1d}(k) = \left( u+\cos k_y \right) I_2 \otimes \sigma_z +\sin k_y I_2 \otimes \sigma_y +ib \sigma_x \otimes \sigma_x. \label{1d-chainH_k}
\end{equation}
This Hamiltonian exhibits the maximum imaginary part of the eigenenergies ${\rm Im}\,E=\sqrt{b^2-u'^2}$ at $k_y=0$ and the corresponding eigenvector $(ib,\mp u'\pm i\sqrt{b^2-u'^2},\pm ib,-u'+i\sqrt{b^2-u'^2})^T$, where $u'=u+1$. By utilizing this eigenvector, we construct the following homogeneous stationary (periodic) solution:
\begin{equation}
   \left(
  \begin{array}{c}
   \!Z_1(y)\!\! \\
   \!Z_2(y)\!\! \\
   \!Z_3(y)\!\! \\
   \!Z_4(y)\!\! 
  \end{array}
  \right) = e^{i\omega_0t+i\theta}\sqrt{\frac{\alpha+\sqrt{b^2-u'^2}}{b^2\beta}}
  \left(
  \begin{array}{c}
   ib \\
   \!\mp u'\pm i\sqrt{b^2-u'^2}\!\! \\
   \pm ib \\
   \!-u'+i\sqrt{b^2-u'^2}\!\!
  \end{array}
  \right), \label{1d-chain-SS}
\end{equation}
where $\omega_0$ is the homogeneous natural frequency. Since the gapped bulk bands require $|u'|<|b|$, the absolute values of the components of this stationary solution are the same, which leads to the same intensity of the nonlinear terms. The stationary solution \eqref{1d-chain-SS} represents the frequency-synchronization of the edge oscillators, where they oscillate at the same frequency $\omega_0$.

To confirm the stability of the frequency-synchronization of the edge oscillators, we conduct linear stability analysis around the stationary solution \eqref{1d-chain-SS}. We linearize the Stuart-Landau equation $\dot{Z} = (i\omega+\alpha-\beta {Z}^2) Z$ and obtain
\begin{eqnarray}
  &{}& \frac{d}{dt}
  \left(
  \begin{array}{c}
   \delta X\! \\
   \delta Y\!
  \end{array}
  \right) = \nonumber \\
  &{}& \left(
  \begin{array}{cc}
   \!i\omega+\alpha-3\beta X_0^2-\beta Y_0^2\!\!\! & -2\beta X_0 Y_0 \\
   -2\beta X_0 Y_0 & i\omega+\alpha-\beta X_0^2-3\beta Y_0^2\!\!
  \end{array}
  \right) \left(
  \begin{array}{c}
   \!\delta X\!\! \\
   \!\delta Y\!\!
  \end{array}
  \right), \nonumber\\ \label{linearized_Stuart}
\end{eqnarray}
where $Z=X+iY$ ($X$, $Y$ are real) and $Z_0=X_0+iY_0$ is the state around which the equation is linearized. The eigenvalues of the coefficient matrix in this equation are $i\omega+\alpha-X_0^2-Y_0^2$, $i\omega+\alpha-3X_0^2-3Y_0^2$ and have negative real parts when we linearize the equation around the stationary solution \eqref{1d-chain-SS}. Therefore, these terms attenuate the fluctuations from the stationary solution, which leads to the linear stability of the one-dimensional chain of nonlinear oscillators. We numerically diagonalize the linearized equation of Eq.~\eqref{1d-chain} and obtain the dispersion relation in Fig.~\ref{fig17}. We confirm that all the eigenvalues have nonpositive imaginary parts, which indicates the linear stability of the stationary solution. Therefore, the synchronized state in the one-dimensional model \eqref{1d-chain} is robust against the perturbations, such as the inhomogeneity of the natural frequencies and the perturbative interactions from bulk ones. Since the one-dimensional chain of oscillators and its stationary solution describes the synchronized edge oscillators in our first model (Eqs.~\eqref{model1}, \eqref{hamiltonian_TIL_wavenumber}), the TSS is also stable against the existence of disorders.

\subsection{Robustness of the electrical circuit realization of the TSS}\label{appx_D2}
\begin{figure}[t]
\includegraphics[width=86mm,bb=0 0 610 280,clip]{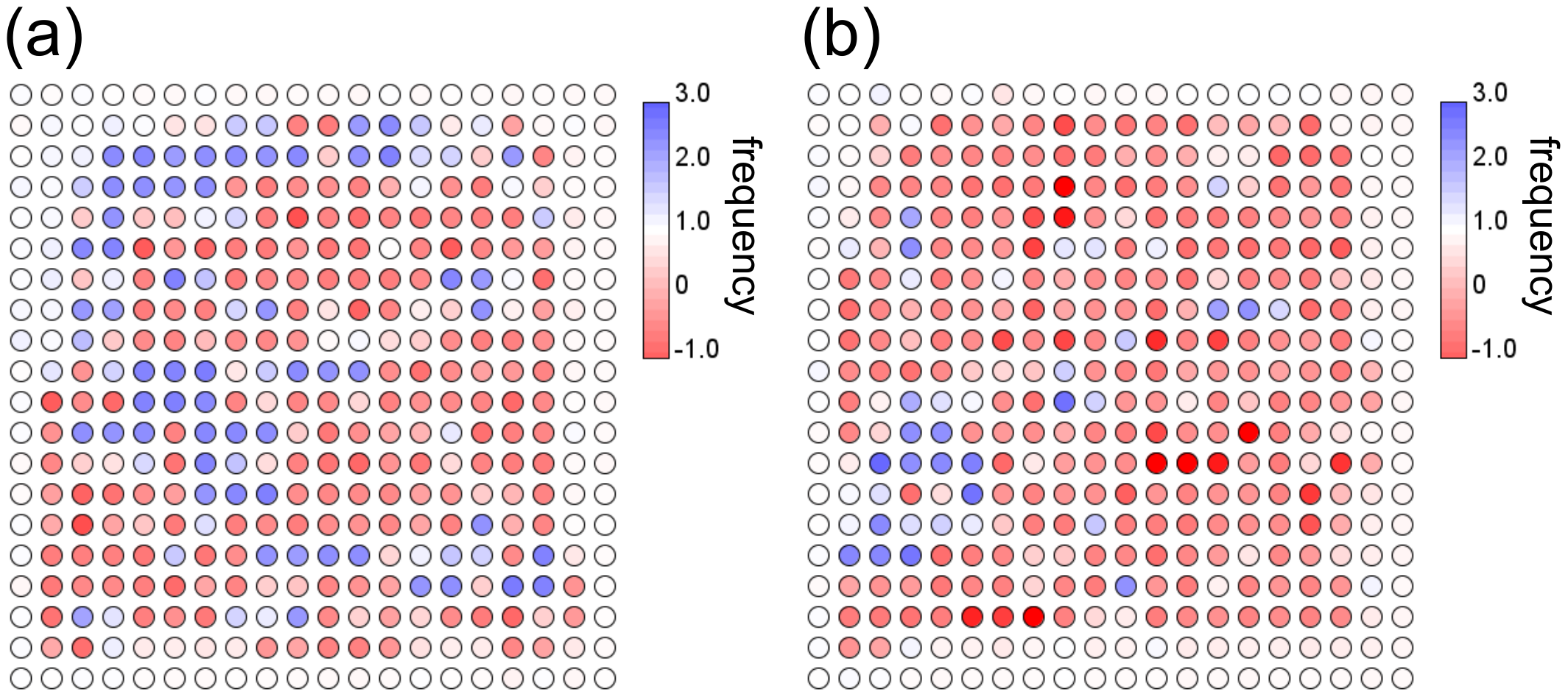}
\caption{\label{fig18} Robustness of TSS against disorders in linear couplings. We numerically simulate the model of the TSS (Eqs.~\eqref{model1}, \eqref{hamiltonian_TIL_wavenumber}) with the inhomogeneous linear couplings. Panels (a) and (b) show the frequency distributions at time $t=100$ and $t=200$ for each. The edge oscillators exhibit homogeneous and constant frequencies, while the bulk ones oscillate at time- and space-varying frequencies. Therefore, we obtain the TSS robustly against the noise in linear couplings. The parameters used are $u=-1$, $b=0.5$, $\alpha=0.5$, $\beta=1$, $\omega_0=1$, and $\Delta\omega=0.2$.
}
\end{figure}

TSS is robust against disorder due to its topological nature and thus is realizable without a fine-tuning of parameters. Especially, while there should be inhomogeneity in circuit constants of elements constructing the TSS circuit (cf. Fig.~\ref{fig12}), TSS remains stable against such disorders. To confirm the stability of TSS against the inhomogeneity of circuit constants, we numerically calculate the dynamics of our first model (Eqs.~\eqref{model1}, \eqref{hamiltonian_TIL_wavenumber}) under the fluctuating coupling strengths. We introduce the fluctuations of parameters corresponding to the inhomogeneous circuit elements by setting the proportion of deviation in the coupling strength from a site $(\mathbf{x}; a)$ to a site $(\mathbf{x}'; b)$ to be the same as that from a site $(\mathbf{x}'; b)$ to a site $(\mathbf{x}; a)$. We also consider the fluctuation of the parameters $\alpha$ and $\beta$ in Eq.~\eqref{model1}, which is determined from the characteristics of the nonlinear current resources in the circuit of TSS (cf. Fig.~\ref{fig12}). We set the maximum width of these fluctuations as 10 percent of the mean values, and determine them as the distributions follow the uniform distribution. Figure \ref{fig18} shows the frequency distribution obtained from the simulation. One can confirm the emergence of the TSS, that is, the edge oscillators synchronize while the bulk ones exhibit chaotic behavior. Therefore, the TSS is robust against the fluctuations of parameters in Eq.~\eqref{model1}, which should be of practical advantage to experimentally realize the TSS.

\section{Dispersion relation of the effective Hamiltonian}\label{appx_E}
\begin{figure}[t]
\includegraphics[width=70mm,bb=0 0 390 615,clip]{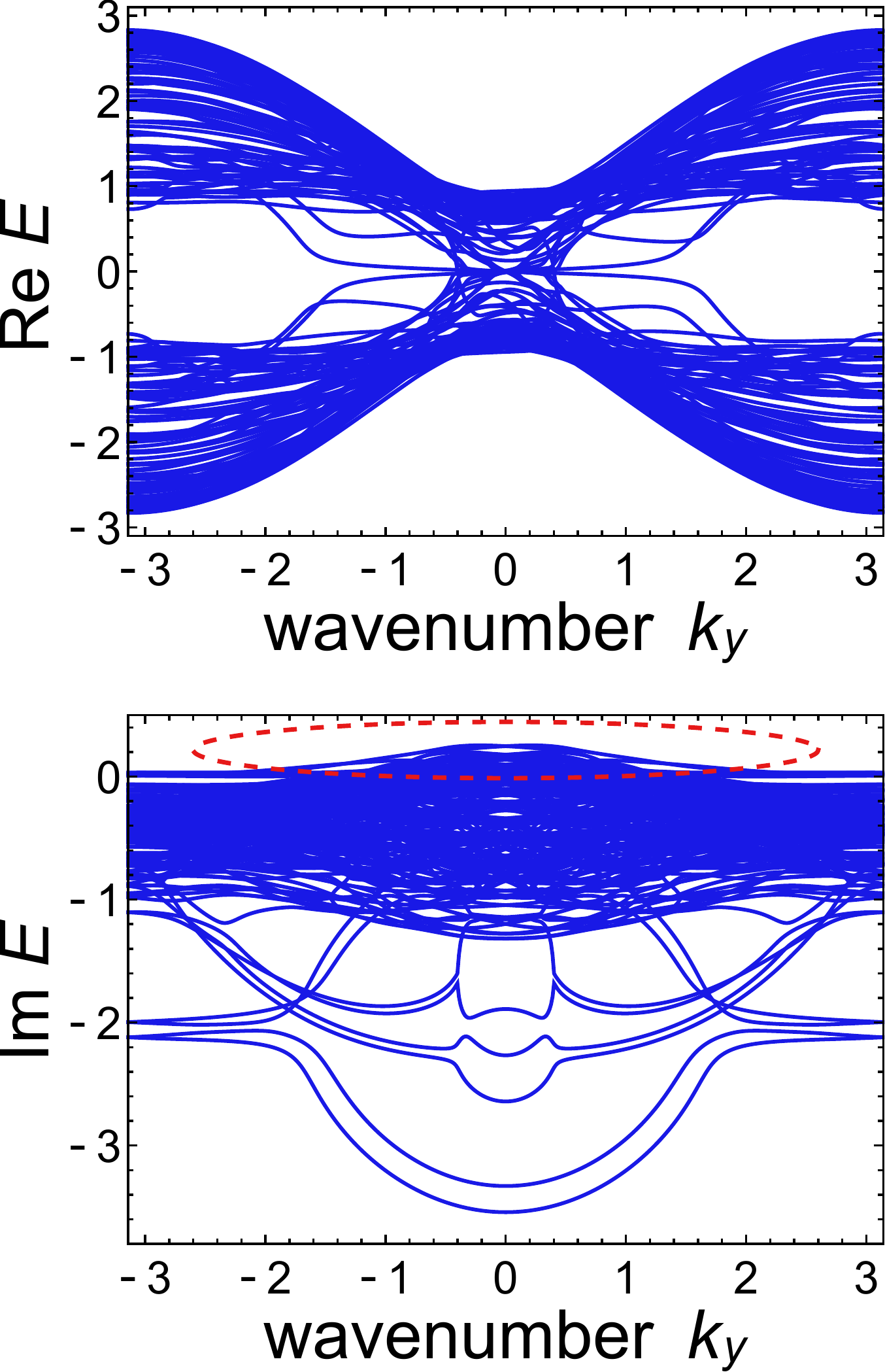}
\caption{\label{fig19} Dispersion relation of the effective Hamiltonian. We calculate the dispersion relation of the effective Hamiltonian obtained from the linearization of the first model (cf. Eqs.~\eqref{model1}, \eqref{hamiltonian_TIL_wavenumber}). We consider the open (periodic) boundary condition in the $x$ ($y$) direction and arrange $20$ sites in the $x$ direction. While the bulk bands are gapless as opposed to conventional topological systems, the edge localized modes still exist in this dispersion relation. The edge modes exhibit negative imaginary parts of the eigenvalues. In contrast, some bulk modes possess positive imaginary parts of the eigenvalues depicted by the red dashed circle, which indicates instability of the bulk oscillators. The parameters used are $u=-1$, $b=0.5$, $\alpha=0.5$, $\beta=1$, $\omega_0=1$, and $\Delta\omega=0$.
}
\end{figure}

Here we calculate the dispersion relation of the effective Hamiltonian derived from the linearization of the equation of our first model (Eqs.~\eqref{model1}, \eqref{hamiltonian_TIL_wavenumber}). Such effective Hamiltonian includes the on-site loss induced by the nonlinear terms, which depends on the state around which the linearization is performed. Specifically, such on-site loss terms are obtained from the linearization of the Stuart-Landau oscillators $\dot{Z} = (i\omega+\alpha-\beta {Z}^2) Z$ as
\begin{eqnarray}
  &{}& \frac{d}{dt}
  \left(
  \begin{array}{c}
   \delta X\! \\
   \delta Y\!
  \end{array}
  \right) = \nonumber \\
  &{}& \left(
  \begin{array}{cc}
   \!i\omega+\alpha-3\beta X_0^2-\beta Y_0^2\!\!\! & -2\beta X_0 Y_0 \\
   -2\beta X_0 Y_0 & i\omega+\alpha-\beta X_0^2-3\beta Y_0^2\!\!
  \end{array}
  \right) \left(
  \begin{array}{c}
   \!\delta X\!\! \\
   \!\delta Y\!\!
  \end{array}
  \right), \nonumber\\ \label{linearized_Stuart}
\end{eqnarray}
where $Z=X+iY$ ($X$, $Y$ are real) and $Z_0=X_0+iY_0$ is the state around which the equation is linearized. We note that the coefficient matrix in this equation has eigenvalues $i\omega+\alpha-X_0^2-Y_0^2$, $i\omega+\alpha-3X_0^2-3Y_0^2$. Thus, if the square of the amplitude is $X_0^2+Y_0^2=\alpha/\beta$ (equal to the amplitude of the isolated Stuart-Landau oscillator), the obtained on-site term leads to attenuation of the fluctuation.

To determine the intensities of those on-site loss terms, we first simulate the dynamics of our first model (Eqs.~\eqref{model1}, \eqref{hamiltonian_TIL_wavenumber}) in a $20\times20$ square lattice under the periodic (open) boundary condition in the $x$ ($y$) direction. Here we consider the random initial condition as in Fig.~\ref{fig1}. We obtain the state variables of the oscillators in the $10$th row at $t=200$ and linearize the equation around them. In the calculation of the dispersion relation, we consider a $1\times20$ supercell structure. Finally, we numerically diagonalize the obtained effective Hamiltonian and plot the dispersion relation as shown in Fig.~\ref{fig19}. We find bulk modes exhibiting positive imaginary parts of eigenvalues, which lead to the chaotic dynamics of the bulk oscillators. There are also gapless modes with negative imaginary parts of eigenvalues that are localized at the edge of the sample. These edge modes should correspond to the edge-localized Lyapunov vectors in our first model (cf. Figs.~\ref{fig3}, \ref{fig23}). We note that the non-Hermiticity of the effective Hamiltonian plays an important role, because nonzero imaginary parts of eigenvalues correspond to nonzero Lyapunov exponents and are unique to non-Hermitian Hamiltonians. It is also noteworthy that the bulk bands with the positive imaginary parts of the eigenvalues are derived from the repulsion of the bulk bands of the original linear couplings with zero imaginary parts (cf. Figs.~\ref{fig24}, \ref{fig25}). To realize such degenerated imaginary bulk bands and lasing edge modes, (at least) four oscillators seem to be necessary at each site.

While nonlinearity generates random on-site loss terms in the effective Hamiltonian, topological modes are robust against such disorder. Especially, previous research \cite{Sone2020} shows that the edge modes of the Hamiltonian describing the linear coupling of our first model \eqref{hamiltonian_TIL_wavenumber} are robust against perturbative on-site gain and loss. In contrast to conventional topological insulators, the bulk bands are also gapless in Fig.~\ref{fig21}. However, the edge and bulk modes are not degenerate, and thus edge modes remain in the effective Hamiltonian. We note that the state-dependent Hamiltonian analyzed in Fig.~\ref{fig4}(a),(b) also includes random on-site loss terms, while they have no effects on the robust existence of the topological modes in the state-dependent Hamiltonian.

We note that the imaginary parts of the eigenvalues of the effective Hamiltonian represent the short-term stability or instability of the system, while the Lyapunov exponents represent the long-term stability or instability. The Lyapunov exponents and vectors correspond to the time-averaged behavior of the effective Hamiltonian and thus are different from the eigenvalues and eigenvectors of the effective Hamiltonian obtained at each period. However, since the edge modes are robust against the unsteady disordered effect of nonlinear terms, the edge-localized modes and their dissipative behaviors remain even after the time averaging.

\begin{figure}[b]
\includegraphics[width=60mm,bb=0 0 300 250,clip]{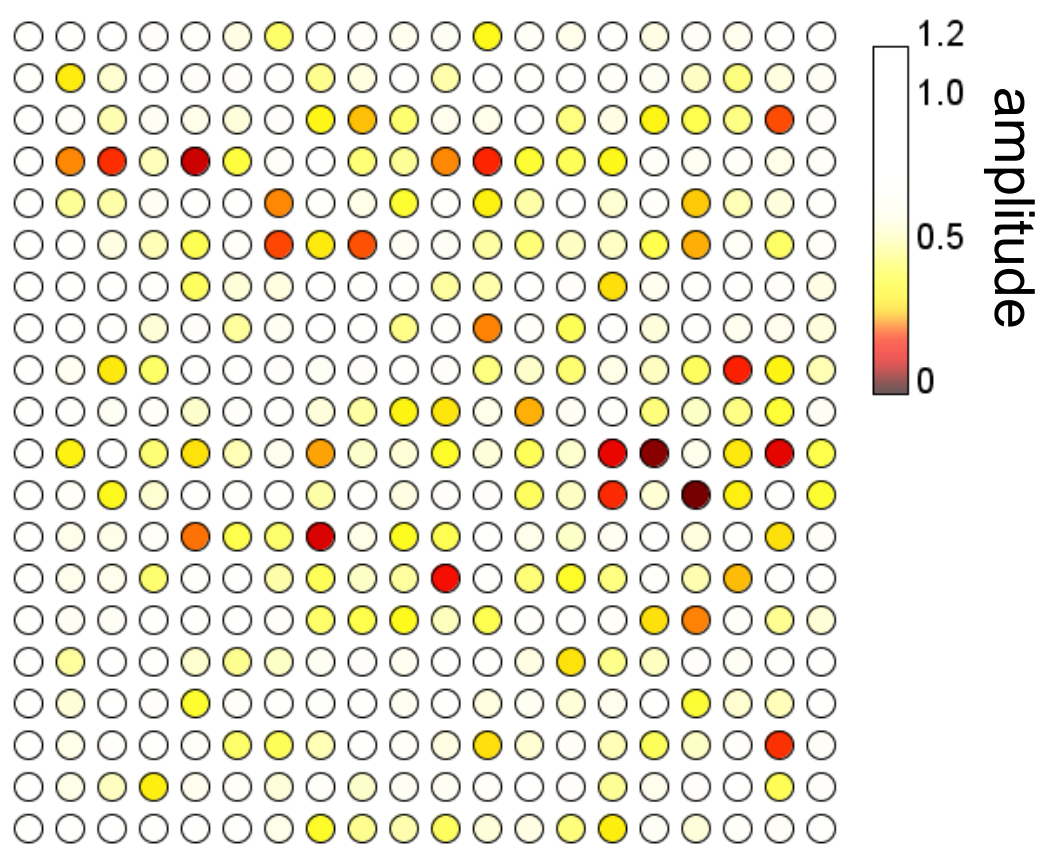}
\caption{\label{fig20} Amplitude distribution indicating amplitude chaos. The amplitudes of oscillators at the first components of lattice points are shown. To obtain the amplitude distribution, we numerically calculate the dynamics of the first model of TSS (cf. Eqs.~\eqref{model1}, \eqref{hamiltonian_TIL_wavenumber}). The red-filled circles represent the site with small amplitudes. The existence of such small-amplitude oscillators indicates amplitude chaos. The parameters used are $u=-1$, $b=0.5$, $\alpha=0.5$, $\beta=1$, $\omega_0=1$, and $\Delta\omega=0.2$.
}
\end{figure}
\section{Detailed analyses on the chaotic behavior in the TSS}\label{appx_F}
\subsection{Amplitude chaos in the TSS}\label{appx_F1}

Amplitude oscillators such as Stuart-Landau oscillators can exhibit two types of chaos, amplitude chaos and phase chaos. Amplitude chaos \cite{Shraiman1992} accompanies phase slips where the amplitudes of oscillators go to zero, and thus the phases jump. Such amplitude-dependent behavior is unique to amplitude oscillators and thus cannot be described by their phase equations. 

We confirm the amplitude chaos in our first model (cf. Eqs.~\eqref{model1}, \eqref{hamiltonian_TIL_wavenumber}). We numerically calculate the dynamics of our model as in Fig.~\ref{fig1}. Figure \ref{fig20} plots the amplitudes of the first component at each site and emphasizes the sites with small amplitudes. A part of bulk oscillators exhibit almost zero amplitudes, and thus phase slips can occur. Therefore, amplitude chaos appears in the bulk of our model. We note that our model can also exhibit the phase chaotic behavior (i.e. chaos irrelevant to the phase slips observed in the numerical calculation).

\subsection{Inverse participation ratios of the Lyapunov vectors in the wavenumber space}\label{appx_F2}
\begin{figure}[t]
\includegraphics[width=70mm,bb=0 0 400 280,clip]{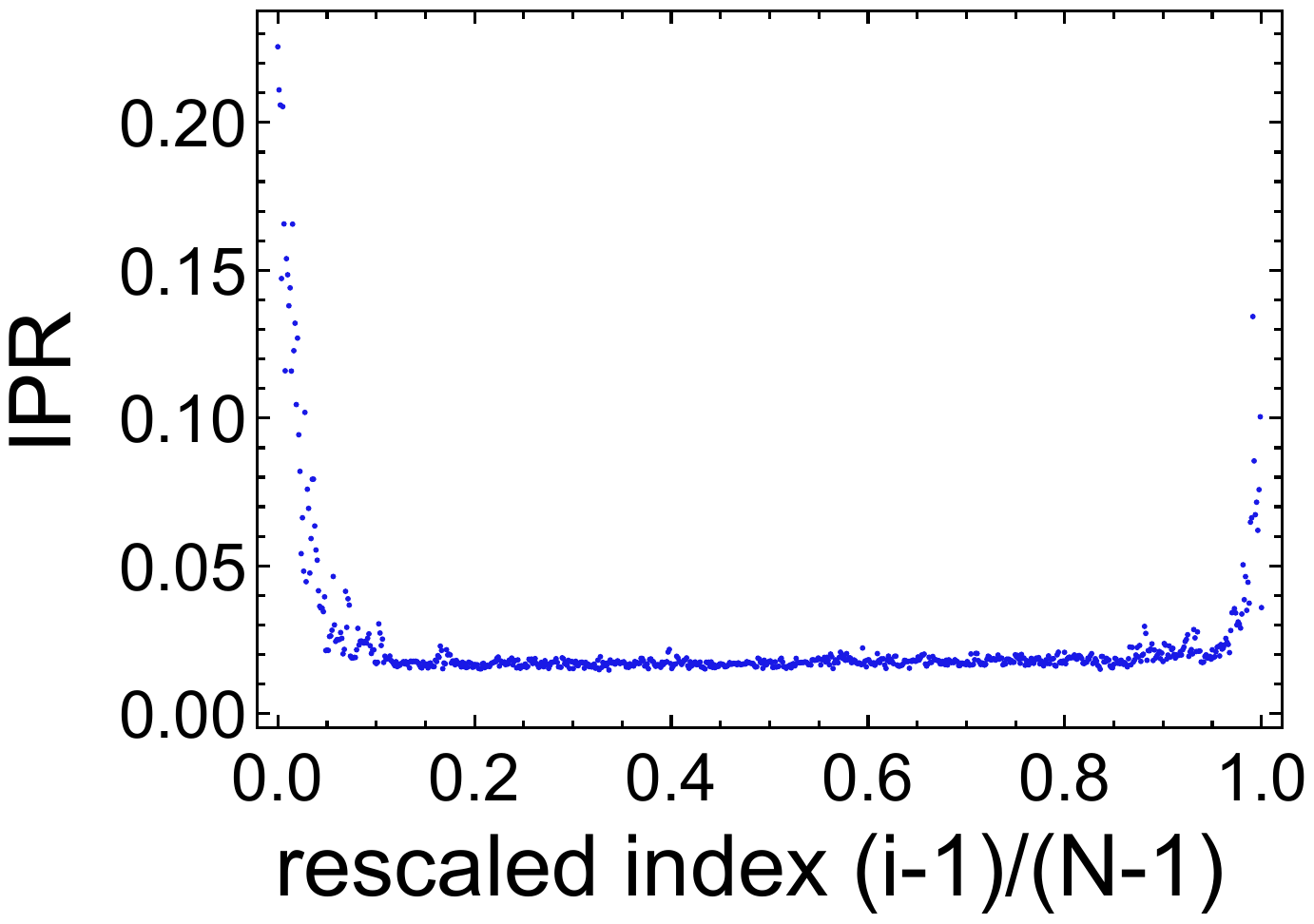}
\caption{\label{fig21} Inverse participation ratios of the Lyapunov vectors in the wavenumber space. We numerically calculate and plot the inverse participation ratios of the Fourier components of the Lyapunov vectors obtained from the model of the TSS (Eqs.~\eqref{model1}, \eqref{hamiltonian_TIL_wavenumber}). The relative indices correspond to those in Fig.~\ref{fig3}. We confirm the large IPRs at small relative indices (smaller than about $0.05$) and large relative indices (larger than about $0.95$). The parameters used are $u=-1$, $b=0.5$, $\alpha=0.5$, $\beta=1$, $\omega_0=0.2$, and $\Delta\omega=0.2$.
}
\end{figure}

While in the main text we analyze the inverse participation ratios (IPRs) of the Lyapunov vectors in the real space, here we calculate and discuss the IPRs of our first model (cf. Eqs.~\eqref{model1}, \eqref{hamiltonian_TIL_wavenumber}) in the wavenumber space. Especially, we find that the IPRs in the wavenumber space classify the bulk-extended Lyapunov vectors into two groups. To obtain the IPRs in the wavenumber space, we conduct the Fourier transformation of the Lyapunov vectors of our model, and calculate the following value:
\begin{equation}
 {\rm IPR}_{\rm w} = \sum_{i,\mathbf{k}} |\tilde{v}_i(\mathbf{k})|^4, \label{wavenumber_ipr}
\end{equation}
where $\tilde{v}_i(\mathbf{k})$ is the Fourier component of the $i$th component of the Lyapunov vectors corresponding to the wavenumber $\mathbf{k}$. Figure \ref{fig21} presents the IPRs of the Lyapunov vectors in the wavenumber space. We find that the first and last $10$ percents of the Lyapunov vectors exhibit large IPRs in the wavenumber space, while the others exhibit small IPRs. We note that the Lyapunov vectors corresponding to the small Lyapunov exponents also exhibit large IPRs in the real space. The Lyapunov vectors exhibiting large Lyapunov exponents and large IPRs in the wavenumber space should belong to a different class from the other bulk-extended ones.

The large IPRs in the wavenumber space are related to the band structure of the effective Hamiltonian obtained from the linearization of the equations of motion (see Appendix \ref{appx_E} and Fig.~\ref{fig19}). The bulk modes with large positive imaginary parts of eigenenergies are concentrated around the wavenumber $k_y=0$. Therefore, we can describe the Lyapunov vectors corresponding to the large Lyapunov exponents as superpositions of the bulk modes only around $k_y=0$, which leads to the large IPRs in the wavenumber space. We can discuss the large IPRs of some edge-localized Lyapunov vectors and the small IPRs of the other ones in a similar way.

Previous research of Hamiltonian chaos \cite{Stratt1979} has studied the participation ratio of coefficients of the eigenstates in a quantum chaotic system when they are described as the superposition of the eigenstates of decoupled harmonic oscillators. Such participation ratios classify the eigenstates in chaotic systems into regular and ergodic ones. In our model, the Lyapunov vectors exhibiting large (small) IPRs in the wavenumber space corresponds to regular (ergodic) ones. However, we leave the full understanding of the topological synchronized state in terms of Hamiltonian chaos to a future work.

\section{Real-space distribution of frequencies and Lyapunov vectors in a $10\times 10$ square lattice}\label{appx_G}
\begin{figure}[t]
\includegraphics[width=86mm,bb=0 0 435 180,clip]{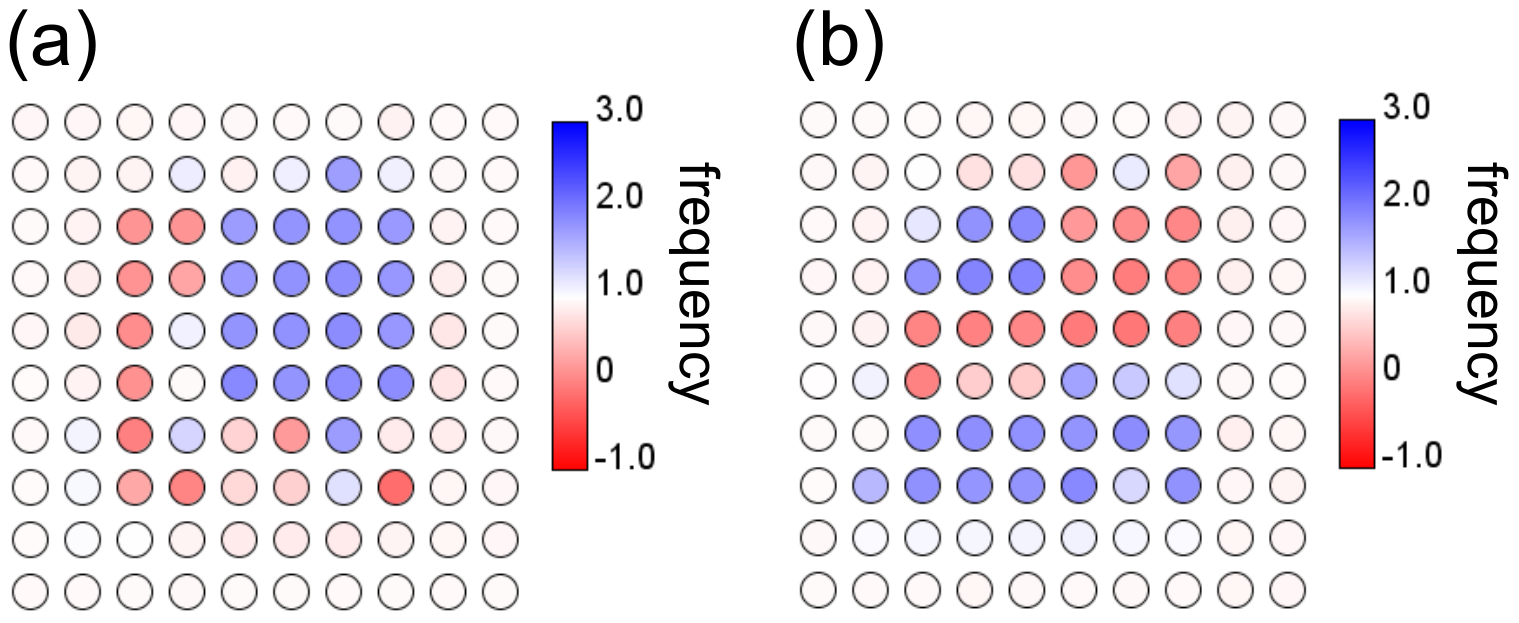}
\caption{\label{fig22} Frequency distribution of the model of the TSS in a $10\times10$ lattice. We numerically simulate the first model of the TSS (Eqs.~\eqref{model1}, \eqref{hamiltonian_TIL_wavenumber}) in a smaller lattice than that considered in Fig.~\ref{fig1}. Here, we arrange $10\times10$ sites and consider the open boundary condition. Panels (a) and (b) show the snapshots of the frequency distributions of the first component of oscillators at the time $t=100$ and $t=200$ for each. As in the $20\times20$ lattices (cf. Fig.~\ref{fig1}), the edge oscillators exhibit homogeneous and constant frequencies, while the bulk ones vibrate at space- and time-varying frequencies. Therefore, the edge oscillators synchronize, while the bulk ones desynchronize, which indicates the emergence of the TSS. The parameters used are $u=-1$, $b=0.5$, $\alpha=0.5$, $\beta=1$, $\omega_0=1$, and $\Delta\omega=0.2$.
}
\end{figure}
\begin{figure*}[t]
\includegraphics[width=120mm,bb=0 0 570 180,clip]{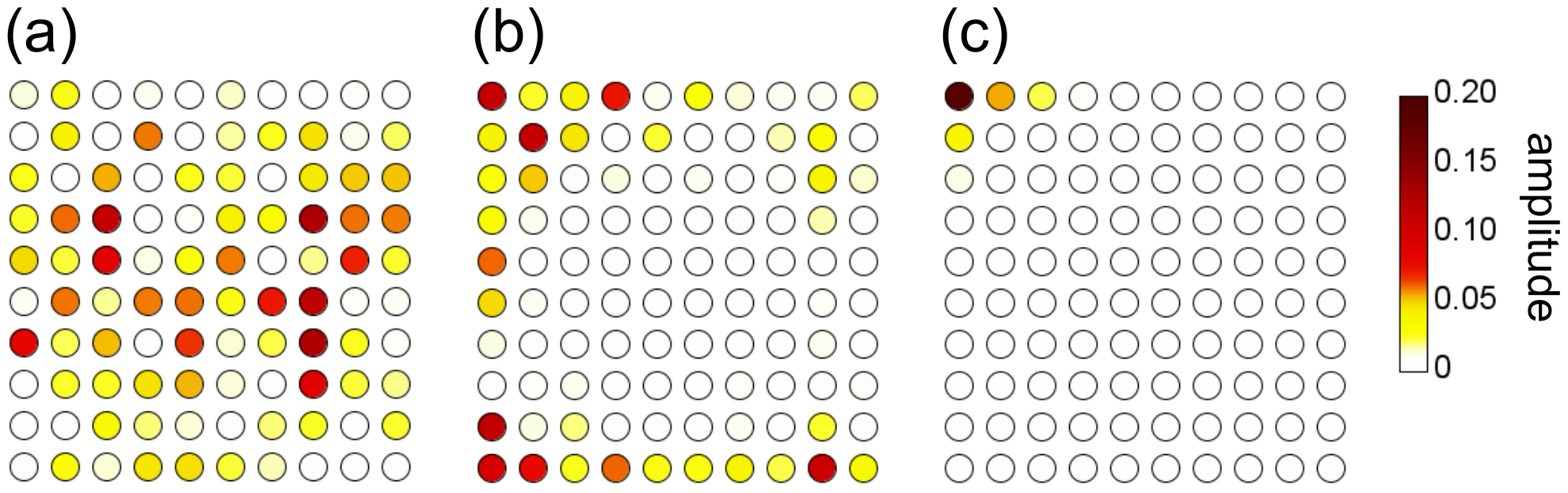}
\caption{\label{fig23} Amplitude distributions of the Lyapunov vectors. The absolute values of the first components of the Lyapunov vectors at each site are plotted in $10\times10$ square lattices. In panel (a), the Lyapunov vector associated with a positive Lyapunov exponent (the index $n=50$) is shown. One can confirm that the bulk sites exhibit large amplitudes. In panel (b), the Lyapunov vector of the index $n=700$ is shown. We can see its localization to the edge of the system. In panel (c), the Lyapunov vector associated with the smallest Lyapunov exponent (the index $n=800$) is shown. It is strongly localized to the upper-left corner. Thus, most of the sites have small amplitudes.
}
\end{figure*}

Related to the calculation in Fig.~\ref{fig3}, we calculate the dynamics of the first model (cf. Eqs.~\eqref{model1}, \eqref{hamiltonian_TIL_wavenumber}) of the TSS in a $10\times10$ square lattice. Figure \ref{fig22} shows snapshots of the frequencies of oscillators. We can confirm the frequency-synchronization of the edge oscillators and the desynchronization of the bulk oscillators. We thus expect that the TSS can appear independently of the system size.

\begin{figure}[t]
\includegraphics[width=70mm,bb=0 0 410 610,clip]{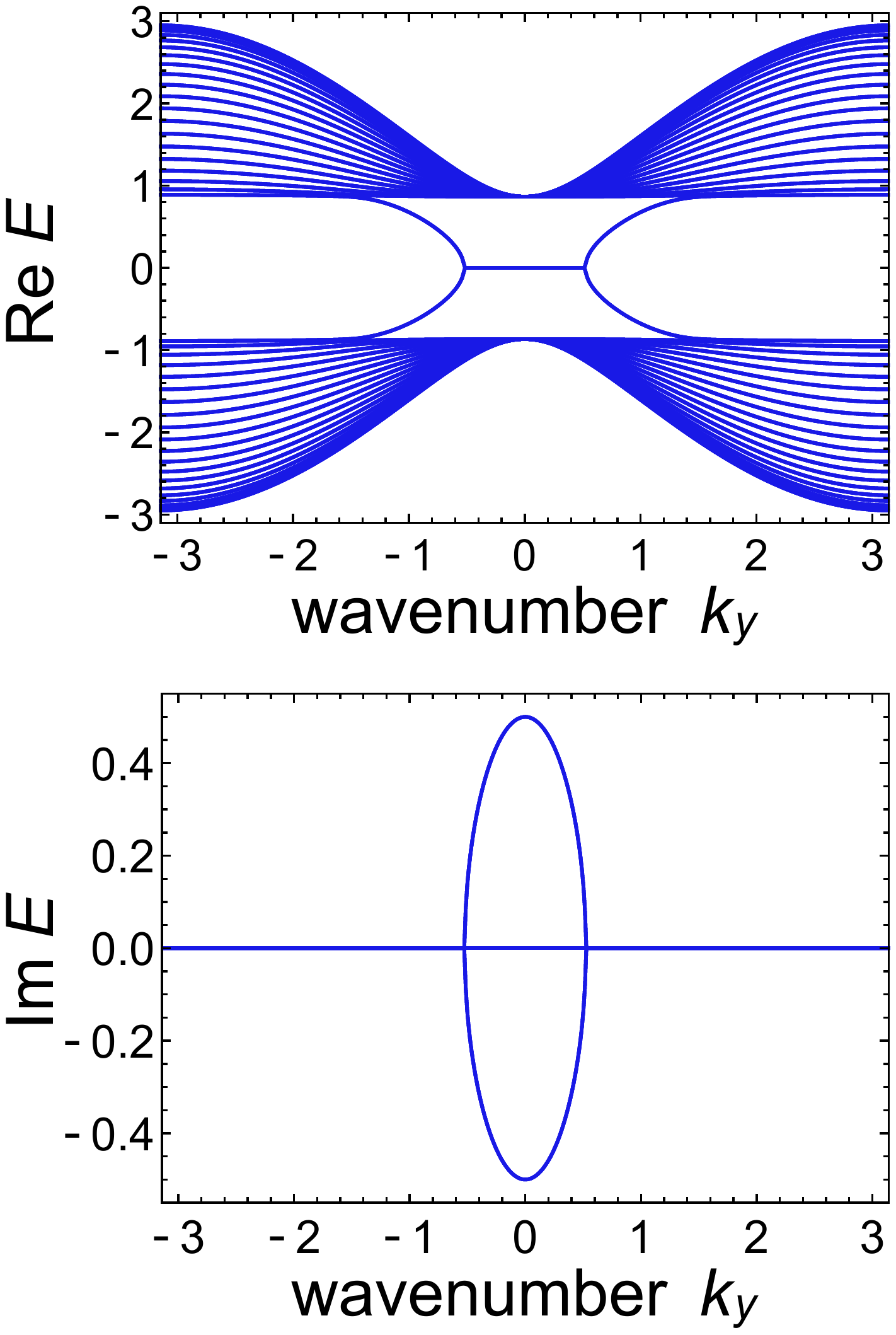}
\caption{\label{fig24} Dispersion relation of the Hamiltonian of topological lasing modes. We calculate the dispersion relation of the Hamiltonian of topological lasing modes utilized in our model \eqref{hamiltonian_TIL_wavenumber}. We consider the open (periodic) boundary condition in the $x$ ($y$) direction and arrange $20$ sites in the $x$ direction. We obtain doubly-degenerate gapless edge modes. Thus, the number of topological modes is four in this Hamiltonian, which is half of that in the state-dependent Hamiltonian. The parameter used are $u=-1$ and $b=0.5$.
}
\end{figure}

We also calculate the Lyapunov vectors \cite{Ginelli2007,Ginelli2013} in the $10\times10$ lattice model and find that some of them are localized to the edge of the system (cf. Fig.~\ref{fig3}(b),(c)). Here we plot the Lyapunov vectors in the real space and directly demonstrate such localization of the Lyapunov vectors. Figure \ref{fig23} shows some examples of Lyapunov vectors. One can see that the Lyapunov vector associated with a positive Lyapunov exponent (corresponding to Fig.~\ref{fig23}(a)) has large amplitudes at a large number of bulk sites. Meanwhile, most of the bulk sites have small amplitudes in the Lyapunov vector of the relative index around $0.875$ (corresponding to Fig.~\ref{fig23}(b)), which indicates its localization to the edge. Furthermore, the Lyapunov vector of the largest index is strongly localized to corner sites as shown in Fig.~\ref{fig23}(c), which leads to the large value of its inverse participation ratio (see Eq.~\eqref{ipr} and Fig.~\ref{fig3}(c)). Thus, these observations of the Lyapunov vectors in the real space are consistent with the results in Fig.~\ref{fig3}.

\section{Dispersion relations of the Hamiltonians of topological insulator lasers}\label{appx_H}

We compare the number of the gapless topological modes in the state-dependent Hamiltonian (cf. Fig.~\ref{fig4}(a)) and the original Hamiltonian of topological lasing modes \eqref{hamiltonian_TIL_wavenumber} utilized as the linear coupling in our first model. Here, we explicitly show the dispersion relation of the Hamiltonian of topological lasing modes without nonlinear loss terms. We consider the periodic boundary condition in the $y$ direction. We arrange $20$ sites and assume the open boundary condition in the $x$ direction. We numerically diagonalize the Hamiltonian and obtain the dispersion relation shown in Fig.~\ref{fig24}. Since all the gapless modes are doubly-degenerated, there are four topological modes in this Hamiltonian. Meanwhile, the state-dependent Hamiltonian exhibits eight topological modes (note that the gapless modes in Fig.~\ref{fig4}(a) are also doubly-degenerated). Therefore, the nonlinearity-induced on-site loss increases the number of topological boundary modes in the state-dependent Hamiltonian.

\begin{figure}[t]
\includegraphics[width=70mm,bb=0 0 420 610,clip]{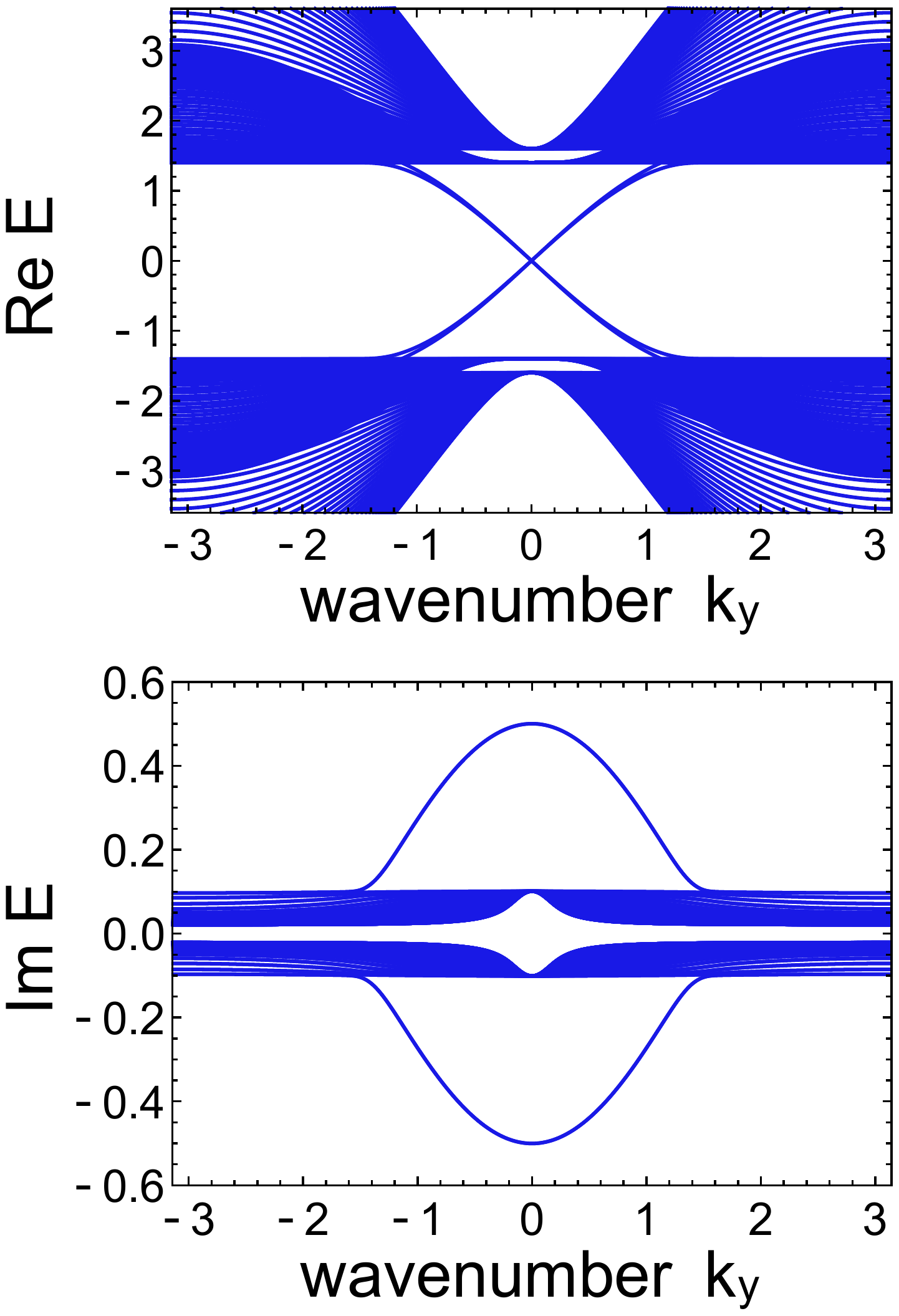}
\caption{\label{fig25} Band structure of lasing edge modes protected by conventional bulk topology. The band structure of the Hamiltonian of lasing edge modes (Eqs.~\eqref{hamiltonian_conventionaltopo_TIL}, \eqref{lasing_2layer_QWZ}) is shown. There exist gapless edge modes with larger imaginary parts of eigenvalues than those of the bulk modes. The parameters used are $u=-1$, $u'=0.02$, $a=2$, and $b=0.5$.
}
\end{figure}

To confirm the existence of lasing edge modes in the Hamiltonian used in our second model \eqref{hamiltonian_conventionaltopo_TIL}, we also calculate its dispersion relation. The Hamiltonian used to describe the linear couplings is
\begin{eqnarray}
 H(\mathbf{k}) &=& \left(
  \begin{array}{cc}
   aH_{\rm QWZ} +iu' I_2 & ibI_2 \\
   ibI_2 & H_{\rm QWZ} -iu' I_2
  \end{array}
  \right) \\
  &=&  \left(u+\cos k_x+\cos k_y\right) \nonumber\\ 
  &{}& \times \left[ \frac{a+1}{2}(I_2 \otimes \sigma_z) + \frac{a-1}{2}(\sigma_z \otimes \sigma_z) \right] \nonumber\\
 &{}& + \sin k_y \left[ \frac{a+1}{2}(I_2 \otimes \sigma_y) + \frac{a-1}{2}(\sigma_z \otimes \sigma_y) \right] \nonumber\\
 &{}& + \sin k_x \left[ \frac{a+1}{2}(I_2 \otimes \sigma_x) + \frac{a-1}{2}(\sigma_z \otimes \sigma_x) \right]  \nonumber\\
 &{}& + ib (\sigma_x \otimes I_2) + iu' (\sigma_z \otimes I_2) \label{lasing_2layer_QWZ}
\end{eqnarray}
in the wavenumber space, where $H_{\rm QWZ}$ is the Hamiltonian of the Qi-Wu-Zhang model \cite{Qi2006} and $I_2$, $\sigma_{x,y,z}$ are the $2\times2$ identity matrix and the Pauli matrices respectively. This Hamiltonian exhibits nonzero Chern numbers and lasing edge modes, i.e., gapless localized modes with the positive imaginary parts of the eigenvalues. We check the existence of such edge modes by calculating the dispersion relation under the open (periodic) boundary condition in the $x$ ($y$) direction. Figure \ref{fig25} shows the dispersion relation of the lasing edge modes. One can find the gapless modes with positive imaginary parts of eigenvalues larger than those of the bulk modes. As discussed in the main text, these edge modes lead to the TSS in the linearly coupled Stuart-Landau oscillators (cf. Fig.~\ref{fig5}).

\section{Absence of synchronized pattern in a topologically trivial system}\label{appx_I}

\begin{figure}[t]
\includegraphics[width=86mm,bb=0 0 690 635,clip]{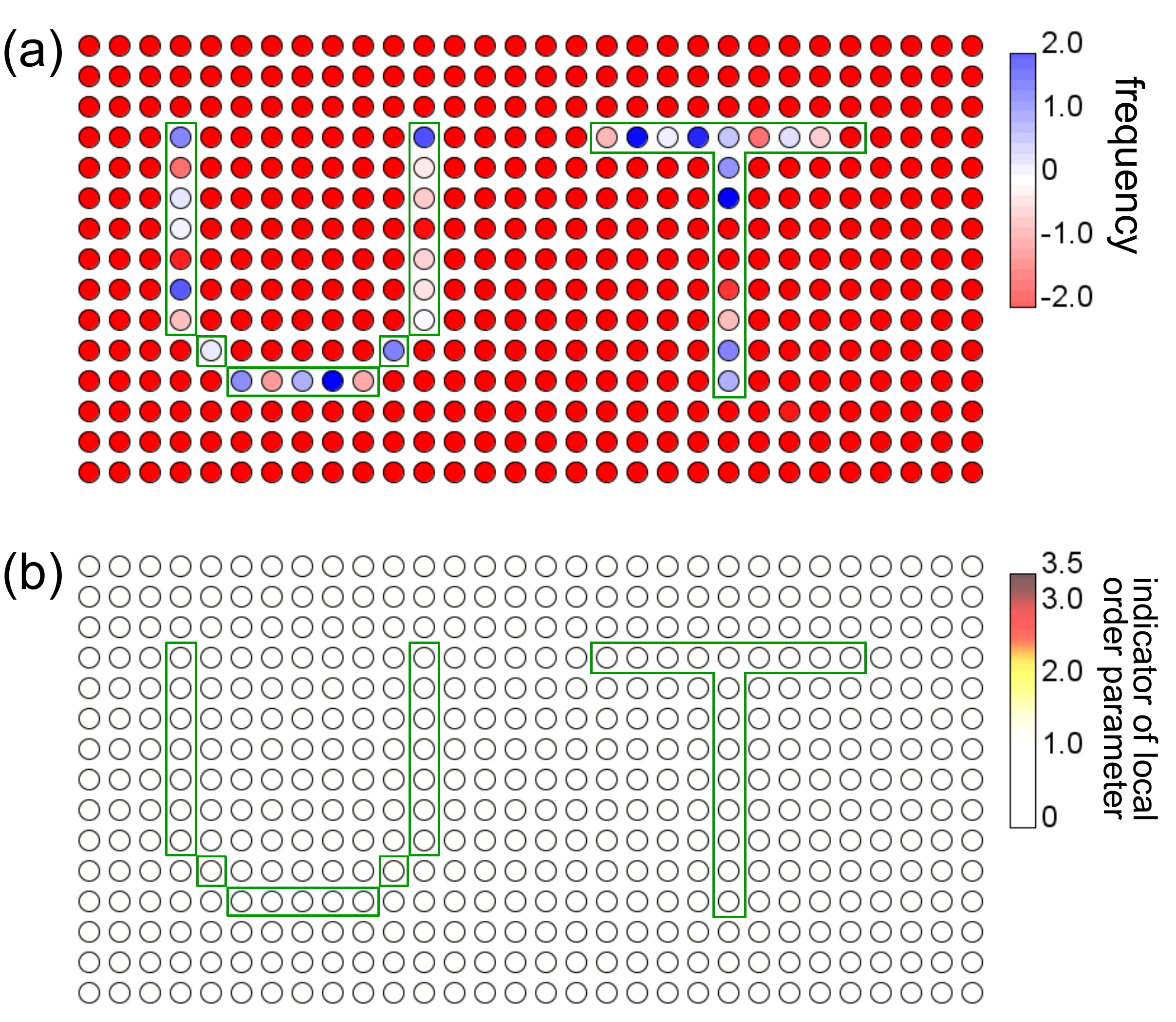}
\caption{\label{fig26} Absence of synchronization pattern in a topologically trivial parameter regime. We numerically simulate the first model of the TSS (Eqs.~\eqref{model1}, \eqref{hamiltonian_TIL_wavenumber}) in a topologically trivial parameter regime and confirm the absence of the TSS and its synchronization pattern. We damp the oscillators encircled by the green boxes. The parameters used are $u=-3$, $b=0.3$, $\alpha=0.5$, $\beta=1$, $\omega_0=0.1$, and $\Delta\omega=0.2$. In panel (a), the frequency of the first component of the oscillator at each site at time $t=200$ is depicted. The undamped oscillators exhibit inhomogeneous frequencies around $-2.0$. In panel (b), the indicator of the local order parameter of the first component of the oscillator at each site is shown. The indicator is small at every site, which indicates the absence of the synchronized oscillators.
}
\end{figure}

The synchronization pattern discussed in Fig.~\ref{fig10} requires the topology of linear coupling. To clearly demonstrate this, we simulate the dynamics of our first model (Eqs.~\eqref{model1}, \eqref{hamiltonian_TIL_wavenumber}) in a topologically trivial parameter regime where the linear couplings exhibit no edge modes. We set the parameter $u=-3$ (cf. Eq.~\eqref{hamiltonian_TIL_wavenumber}) and the other parameters to be equal to those used in Fig.~\ref{fig10}. Figure \ref{fig26} shows the frequencies and the indicators of order parameters in the nontopological parameter region. We can confirm that the frequencies of the oscillators around damped ones are close to those away from damped regions and the indicators of order parameters are small, which implies the absence of TSS. Therefore, the topology and edge modes of the linear couplings play a vital role in the synchronization pattern designing.

\section{Circuit equation for realizing the TSS}\label{appx_J}
Previous studies \cite{Ningyuan2015,Albert2015,Lee2018,Imhof2018,Ezawa2019,Hofmann2019} have proposed and realized topological edge modes in electrical circuits. The relation between the input current $I_a$ and the electrical potential $V_a$ at each node is described by Kirchhoff's law
\begin{eqnarray}
 I_a = \sum_b C_{ab} (V_a-V_b) + C_a V_a = \sum_b J_{ab} V_b, \label{Kirchhoff}
\end{eqnarray}
where $C_{ab}$ ($C_a$) is the inverse of the impedance between site $a$ and $b$ (site $a$ and ground). $iJ_{ab}$ plays the role of the effective Hamiltonian in electrical circuits. Resistors play the role of imaginary couplings, while capacitors and inductors lead to real terms in the effective Hamiltonian. By tuning the impedances of circuit elements, one can realize the effective Hamiltonian imitating topological materials. If we introduce capacitors with capacitance $C$ as current resources, we can rewrite Eq.~\eqref{Kirchhoff} as
\begin{eqnarray}
 \frac{dV_a}{dt} = \frac{1}{C} \sum_b J_{ab} V_b, \label{Kirchhoff2}
\end{eqnarray}
which governs the time evolution of the voltage at each node. 

To realize the electrical circuit of topological lasing modes, we utilize negative impedance converters with current inversion (ICINs), which is discussed in a previous study \cite{Hofmann2019}. The ICINs realize nonreciprocal couplings ($C_{ab} \neq C_{ba}$) that cannot be realized by using resistors, inductors, and capacitors. Input current $I_{\rm in}$ and output current $I_{\rm out}$ are described as 
\begin{eqnarray}
 I_{\rm out} &=& \frac{1}{R} (V_{\rm in}-V_{\rm out}), \\
 I_{\rm in} &=& - \frac{1}{R'} (V_{\rm in}-V_{\rm out}),
\end{eqnarray}
where $V_{\rm in}$ ($V_{\rm out}$) is the electric potentials at the input (output) side. We assume $R=R'$ in the electrical circuit of the TSS (cf. Fig.~\ref{fig12}). Then by tuning the parameters of circuit elements, one can construct the effective Hamiltonian $iJ$ that is equal to the Hamiltonian of the topological insulator laser \cite{Song2020,Sone2020} used in our first model \eqref{hamiltonian_TIL_wavenumber}. It is noteworthy that impedances of capacitors and inductors and thus the effective Hamiltonian $iJ$ depends on the driving voltage frequency.

To directly realize an electrical circuit whose voltage dynamics is described by nonlinear equations similar to our model, we next omit capacitors and inductors and construct a frequency-independent circuit of topological lasing modes. To effectively realize imaginary terms without capacitors and inductors, we prepare twice the number of nodes compared to the circuit in Fig.~\ref{fig12} and express one complex-valued state variable by using the voltages at two nodes as $Z=V_1+iV_2$. Then, real and imaginary couplings are substituted as
\begin{eqnarray}
 CZ_j &{\rightarrow}& \left(
  \begin{array}{cc}
   C & 0 \\
   0 & C
  \end{array}
  \right)
  \left(
  \begin{array}{c}
   V_{j1} \\
   V_{j2}
  \end{array}
  \right), \\
 iCZ_j &{\rightarrow}& \left(
  \begin{array}{cc}
   0 & -C \\
   C & 0
  \end{array}
  \right)
  \left(
  \begin{array}{c}
   V_{j1} \\
   V_{j2}
  \end{array}
  \right), 
\end{eqnarray}
where $C$ is real. These real couplings are realized by the substitution of circuit elements shown in Fig.~\ref{fig12}(b).

Each node in the topological circuit of the TSS consists of van der Pol circuits \cite{Pol1926,Johnson2014}, which realize nonlinear oscillators. To construct the van der Pol circuits, we utilize nonlinear resistors whose conductance is described as $\tilde{C}(V_i)$. The dynamics of the van der Pol circuits is derived by Kirchhoff's law as
\begin{eqnarray}
 C\frac{dV_i}{dt} = \tilde{C}(V_i) V_i + I_{\rm in}, \label{van_der_Pol}
\end{eqnarray}
where $C$ is the capacitance of the capacitor and $I_{\rm in}$ is the input current. When we can expand $\tilde{C}(V_i)$ as $\tilde{C}(V_i) = \alpha - \beta V_i^2 + \mathcal{O}(V_i^3)$, the dynamics of the van der Pol circuits can be approximated by the Stuart-Landau oscillators in the leading order. Therefore, the entire dynamics of the circuit imitates the first model of the TSS (Eqs.~\eqref{model1}, \eqref{hamiltonian_TIL_wavenumber}).

\bibliography{reference}

\begin{thebibliography}{66}%
\makeatletter
\providecommand \@ifxundefined [1]{%
 \@ifx{#1\undefined}
}%
\providecommand \@ifnum [1]{%
 \ifnum #1\expandafter \@firstoftwo
 \else \expandafter \@secondoftwo
 \fi
}%
\providecommand \@ifx [1]{%
 \ifx #1\expandafter \@firstoftwo
 \else \expandafter \@secondoftwo
 \fi
}%
\providecommand \natexlab [1]{#1}%
\providecommand \enquote  [1]{``#1''}%
\providecommand \bibnamefont  [1]{#1}%
\providecommand \bibfnamefont [1]{#1}%
\providecommand \citenamefont [1]{#1}%
\providecommand \href@noop [0]{\@secondoftwo}%
\providecommand \href [0]{\begingroup \@sanitize@url \@href}%
\providecommand \@href[1]{\@@startlink{#1}\@@href}%
\providecommand \@@href[1]{\endgroup#1\@@endlink}%
\providecommand \@sanitize@url [0]{\catcode `\\12\catcode `\$12\catcode
  `\&12\catcode `\#12\catcode `\^12\catcode `\_12\catcode `\%12\relax}%
\providecommand \@@startlink[1]{}%
\providecommand \@@endlink[0]{}%
\providecommand \url  [0]{\begingroup\@sanitize@url \@url }%
\providecommand \@url [1]{\endgroup\@href {#1}{\urlprefix }}%
\providecommand \urlprefix  [0]{URL }%
\providecommand \Eprint [0]{\href }%
\providecommand \doibase [0]{http://dx.doi.org/}%
\providecommand \selectlanguage [0]{\@gobble}%
\providecommand \bibinfo  [0]{\@secondoftwo}%
\providecommand \bibfield  [0]{\@secondoftwo}%
\providecommand \translation [1]{[#1]}%
\providecommand \BibitemOpen [0]{}%
\providecommand \bibitemStop [0]{}%
\providecommand \bibitemNoStop [0]{.\EOS\space}%
\providecommand \EOS [0]{\spacefactor3000\relax}%
\providecommand \BibitemShut  [1]{\csname bibitem#1\endcsname}%
\let\auto@bib@innerbib\@empty
%</preamble>
\bibitem [{\citenamefont {Buck}(1988)}]{Buck1988}%
  \BibitemOpen
  \bibfield  {author} {\bibinfo {author} {\bibfnamefont {J.}~\bibnamefont
  {Buck}},\ }\href {\doibase 10.1086/415929} {\bibfield  {journal} {\bibinfo
  {journal} {Quart. Rev. Biol.}\ }\textbf {\bibinfo {volume} {63}},\ \bibinfo
  {pages} {265} (\bibinfo {year} {1988})}\BibitemShut {NoStop}%
\bibitem [{\citenamefont {Luo}\ and\ \citenamefont {Rudy}(1991)}]{Luo1991}%
  \BibitemOpen
  \bibfield  {author} {\bibinfo {author} {\bibfnamefont {C.~H.}\ \bibnamefont
  {Luo}}\ and\ \bibinfo {author} {\bibfnamefont {Y.}~\bibnamefont {Rudy}},\
  }\href {\doibase 10.1161/01.RES.68.6.1501} {\bibfield  {journal} {\bibinfo
  {journal} {Circ. Res.}\ }\textbf {\bibinfo {volume} {68}},\ \bibinfo {pages}
  {1501} (\bibinfo {year} {1991})}\BibitemShut {NoStop}%
\bibitem [{\citenamefont {Kozyreff}\ \emph {et~al.}(2000)\citenamefont
  {Kozyreff}, \citenamefont {Vladimirov},\ and\ \citenamefont
  {Mandel}}]{Kozyreff2000}%
  \BibitemOpen
  \bibfield  {author} {\bibinfo {author} {\bibfnamefont {G.}~\bibnamefont
  {Kozyreff}}, \bibinfo {author} {\bibfnamefont {A.~G.}\ \bibnamefont
  {Vladimirov}}, \ and\ \bibinfo {author} {\bibfnamefont {P.}~\bibnamefont
  {Mandel}},\ }\href {\doibase 10.1103/PhysRevLett.85.3809} {\bibfield
  {journal} {\bibinfo  {journal} {Phys. Rev. Lett.}\ }\textbf {\bibinfo
  {volume} {85}},\ \bibinfo {pages} {3809} (\bibinfo {year}
  {2000})}\BibitemShut {NoStop}%
\bibitem [{\citenamefont {Ludwig}\ and\ \citenamefont
  {Marquardt}(2013)}]{Ludwig2013}%
  \BibitemOpen
  \bibfield  {author} {\bibinfo {author} {\bibfnamefont {M.}~\bibnamefont
  {Ludwig}}\ and\ \bibinfo {author} {\bibfnamefont {F.}~\bibnamefont
  {Marquardt}},\ }\href {\doibase 10.1103/PhysRevLett.111.073603} {\bibfield
  {journal} {\bibinfo  {journal} {Phys. Rev. Lett.}\ }\textbf {\bibinfo
  {volume} {111}},\ \bibinfo {pages} {073603} (\bibinfo {year}
  {2013})}\BibitemShut {NoStop}%
\bibitem [{\citenamefont {Blaabjerg}\ \emph {et~al.}(2006)\citenamefont
  {Blaabjerg}, \citenamefont {Teodorescu}, \citenamefont {Liserre},\ and\
  \citenamefont {Timbus}}]{Blaabjerg2006}%
  \BibitemOpen
  \bibfield  {author} {\bibinfo {author} {\bibfnamefont {F.}~\bibnamefont
  {Blaabjerg}}, \bibinfo {author} {\bibfnamefont {R.}~\bibnamefont
  {Teodorescu}}, \bibinfo {author} {\bibfnamefont {M.}~\bibnamefont {Liserre}},
  \ and\ \bibinfo {author} {\bibfnamefont {A.}~\bibnamefont {Timbus}},\ }\href
  {\doibase 10.1109/TIE.2006.881997} {\bibfield  {journal} {\bibinfo  {journal}
  {IEEE Trans. Ind. Electron.}\ }\textbf {\bibinfo {volume} {53}},\ \bibinfo
  {pages} {1398} (\bibinfo {year} {2006})}\BibitemShut {NoStop}%
\bibitem [{\citenamefont {Wiesenfeld}\ \emph {et~al.}(1996)\citenamefont
  {Wiesenfeld}, \citenamefont {Colet},\ and\ \citenamefont
  {Strogatz}}]{Wiesenfeld1996}%
  \BibitemOpen
  \bibfield  {author} {\bibinfo {author} {\bibfnamefont {K.}~\bibnamefont
  {Wiesenfeld}}, \bibinfo {author} {\bibfnamefont {P.}~\bibnamefont {Colet}}, \
  and\ \bibinfo {author} {\bibfnamefont {S.~H.}\ \bibnamefont {Strogatz}},\
  }\href {\doibase 10.1103/PhysRevLett.76.404} {\bibfield  {journal} {\bibinfo
  {journal} {Phys. Rev. Lett.}\ }\textbf {\bibinfo {volume} {76}},\ \bibinfo
  {pages} {404} (\bibinfo {year} {1996})}\BibitemShut {NoStop}%
\bibitem [{\citenamefont {Acebr\'on}\ \emph {et~al.}(2005)\citenamefont
  {Acebr\'on}, \citenamefont {Bonilla}, \citenamefont {P\'erez}, \citenamefont
  {Ritort},\ and\ \citenamefont {Spigler}}]{Acebron2005}%
  \BibitemOpen
  \bibfield  {author} {\bibinfo {author} {\bibfnamefont {J.~A.}\ \bibnamefont
  {Acebr\'on}}, \bibinfo {author} {\bibfnamefont {L.~L.}\ \bibnamefont
  {Bonilla}}, \bibinfo {author} {\bibfnamefont {V.~C.~J.}\ \bibnamefont
  {P\'erez}}, \bibinfo {author} {\bibfnamefont {F.}~\bibnamefont {Ritort}}, \
  and\ \bibinfo {author} {\bibfnamefont {R.}~\bibnamefont {Spigler}},\ }\href
  {\doibase 10.1103/RevModPhys.77.137} {\bibfield  {journal} {\bibinfo
  {journal} {Rev. Mod. Phys.}\ }\textbf {\bibinfo {volume} {77}},\ \bibinfo
  {pages} {137} (\bibinfo {year} {2005})}\BibitemShut {NoStop}%
\bibitem [{\citenamefont {Dibner}\ \emph {et~al.}(2010)\citenamefont {Dibner},
  \citenamefont {Schibler},\ and\ \citenamefont {Albrecht}}]{Dibner2010}%
  \BibitemOpen
  \bibfield  {author} {\bibinfo {author} {\bibfnamefont {C.}~\bibnamefont
  {Dibner}}, \bibinfo {author} {\bibfnamefont {U.}~\bibnamefont {Schibler}}, \
  and\ \bibinfo {author} {\bibfnamefont {U.}~\bibnamefont {Albrecht}},\ }\href
  {\doibase 10.1146/annurev-physiol-021909-135821} {\bibfield  {journal}
  {\bibinfo  {journal} {Annu. Rev. Physiol.}\ }\textbf {\bibinfo {volume}
  {72}},\ \bibinfo {pages} {517} (\bibinfo {year} {2010})}\BibitemShut
  {NoStop}%
\bibitem [{\citenamefont {Lee}\ and\ \citenamefont
  {Sadeghpour}(2013)}]{Lee2013}%
  \BibitemOpen
  \bibfield  {author} {\bibinfo {author} {\bibfnamefont {T.~E.}\ \bibnamefont
  {Lee}}\ and\ \bibinfo {author} {\bibfnamefont {H.~R.}\ \bibnamefont
  {Sadeghpour}},\ }\href {\doibase 10.1103/PhysRevLett.111.234101} {\bibfield
  {journal} {\bibinfo  {journal} {Phys. Rev. Lett.}\ }\textbf {\bibinfo
  {volume} {111}},\ \bibinfo {pages} {234101} (\bibinfo {year}
  {2013})}\BibitemShut {NoStop}%
\bibitem [{\citenamefont {Laskar}\ \emph {et~al.}(2020)\citenamefont {Laskar},
  \citenamefont {Adhikary}, \citenamefont {Mondal}, \citenamefont {Katiyar},
  \citenamefont {Vinjanampathy},\ and\ \citenamefont {Ghosh}}]{Laskar2020}%
  \BibitemOpen
  \bibfield  {author} {\bibinfo {author} {\bibfnamefont {A.~W.}\ \bibnamefont
  {Laskar}}, \bibinfo {author} {\bibfnamefont {P.}~\bibnamefont {Adhikary}},
  \bibinfo {author} {\bibfnamefont {S.}~\bibnamefont {Mondal}}, \bibinfo
  {author} {\bibfnamefont {P.}~\bibnamefont {Katiyar}}, \bibinfo {author}
  {\bibfnamefont {S.}~\bibnamefont {Vinjanampathy}}, \ and\ \bibinfo {author}
  {\bibfnamefont {S.}~\bibnamefont {Ghosh}},\ }\href {\doibase
  10.1103/PhysRevLett.125.013601} {\bibfield  {journal} {\bibinfo  {journal}
  {Phys. Rev. Lett.}\ }\textbf {\bibinfo {volume} {125}},\ \bibinfo {pages}
  {013601} (\bibinfo {year} {2020})}\BibitemShut {NoStop}%
\bibitem [{\citenamefont {Sieber}\ \emph {et~al.}(2014)\citenamefont {Sieber},
  \citenamefont {Omel'chenko},\ and\ \citenamefont {Wolfrum}}]{Sieber2014}%
  \BibitemOpen
  \bibfield  {author} {\bibinfo {author} {\bibfnamefont {J.}~\bibnamefont
  {Sieber}}, \bibinfo {author} {\bibfnamefont {O.~E.}\ \bibnamefont
  {Omel'chenko}}, \ and\ \bibinfo {author} {\bibfnamefont {M.}~\bibnamefont
  {Wolfrum}},\ }\href {\doibase 10.1103/PhysRevLett.112.054102} {\bibfield
  {journal} {\bibinfo  {journal} {Phys. Rev. Lett.}\ }\textbf {\bibinfo
  {volume} {112}},\ \bibinfo {pages} {054102} (\bibinfo {year}
  {2014})}\BibitemShut {NoStop}%
\bibitem [{\citenamefont {Isele}\ \emph {et~al.}(2016)\citenamefont {Isele},
  \citenamefont {Hizanidis}, \citenamefont {Provata},\ and\ \citenamefont
  {H\"ovel}}]{Isele2016}%
  \BibitemOpen
  \bibfield  {author} {\bibinfo {author} {\bibfnamefont {T.}~\bibnamefont
  {Isele}}, \bibinfo {author} {\bibfnamefont {J.}~\bibnamefont {Hizanidis}},
  \bibinfo {author} {\bibfnamefont {A.}~\bibnamefont {Provata}}, \ and\
  \bibinfo {author} {\bibfnamefont {P.}~\bibnamefont {H\"ovel}},\ }\href
  {\doibase 10.1103/PhysRevE.93.022217} {\bibfield  {journal} {\bibinfo
  {journal} {Phys. Rev. E}\ }\textbf {\bibinfo {volume} {93}},\ \bibinfo
  {pages} {022217} (\bibinfo {year} {2016})}\BibitemShut {NoStop}%
\bibitem [{\citenamefont {Zhang}\ \emph
  {et~al.}(2020{\natexlab{a}})\citenamefont {Zhang}, \citenamefont {Nicolaou},
  \citenamefont {Hart}, \citenamefont {Roy},\ and\ \citenamefont
  {Motter}}]{Zhang2020a}%
  \BibitemOpen
  \bibfield  {author} {\bibinfo {author} {\bibfnamefont {Y.}~\bibnamefont
  {Zhang}}, \bibinfo {author} {\bibfnamefont {Z.~G.}\ \bibnamefont {Nicolaou}},
  \bibinfo {author} {\bibfnamefont {J.~D.}\ \bibnamefont {Hart}}, \bibinfo
  {author} {\bibfnamefont {R.}~\bibnamefont {Roy}}, \ and\ \bibinfo {author}
  {\bibfnamefont {A.~E.}\ \bibnamefont {Motter}},\ }\href {\doibase
  10.1103/PhysRevX.10.011044} {\bibfield  {journal} {\bibinfo  {journal} {Phys.
  Rev. X}\ }\textbf {\bibinfo {volume} {10}},\ \bibinfo {pages} {011044}
  (\bibinfo {year} {2020}{\natexlab{a}})}\BibitemShut {NoStop}%
\bibitem [{\citenamefont {Kane}\ and\ \citenamefont {Mele}(2005)}]{Kane2005}%
  \BibitemOpen
  \bibfield  {author} {\bibinfo {author} {\bibfnamefont {C.~L.}\ \bibnamefont
  {Kane}}\ and\ \bibinfo {author} {\bibfnamefont {E.~J.}\ \bibnamefont
  {Mele}},\ }\href {\doibase 10.1103/PhysRevLett.95.146802} {\bibfield
  {journal} {\bibinfo  {journal} {Phys. Rev. Lett.}\ }\textbf {\bibinfo
  {volume} {95}},\ \bibinfo {pages} {146802} (\bibinfo {year}
  {2005})}\BibitemShut {NoStop}%
\bibitem [{\citenamefont {Hasan}\ and\ \citenamefont {Kane}(2010)}]{Hasan2010}%
  \BibitemOpen
  \bibfield  {author} {\bibinfo {author} {\bibfnamefont {M.~Z.}\ \bibnamefont
  {Hasan}}\ and\ \bibinfo {author} {\bibfnamefont {C.~L.}\ \bibnamefont
  {Kane}},\ }\href {\doibase 10.1103/RevModPhys.82.3045} {\bibfield  {journal}
  {\bibinfo  {journal} {Rev. Mod. Phys.}\ }\textbf {\bibinfo {volume} {82}},\
  \bibinfo {pages} {3045} (\bibinfo {year} {2010})}\BibitemShut {NoStop}%
\bibitem [{\citenamefont {Qi}\ and\ \citenamefont {Zhang}(2011)}]{Qi2011}%
  \BibitemOpen
  \bibfield  {author} {\bibinfo {author} {\bibfnamefont {X.~L.}\ \bibnamefont
  {Qi}}\ and\ \bibinfo {author} {\bibfnamefont {S.~C.}\ \bibnamefont {Zhang}},\
  }\href {\doibase 10.1103/RevModPhys.83.1057} {\bibfield  {journal} {\bibinfo
  {journal} {Rev. Mod. Phys.}\ }\textbf {\bibinfo {volume} {83}},\ \bibinfo
  {pages} {1057} (\bibinfo {year} {2011})}\BibitemShut {NoStop}%
\bibitem [{\citenamefont {Smirnova}\ \emph {et~al.}(2020)\citenamefont
  {Smirnova}, \citenamefont {Leykam}, \citenamefont {Chong},\ and\
  \citenamefont {Kivshar}}]{Smirnova2020}%
  \BibitemOpen
  \bibfield  {author} {\bibinfo {author} {\bibfnamefont {D.}~\bibnamefont
  {Smirnova}}, \bibinfo {author} {\bibfnamefont {D.}~\bibnamefont {Leykam}},
  \bibinfo {author} {\bibfnamefont {Y.}~\bibnamefont {Chong}}, \ and\ \bibinfo
  {author} {\bibfnamefont {Y.}~\bibnamefont {Kivshar}},\ }\href {\doibase
  10.1063/1.5142397} {\bibfield  {journal} {\bibinfo  {journal} {Appl. Phys.
  Rev.}\ }\textbf {\bibinfo {volume} {7}},\ \bibinfo {pages} {021306} (\bibinfo
  {year} {2020})}\BibitemShut {NoStop}%
\bibitem [{\citenamefont {Ota}\ \emph {et~al.}(2020)\citenamefont {Ota},
  \citenamefont {Takata}, \citenamefont {Ozawa}, \citenamefont {Amo},
  \citenamefont {Jia}, \citenamefont {Kante}, \citenamefont {Notomi},
  \citenamefont {Arakawa},\ and\ \citenamefont {Iwamoto}}]{Ota2020}%
  \BibitemOpen
  \bibfield  {author} {\bibinfo {author} {\bibfnamefont {Y.}~\bibnamefont
  {Ota}}, \bibinfo {author} {\bibfnamefont {K.}~\bibnamefont {Takata}},
  \bibinfo {author} {\bibfnamefont {T.}~\bibnamefont {Ozawa}}, \bibinfo
  {author} {\bibfnamefont {A.}~\bibnamefont {Amo}}, \bibinfo {author}
  {\bibfnamefont {Z.}~\bibnamefont {Jia}}, \bibinfo {author} {\bibfnamefont
  {B.}~\bibnamefont {Kante}}, \bibinfo {author} {\bibfnamefont
  {M.}~\bibnamefont {Notomi}}, \bibinfo {author} {\bibfnamefont
  {Y.}~\bibnamefont {Arakawa}}, \ and\ \bibinfo {author} {\bibfnamefont
  {S.}~\bibnamefont {Iwamoto}},\ }\href {\doibase doi:10.1515/nanoph-2019-0376}
  {\bibfield  {journal} {\bibinfo  {journal} {Nanophotonics}\ }\textbf
  {\bibinfo {volume} {9}},\ \bibinfo {pages} {547} (\bibinfo {year}
  {2020})}\BibitemShut {NoStop}%
\bibitem [{\citenamefont {Kotwal}\ \emph {et~al.}(2021)\citenamefont {Kotwal},
  \citenamefont {Moseley}, \citenamefont {Stegmaier}, \citenamefont {Imhof},
  \citenamefont {Brand}, \citenamefont {Kie{\ss}ling}, \citenamefont {Thomale},
  \citenamefont {Ronellenfitsch},\ and\ \citenamefont {Dunkel}}]{Kotwal2021}%
  \BibitemOpen
  \bibfield  {author} {\bibinfo {author} {\bibfnamefont {T.}~\bibnamefont
  {Kotwal}}, \bibinfo {author} {\bibfnamefont {F.}~\bibnamefont {Moseley}},
  \bibinfo {author} {\bibfnamefont {A.}~\bibnamefont {Stegmaier}}, \bibinfo
  {author} {\bibfnamefont {S.}~\bibnamefont {Imhof}}, \bibinfo {author}
  {\bibfnamefont {H.}~\bibnamefont {Brand}}, \bibinfo {author} {\bibfnamefont
  {T.}~\bibnamefont {Kie{\ss}ling}}, \bibinfo {author} {\bibfnamefont
  {R.}~\bibnamefont {Thomale}}, \bibinfo {author} {\bibfnamefont
  {H.}~\bibnamefont {Ronellenfitsch}}, \ and\ \bibinfo {author} {\bibfnamefont
  {J.}~\bibnamefont {Dunkel}},\ }\href
  {https://www.pnas.org/content/118/32/e2106411118} {\bibfield  {journal}
  {\bibinfo  {journal} {Proc. Natl. Acad. Sci. U.S.A.}\ }\textbf {\bibinfo
  {volume} {118}},\ \bibinfo {pages} {e2106411118} (\bibinfo {year}
  {2021})}\BibitemShut {NoStop}%
\bibitem [{\citenamefont {Zhang}\ \emph
  {et~al.}(2020{\natexlab{b}})\citenamefont {Zhang}, \citenamefont {Wang},
  \citenamefont {Zhang}, \citenamefont {Kartashov}, \citenamefont {Li},
  \citenamefont {Zhong}, \citenamefont {Guan}, \citenamefont {Gao},
  \citenamefont {Li}, \citenamefont {Zhang},\ and\ \citenamefont
  {Xiao}}]{Zhang2020b}%
  \BibitemOpen
  \bibfield  {author} {\bibinfo {author} {\bibfnamefont {Z.}~\bibnamefont
  {Zhang}}, \bibinfo {author} {\bibfnamefont {R.}~\bibnamefont {Wang}},
  \bibinfo {author} {\bibfnamefont {Y.}~\bibnamefont {Zhang}}, \bibinfo
  {author} {\bibfnamefont {Y.~V.}\ \bibnamefont {Kartashov}}, \bibinfo {author}
  {\bibfnamefont {F.}~\bibnamefont {Li}}, \bibinfo {author} {\bibfnamefont
  {H.}~\bibnamefont {Zhong}}, \bibinfo {author} {\bibfnamefont
  {H.}~\bibnamefont {Guan}}, \bibinfo {author} {\bibfnamefont {K.}~\bibnamefont
  {Gao}}, \bibinfo {author} {\bibfnamefont {F.}~\bibnamefont {Li}}, \bibinfo
  {author} {\bibfnamefont {Y.}~\bibnamefont {Zhang}}, \ and\ \bibinfo {author}
  {\bibfnamefont {M.}~\bibnamefont {Xiao}},\ }\href {\doibase
  10.1038/s41467-020-15635-9} {\bibfield  {journal} {\bibinfo  {journal} {Nat.
  Commun.}\ }\textbf {\bibinfo {volume} {11}},\ \bibinfo {pages} {1902}
  (\bibinfo {year} {2020}{\natexlab{b}})}\BibitemShut {NoStop}%
\bibitem [{\citenamefont {Chen}\ \emph {et~al.}(2014)\citenamefont {Chen},
  \citenamefont {Upadhyaya},\ and\ \citenamefont {Vitelli}}]{Chen2014}%
  \BibitemOpen
  \bibfield  {author} {\bibinfo {author} {\bibfnamefont {B.~G.}\ \bibnamefont
  {Chen}}, \bibinfo {author} {\bibfnamefont {N.}~\bibnamefont {Upadhyaya}}, \
  and\ \bibinfo {author} {\bibfnamefont {V.}~\bibnamefont {Vitelli}},\ }\href
  {https://www.pnas.org/content/111/36/13004} {\bibfield  {journal} {\bibinfo
  {journal} {Proc. Natl. Acad. Sci. U.S.A.}\ }\textbf {\bibinfo {volume}
  {111}},\ \bibinfo {pages} {13004} (\bibinfo {year} {2014})}\BibitemShut
  {NoStop}%
\bibitem [{\citenamefont {Leykam}\ and\ \citenamefont
  {Chong}(2016)}]{Leykam2016}%
  \BibitemOpen
  \bibfield  {author} {\bibinfo {author} {\bibfnamefont {D.}~\bibnamefont
  {Leykam}}\ and\ \bibinfo {author} {\bibfnamefont {Y.~D.}\ \bibnamefont
  {Chong}},\ }\href {\doibase 10.1103/PhysRevLett.117.143901} {\bibfield
  {journal} {\bibinfo  {journal} {Phys. Rev. Lett.}\ }\textbf {\bibinfo
  {volume} {117}},\ \bibinfo {pages} {143901} (\bibinfo {year}
  {2016})}\BibitemShut {NoStop}%
\bibitem [{\citenamefont {Lumer}\ \emph {et~al.}(2013)\citenamefont {Lumer},
  \citenamefont {Plotnik}, \citenamefont {Rechtsman},\ and\ \citenamefont
  {Segev}}]{Lumer2013}%
  \BibitemOpen
  \bibfield  {author} {\bibinfo {author} {\bibfnamefont {Y.}~\bibnamefont
  {Lumer}}, \bibinfo {author} {\bibfnamefont {Y.}~\bibnamefont {Plotnik}},
  \bibinfo {author} {\bibfnamefont {M.~C.}\ \bibnamefont {Rechtsman}}, \ and\
  \bibinfo {author} {\bibfnamefont {M.}~\bibnamefont {Segev}},\ }\href
  {\doibase 10.1103/PhysRevLett.111.243905} {\bibfield  {journal} {\bibinfo
  {journal} {Phys. Rev. Lett.}\ }\textbf {\bibinfo {volume} {111}},\ \bibinfo
  {pages} {243905} (\bibinfo {year} {2013})}\BibitemShut {NoStop}%
\bibitem [{\citenamefont {Oseledets}(1968)}]{Oseledets1968}%
  \BibitemOpen
  \bibfield  {author} {\bibinfo {author} {\bibfnamefont {V.~I.}\ \bibnamefont
  {Oseledets}},\ }\href {http://mi.mathnet.ru/mmo214} {\bibfield  {journal}
  {\bibinfo  {journal} {Trans. Moscow. Math. Soc.}\ }\textbf {\bibinfo {volume}
  {19}},\ \bibinfo {pages} {197} (\bibinfo {year} {1968})}\BibitemShut
  {NoStop}%
\bibitem [{\citenamefont {Takeuchi}\ \emph {et~al.}(2009)\citenamefont
  {Takeuchi}, \citenamefont {Ginelli},\ and\ \citenamefont
  {Chat\'e}}]{Takeuchi2009}%
  \BibitemOpen
  \bibfield  {author} {\bibinfo {author} {\bibfnamefont {K.~A.}\ \bibnamefont
  {Takeuchi}}, \bibinfo {author} {\bibfnamefont {F.}~\bibnamefont {Ginelli}}, \
  and\ \bibinfo {author} {\bibfnamefont {H.}~\bibnamefont {Chat\'e}},\ }\href
  {\doibase 10.1103/PhysRevLett.103.154103} {\bibfield  {journal} {\bibinfo
  {journal} {Phys. Rev. Lett.}\ }\textbf {\bibinfo {volume} {103}},\ \bibinfo
  {pages} {154103} (\bibinfo {year} {2009})}\BibitemShut {NoStop}%
\bibitem [{\citenamefont {Ginelli}\ \emph {et~al.}(2013)\citenamefont
  {Ginelli}, \citenamefont {Chat{\'{e}}}, \citenamefont {Livi},\ and\
  \citenamefont {Politi}}]{Ginelli2013}%
  \BibitemOpen
  \bibfield  {author} {\bibinfo {author} {\bibfnamefont {F.}~\bibnamefont
  {Ginelli}}, \bibinfo {author} {\bibfnamefont {H.}~\bibnamefont
  {Chat{\'{e}}}}, \bibinfo {author} {\bibfnamefont {R.}~\bibnamefont {Livi}}, \
  and\ \bibinfo {author} {\bibfnamefont {A.}~\bibnamefont {Politi}},\ }\href
  {\doibase 10.1088/1751-8113/46/25/254005} {\bibfield  {journal} {\bibinfo
  {journal} {J. Phys. A}\ }\textbf {\bibinfo {volume} {46}},\ \bibinfo {pages}
  {254005} (\bibinfo {year} {2013})}\BibitemShut {NoStop}%
\bibitem [{\citenamefont {Kuramoto}\ and\ \citenamefont
  {Battogtokh}(2002)}]{Kuramoto2002}%
  \BibitemOpen
  \bibfield  {author} {\bibinfo {author} {\bibfnamefont {Y.}~\bibnamefont
  {Kuramoto}}\ and\ \bibinfo {author} {\bibfnamefont {D.}~\bibnamefont
  {Battogtokh}},\ }\href@noop {} {\bibfield  {journal} {\bibinfo  {journal}
  {Nonlinear Phenom. Complex. Syst.}\ }\textbf {\bibinfo {volume} {5}},\
  \bibinfo {pages} {380} (\bibinfo {year} {2002})}\BibitemShut {NoStop}%
\bibitem [{\citenamefont {Abrams}\ and\ \citenamefont
  {Strogatz}(2004)}]{Abrams2004}%
  \BibitemOpen
  \bibfield  {author} {\bibinfo {author} {\bibfnamefont {D.~M.}\ \bibnamefont
  {Abrams}}\ and\ \bibinfo {author} {\bibfnamefont {S.~H.}\ \bibnamefont
  {Strogatz}},\ }\href {\doibase 10.1103/PhysRevLett.93.174102} {\bibfield
  {journal} {\bibinfo  {journal} {Phys. Rev. Lett.}\ }\textbf {\bibinfo
  {volume} {93}},\ \bibinfo {pages} {174102} (\bibinfo {year}
  {2004})}\BibitemShut {NoStop}%
\bibitem [{\citenamefont {Panaggio}\ and\ \citenamefont
  {Abrams}(2015)}]{Panaggio2015}%
  \BibitemOpen
  \bibfield  {author} {\bibinfo {author} {\bibfnamefont {M.~J.}\ \bibnamefont
  {Panaggio}}\ and\ \bibinfo {author} {\bibfnamefont {D.~M.}\ \bibnamefont
  {Abrams}},\ }\href {\doibase 10.1088/0951-7715/28/3/r67} {\bibfield
  {journal} {\bibinfo  {journal} {Nonlinearity}\ }\textbf {\bibinfo {volume}
  {28}},\ \bibinfo {pages} {R67} (\bibinfo {year} {2015})}\BibitemShut
  {NoStop}%
\bibitem [{\citenamefont {Wolfrum}\ \emph {et~al.}(2011)\citenamefont
  {Wolfrum}, \citenamefont {Omel’chenko}, \citenamefont {Yanchuk},\ and\
  \citenamefont {Maistrenko}}]{Wolfrum2011}%
  \BibitemOpen
  \bibfield  {author} {\bibinfo {author} {\bibfnamefont {M.}~\bibnamefont
  {Wolfrum}}, \bibinfo {author} {\bibfnamefont {O.~E.}\ \bibnamefont
  {Omel’chenko}}, \bibinfo {author} {\bibfnamefont {S.}~\bibnamefont
  {Yanchuk}}, \ and\ \bibinfo {author} {\bibfnamefont {Y.~L.}\ \bibnamefont
  {Maistrenko}},\ }\href {\doibase 10.1063/1.3563579} {\bibfield  {journal}
  {\bibinfo  {journal} {Chaos}\ }\textbf {\bibinfo {volume} {21}},\ \bibinfo
  {pages} {013112} (\bibinfo {year} {2011})}\BibitemShut {NoStop}%
\bibitem [{\citenamefont {H\"ohlein}\ \emph {et~al.}(2019)\citenamefont
  {H\"ohlein}, \citenamefont {Kemeth},\ and\ \citenamefont
  {Krischer}}]{Hohlein2019}%
  \BibitemOpen
  \bibfield  {author} {\bibinfo {author} {\bibfnamefont {K.}~\bibnamefont
  {H\"ohlein}}, \bibinfo {author} {\bibfnamefont {F.~P.}\ \bibnamefont
  {Kemeth}}, \ and\ \bibinfo {author} {\bibfnamefont {K.}~\bibnamefont
  {Krischer}},\ }\href {\doibase 10.1103/PhysRevE.100.022217} {\bibfield
  {journal} {\bibinfo  {journal} {Phys. Rev. E}\ }\textbf {\bibinfo {volume}
  {100}},\ \bibinfo {pages} {022217} (\bibinfo {year} {2019})}\BibitemShut
  {NoStop}%
\bibitem [{\citenamefont {Majhi}\ \emph {et~al.}(2019)\citenamefont {Majhi},
  \citenamefont {Bera}, \citenamefont {Ghosh},\ and\ \citenamefont
  {Perc}}]{Majhi2019}%
  \BibitemOpen
  \bibfield  {author} {\bibinfo {author} {\bibfnamefont {S.}~\bibnamefont
  {Majhi}}, \bibinfo {author} {\bibfnamefont {B.~K.}\ \bibnamefont {Bera}},
  \bibinfo {author} {\bibfnamefont {D.}~\bibnamefont {Ghosh}}, \ and\ \bibinfo
  {author} {\bibfnamefont {M.}~\bibnamefont {Perc}},\ }\href {\doibase
  https://doi.org/10.1016/j.plrev.2018.09.003} {\bibfield  {journal} {\bibinfo
  {journal} {Phys. Life Rev.}\ }\textbf {\bibinfo {volume} {28}},\ \bibinfo
  {pages} {100} (\bibinfo {year} {2019})}\BibitemShut {NoStop}%
\bibitem [{\citenamefont {Tinsley}\ \emph {et~al.}(2012)\citenamefont
  {Tinsley}, \citenamefont {Nkomo},\ and\ \citenamefont
  {Showalter}}]{Tinsley2012}%
  \BibitemOpen
  \bibfield  {author} {\bibinfo {author} {\bibfnamefont {M.~R.}\ \bibnamefont
  {Tinsley}}, \bibinfo {author} {\bibfnamefont {S.}~\bibnamefont {Nkomo}}, \
  and\ \bibinfo {author} {\bibfnamefont {K.}~\bibnamefont {Showalter}},\ }\href
  {\doibase 10.1038/nphys2371} {\bibfield  {journal} {\bibinfo  {journal} {Nat.
  Phys.}\ }\textbf {\bibinfo {volume} {8}},\ \bibinfo {pages} {662} (\bibinfo
  {year} {2012})}\BibitemShut {NoStop}%
\bibitem [{\citenamefont {Martens}\ \emph {et~al.}(2013)\citenamefont
  {Martens}, \citenamefont {Thutupalli}, \citenamefont {Fourri{\`e}re},\ and\
  \citenamefont {Hallatschek}}]{Martens2013}%
  \BibitemOpen
  \bibfield  {author} {\bibinfo {author} {\bibfnamefont {E.~A.}\ \bibnamefont
  {Martens}}, \bibinfo {author} {\bibfnamefont {S.}~\bibnamefont {Thutupalli}},
  \bibinfo {author} {\bibfnamefont {A.}~\bibnamefont {Fourri{\`e}re}}, \ and\
  \bibinfo {author} {\bibfnamefont {O.}~\bibnamefont {Hallatschek}},\ }\href
  {https://www.pnas.org/content/110/26/10563} {\bibfield  {journal} {\bibinfo
  {journal} {Proc. Natl. Acad. Sci. U.S.A.}\ }\textbf {\bibinfo {volume}
  {110}},\ \bibinfo {pages} {10563} (\bibinfo {year} {2013})}\BibitemShut
  {NoStop}%
\bibitem [{\citenamefont {Shen}\ \emph {et~al.}(2018)\citenamefont {Shen},
  \citenamefont {Zhen},\ and\ \citenamefont {Fu}}]{Shen2018}%
  \BibitemOpen
  \bibfield  {author} {\bibinfo {author} {\bibfnamefont {H.}~\bibnamefont
  {Shen}}, \bibinfo {author} {\bibfnamefont {B.}~\bibnamefont {Zhen}}, \ and\
  \bibinfo {author} {\bibfnamefont {L.}~\bibnamefont {Fu}},\ }\href {\doibase
  10.1103/PhysRevLett.120.146402} {\bibfield  {journal} {\bibinfo  {journal}
  {Phys. Rev. Lett.}\ }\textbf {\bibinfo {volume} {120}},\ \bibinfo {pages}
  {146402} (\bibinfo {year} {2018})}\BibitemShut {NoStop}%
\bibitem [{\citenamefont {Gong}\ \emph {et~al.}(2018)\citenamefont {Gong},
  \citenamefont {Ashida}, \citenamefont {Kawabata}, \citenamefont {Takasan},
  \citenamefont {Higashikawa},\ and\ \citenamefont {Ueda}}]{Gong2018}%
  \BibitemOpen
  \bibfield  {author} {\bibinfo {author} {\bibfnamefont {Z.}~\bibnamefont
  {Gong}}, \bibinfo {author} {\bibfnamefont {Y.}~\bibnamefont {Ashida}},
  \bibinfo {author} {\bibfnamefont {K.}~\bibnamefont {Kawabata}}, \bibinfo
  {author} {\bibfnamefont {K.}~\bibnamefont {Takasan}}, \bibinfo {author}
  {\bibfnamefont {S.}~\bibnamefont {Higashikawa}}, \ and\ \bibinfo {author}
  {\bibfnamefont {M.}~\bibnamefont {Ueda}},\ }\href {\doibase
  10.1103/PhysRevX.8.031079} {\bibfield  {journal} {\bibinfo  {journal} {Phys.
  Rev. X}\ }\textbf {\bibinfo {volume} {8}},\ \bibinfo {pages} {031079}
  (\bibinfo {year} {2018})}\BibitemShut {NoStop}%
\bibitem [{\citenamefont {Yao}\ and\ \citenamefont {Wang}(2018)}]{Yao2018}%
  \BibitemOpen
  \bibfield  {author} {\bibinfo {author} {\bibfnamefont {S.}~\bibnamefont
  {Yao}}\ and\ \bibinfo {author} {\bibfnamefont {Z.}~\bibnamefont {Wang}},\
  }\href {\doibase 10.1103/PhysRevLett.121.086803} {\bibfield  {journal}
  {\bibinfo  {journal} {Phys. Rev. Lett.}\ }\textbf {\bibinfo {volume} {121}},\
  \bibinfo {pages} {086803} (\bibinfo {year} {2018})}\BibitemShut {NoStop}%
\bibitem [{\citenamefont {Kunst}\ \emph {et~al.}(2018)\citenamefont {Kunst},
  \citenamefont {Edvardsson}, \citenamefont {Budich},\ and\ \citenamefont
  {Bergholtz}}]{Kunst2018}%
  \BibitemOpen
  \bibfield  {author} {\bibinfo {author} {\bibfnamefont {F.~K.}\ \bibnamefont
  {Kunst}}, \bibinfo {author} {\bibfnamefont {E.}~\bibnamefont {Edvardsson}},
  \bibinfo {author} {\bibfnamefont {J.~C.}\ \bibnamefont {Budich}}, \ and\
  \bibinfo {author} {\bibfnamefont {E.~J.}\ \bibnamefont {Bergholtz}},\ }\href
  {\doibase 10.1103/PhysRevLett.121.026808} {\bibfield  {journal} {\bibinfo
  {journal} {Phys. Rev. Lett.}\ }\textbf {\bibinfo {volume} {121}},\ \bibinfo
  {pages} {026808} (\bibinfo {year} {2018})}\BibitemShut {NoStop}%
\bibitem [{\citenamefont {Zhao}\ \emph {et~al.}(2019)\citenamefont {Zhao},
  \citenamefont {Qiao}, \citenamefont {Wu}, \citenamefont {Midya},
  \citenamefont {Longhi},\ and\ \citenamefont {Feng}}]{Zhao2019}%
  \BibitemOpen
  \bibfield  {author} {\bibinfo {author} {\bibfnamefont {H.}~\bibnamefont
  {Zhao}}, \bibinfo {author} {\bibfnamefont {X.}~\bibnamefont {Qiao}}, \bibinfo
  {author} {\bibfnamefont {T.}~\bibnamefont {Wu}}, \bibinfo {author}
  {\bibfnamefont {B.}~\bibnamefont {Midya}}, \bibinfo {author} {\bibfnamefont
  {S.}~\bibnamefont {Longhi}}, \ and\ \bibinfo {author} {\bibfnamefont
  {L.}~\bibnamefont {Feng}},\ }\href {\doibase 10.1126/science.aay1064}
  {\bibfield  {journal} {\bibinfo  {journal} {Science}\ }\textbf {\bibinfo
  {volume} {365}},\ \bibinfo {pages} {1163} (\bibinfo {year}
  {2019})}\BibitemShut {NoStop}%
\bibitem [{\citenamefont {Fruchart}\ \emph {et~al.}(2021)\citenamefont
  {Fruchart}, \citenamefont {Hanai}, \citenamefont {Littlewood},\ and\
  \citenamefont {Vitelli}}]{Fruchart2021}%
  \BibitemOpen
  \bibfield  {author} {\bibinfo {author} {\bibfnamefont {M.}~\bibnamefont
  {Fruchart}}, \bibinfo {author} {\bibfnamefont {R.}~\bibnamefont {Hanai}},
  \bibinfo {author} {\bibfnamefont {P.~B.}\ \bibnamefont {Littlewood}}, \ and\
  \bibinfo {author} {\bibfnamefont {V.}~\bibnamefont {Vitelli}},\ }\href
  {\doibase 10.1038/s41586-021-03375-9} {\bibfield  {journal} {\bibinfo
  {journal} {Nature}\ }\textbf {\bibinfo {volume} {592}},\ \bibinfo {pages}
  {363} (\bibinfo {year} {2021})}\BibitemShut {NoStop}%
\bibitem [{\citenamefont {Harari}\ \emph {et~al.}(2018)\citenamefont {Harari},
  \citenamefont {Bandres}, \citenamefont {Lumer}, \citenamefont {Rechtsman},
  \citenamefont {Chong}, \citenamefont {Khajavikhan}, \citenamefont
  {Christodoulides},\ and\ \citenamefont {Segev}}]{Harari2018}%
  \BibitemOpen
  \bibfield  {author} {\bibinfo {author} {\bibfnamefont {G.}~\bibnamefont
  {Harari}}, \bibinfo {author} {\bibfnamefont {M.~A.}\ \bibnamefont {Bandres}},
  \bibinfo {author} {\bibfnamefont {Y.}~\bibnamefont {Lumer}}, \bibinfo
  {author} {\bibfnamefont {M.~C.}\ \bibnamefont {Rechtsman}}, \bibinfo {author}
  {\bibfnamefont {Y.~D.}\ \bibnamefont {Chong}}, \bibinfo {author}
  {\bibfnamefont {M.}~\bibnamefont {Khajavikhan}}, \bibinfo {author}
  {\bibfnamefont {D.~N.}\ \bibnamefont {Christodoulides}}, \ and\ \bibinfo
  {author} {\bibfnamefont {M.}~\bibnamefont {Segev}},\ }\href {\doibase
  10.1126/science.aar4003} {\bibfield  {journal} {\bibinfo  {journal}
  {Science}\ }\textbf {\bibinfo {volume} {359}},\ \bibinfo {pages} {eaar4003}
  (\bibinfo {year} {2018})}\BibitemShut {NoStop}%
\bibitem [{\citenamefont {Song}\ \emph {et~al.}(2020)\citenamefont {Song},
  \citenamefont {Sun}, \citenamefont {Dutt}, \citenamefont {Minkov},
  \citenamefont {Wojcik}, \citenamefont {Wang}, \citenamefont {Williamson},
  \citenamefont {Orenstein},\ and\ \citenamefont {Fan}}]{Song2020}%
  \BibitemOpen
  \bibfield  {author} {\bibinfo {author} {\bibfnamefont {A.~Y.}\ \bibnamefont
  {Song}}, \bibinfo {author} {\bibfnamefont {X.~Q.}\ \bibnamefont {Sun}},
  \bibinfo {author} {\bibfnamefont {A.}~\bibnamefont {Dutt}}, \bibinfo {author}
  {\bibfnamefont {M.}~\bibnamefont {Minkov}}, \bibinfo {author} {\bibfnamefont
  {C.}~\bibnamefont {Wojcik}}, \bibinfo {author} {\bibfnamefont
  {H.}~\bibnamefont {Wang}}, \bibinfo {author} {\bibfnamefont {I.~A.~D.}\
  \bibnamefont {Williamson}}, \bibinfo {author} {\bibfnamefont
  {M.}~\bibnamefont {Orenstein}}, \ and\ \bibinfo {author} {\bibfnamefont
  {S.}~\bibnamefont {Fan}},\ }\href {\doibase 10.1103/PhysRevLett.125.033603}
  {\bibfield  {journal} {\bibinfo  {journal} {Phys. Rev. Lett.}\ }\textbf
  {\bibinfo {volume} {125}},\ \bibinfo {pages} {033603} (\bibinfo {year}
  {2020})}\BibitemShut {NoStop}%
\bibitem [{\citenamefont {Sone}\ \emph {et~al.}(2020)\citenamefont {Sone},
  \citenamefont {Ashida},\ and\ \citenamefont {Sagawa}}]{Sone2020}%
  \BibitemOpen
  \bibfield  {author} {\bibinfo {author} {\bibfnamefont {K.}~\bibnamefont
  {Sone}}, \bibinfo {author} {\bibfnamefont {Y.}~\bibnamefont {Ashida}}, \ and\
  \bibinfo {author} {\bibfnamefont {T.}~\bibnamefont {Sagawa}},\ }\href
  {\doibase 10.1038/s41467-020-19488-0} {\bibfield  {journal} {\bibinfo
  {journal} {Nat. Commun.}\ }\textbf {\bibinfo {volume} {11}},\ \bibinfo
  {pages} {5745} (\bibinfo {year} {2020})}\BibitemShut {NoStop}%
\bibitem [{\citenamefont {Stuart}(1960)}]{Stuart1960}%
  \BibitemOpen
  \bibfield  {author} {\bibinfo {author} {\bibfnamefont {J.~T.}\ \bibnamefont
  {Stuart}},\ }\href {\doibase 10.1017/S002211206000116X} {\bibfield  {journal}
  {\bibinfo  {journal} {J. Fluid Mech.}\ }\textbf {\bibinfo {volume} {9}},\
  \bibinfo {pages} {353–370} (\bibinfo {year} {1960})}\BibitemShut {NoStop}%
\bibitem [{\citenamefont {Qi}\ \emph {et~al.}(2006)\citenamefont {Qi},
  \citenamefont {Wu},\ and\ \citenamefont {Zhang}}]{Qi2006}%
  \BibitemOpen
  \bibfield  {author} {\bibinfo {author} {\bibfnamefont {X.~L.}\ \bibnamefont
  {Qi}}, \bibinfo {author} {\bibfnamefont {Y.~S.}\ \bibnamefont {Wu}}, \ and\
  \bibinfo {author} {\bibfnamefont {S.~C.}\ \bibnamefont {Zhang}},\ }\href
  {\doibase 10.1103/PhysRevB.74.085308} {\bibfield  {journal} {\bibinfo
  {journal} {Phys. Rev. B}\ }\textbf {\bibinfo {volume} {74}},\ \bibinfo
  {pages} {085308} (\bibinfo {year} {2006})}\BibitemShut {NoStop}%
\bibitem [{\citenamefont {Lee}\ \emph {et~al.}(2018)\citenamefont {Lee},
  \citenamefont {Imhof}, \citenamefont {Berger}, \citenamefont {Bayer},
  \citenamefont {Brehm}, \citenamefont {Molenkamp}, \citenamefont {Kiessling},\
  and\ \citenamefont {Thomale}}]{Lee2018}%
  \BibitemOpen
  \bibfield  {author} {\bibinfo {author} {\bibfnamefont {C.~H.}\ \bibnamefont
  {Lee}}, \bibinfo {author} {\bibfnamefont {S.}~\bibnamefont {Imhof}}, \bibinfo
  {author} {\bibfnamefont {C.}~\bibnamefont {Berger}}, \bibinfo {author}
  {\bibfnamefont {F.}~\bibnamefont {Bayer}}, \bibinfo {author} {\bibfnamefont
  {J.}~\bibnamefont {Brehm}}, \bibinfo {author} {\bibfnamefont {L.~W.}\
  \bibnamefont {Molenkamp}}, \bibinfo {author} {\bibfnamefont {T.}~\bibnamefont
  {Kiessling}}, \ and\ \bibinfo {author} {\bibfnamefont {R.}~\bibnamefont
  {Thomale}},\ }\href {\doibase 10.1038/s42005-018-0035-2} {\bibfield
  {journal} {\bibinfo  {journal} {Commun. Phys.}\ }\textbf {\bibinfo {volume}
  {1}},\ \bibinfo {pages} {39} (\bibinfo {year} {2018})}\BibitemShut {NoStop}%
\bibitem [{\citenamefont {W\"achtler}\ \emph {et~al.}(2020)\citenamefont
  {W\"achtler}, \citenamefont {Bastidas}, \citenamefont {Schaller},\ and\
  \citenamefont {Munro}}]{Wachtler2020}%
  \BibitemOpen
  \bibfield  {author} {\bibinfo {author} {\bibfnamefont {C.~W.}\ \bibnamefont
  {W\"achtler}}, \bibinfo {author} {\bibfnamefont {V.~M.}\ \bibnamefont
  {Bastidas}}, \bibinfo {author} {\bibfnamefont {G.}~\bibnamefont {Schaller}},
  \ and\ \bibinfo {author} {\bibfnamefont {W.~J.}\ \bibnamefont {Munro}},\
  }\href {\doibase 10.1103/PhysRevB.102.014309} {\bibfield  {journal} {\bibinfo
   {journal} {Phys. Rev. B}\ }\textbf {\bibinfo {volume} {102}},\ \bibinfo
  {pages} {014309} (\bibinfo {year} {2020})}\BibitemShut {NoStop}%
\bibitem [{\citenamefont {Shraiman}\ \emph {et~al.}(1992)\citenamefont
  {Shraiman}, \citenamefont {Pumir}, \citenamefont {{van Saarloos}},
  \citenamefont {Hohenberg}, \citenamefont {Chaté},\ and\ \citenamefont
  {Holen}}]{Shraiman1992}%
  \BibitemOpen
  \bibfield  {author} {\bibinfo {author} {\bibfnamefont {B.~I.}\ \bibnamefont
  {Shraiman}}, \bibinfo {author} {\bibfnamefont {A.}~\bibnamefont {Pumir}},
  \bibinfo {author} {\bibfnamefont {W.}~\bibnamefont {{van Saarloos}}},
  \bibinfo {author} {\bibfnamefont {P.~C.}\ \bibnamefont {Hohenberg}}, \bibinfo
  {author} {\bibfnamefont {H.}~\bibnamefont {Chaté}}, \ and\ \bibinfo {author}
  {\bibfnamefont {M.}~\bibnamefont {Holen}},\ }\href {\doibase
  https://doi.org/10.1016/0167-2789(92)90001-4} {\bibfield  {journal} {\bibinfo
   {journal} {Physica D}\ }\textbf {\bibinfo {volume} {57}},\ \bibinfo {pages}
  {241} (\bibinfo {year} {1992})}\BibitemShut {NoStop}%
\bibitem [{\citenamefont {Hata}\ \emph {et~al.}(2014)\citenamefont {Hata},
  \citenamefont {Nakao},\ and\ \citenamefont {Mikhailov}}]{Hata2014}%
  \BibitemOpen
  \bibfield  {author} {\bibinfo {author} {\bibfnamefont {S.}~\bibnamefont
  {Hata}}, \bibinfo {author} {\bibfnamefont {H.}~\bibnamefont {Nakao}}, \ and\
  \bibinfo {author} {\bibfnamefont {A.~S.}\ \bibnamefont {Mikhailov}},\ }\href
  {\doibase 10.1038/srep03585} {\bibfield  {journal} {\bibinfo  {journal} {Sci.
  Rep.}\ }\textbf {\bibinfo {volume} {4}},\ \bibinfo {pages} {3585} (\bibinfo
  {year} {2014})}\BibitemShut {NoStop}%
\bibitem [{\citenamefont {Hata}\ and\ \citenamefont {Nakao}(2017)}]{Hata2017}%
  \BibitemOpen
  \bibfield  {author} {\bibinfo {author} {\bibfnamefont {S.}~\bibnamefont
  {Hata}}\ and\ \bibinfo {author} {\bibfnamefont {H.}~\bibnamefont {Nakao}},\
  }\href {\doibase 10.1038/s41598-017-01010-0} {\bibfield  {journal} {\bibinfo
  {journal} {Sci. Rep.}\ }\textbf {\bibinfo {volume} {7}},\ \bibinfo {pages}
  {1121} (\bibinfo {year} {2017})}\BibitemShut {NoStop}%
\bibitem [{\citenamefont {Tuloup}\ \emph {et~al.}(2020)\citenamefont {Tuloup},
  \citenamefont {Bomantara}, \citenamefont {Lee},\ and\ \citenamefont
  {Gong}}]{Tuloup2020}%
  \BibitemOpen
  \bibfield  {author} {\bibinfo {author} {\bibfnamefont {T.}~\bibnamefont
  {Tuloup}}, \bibinfo {author} {\bibfnamefont {R.~W.}\ \bibnamefont
  {Bomantara}}, \bibinfo {author} {\bibfnamefont {C.~H.}\ \bibnamefont {Lee}},
  \ and\ \bibinfo {author} {\bibfnamefont {J.}~\bibnamefont {Gong}},\ }\href
  {\doibase 10.1103/PhysRevB.102.115411} {\bibfield  {journal} {\bibinfo
  {journal} {Phys. Rev. B}\ }\textbf {\bibinfo {volume} {102}},\ \bibinfo
  {pages} {115411} (\bibinfo {year} {2020})}\BibitemShut {NoStop}%
\bibitem [{\citenamefont {der Pol}(1926)}]{Pol1926}%
  \BibitemOpen
  \bibfield  {author} {\bibinfo {author} {\bibfnamefont {B.~V.}\ \bibnamefont
  {der Pol}},\ }\href {\doibase 10.1080/14786442608564127} {\bibfield
  {journal} {\bibinfo  {journal} {Philos. Mag.}\ }\textbf {\bibinfo {volume}
  {2}},\ \bibinfo {pages} {978} (\bibinfo {year} {1926})}\BibitemShut {NoStop}%
\bibitem [{\citenamefont {Johnson}\ \emph {et~al.}(2014)\citenamefont
  {Johnson}, \citenamefont {Dhople}, \citenamefont {Hamadeh},\ and\
  \citenamefont {Krein}}]{Johnson2014}%
  \BibitemOpen
  \bibfield  {author} {\bibinfo {author} {\bibfnamefont {B.~B.}\ \bibnamefont
  {Johnson}}, \bibinfo {author} {\bibfnamefont {S.~V.}\ \bibnamefont {Dhople}},
  \bibinfo {author} {\bibfnamefont {A.~O.}\ \bibnamefont {Hamadeh}}, \ and\
  \bibinfo {author} {\bibfnamefont {P.~T.}\ \bibnamefont {Krein}},\ }\href
  {\doibase 10.1109/TPEL.2013.2296292} {\bibfield  {journal} {\bibinfo
  {journal} {IEEE Trans. Circuits Syst. I Regul. Pap.}\ }\textbf {\bibinfo
  {volume} {29}},\ \bibinfo {pages} {6124} (\bibinfo {year}
  {2014})}\BibitemShut {NoStop}%
\bibitem [{\citenamefont {Hofmann}\ \emph {et~al.}(2019)\citenamefont
  {Hofmann}, \citenamefont {Helbig}, \citenamefont {Lee}, \citenamefont
  {Greiter},\ and\ \citenamefont {Thomale}}]{Hofmann2019}%
  \BibitemOpen
  \bibfield  {author} {\bibinfo {author} {\bibfnamefont {T.}~\bibnamefont
  {Hofmann}}, \bibinfo {author} {\bibfnamefont {T.}~\bibnamefont {Helbig}},
  \bibinfo {author} {\bibfnamefont {C.~H.}\ \bibnamefont {Lee}}, \bibinfo
  {author} {\bibfnamefont {M.}~\bibnamefont {Greiter}}, \ and\ \bibinfo
  {author} {\bibfnamefont {R.}~\bibnamefont {Thomale}},\ }\href {\doibase
  10.1103/PhysRevLett.122.247702} {\bibfield  {journal} {\bibinfo  {journal}
  {Phys. Rev. Lett.}\ }\textbf {\bibinfo {volume} {122}},\ \bibinfo {pages}
  {247702} (\bibinfo {year} {2019})}\BibitemShut {NoStop}%
\bibitem [{\citenamefont {Ezawa}(2020)}]{Ezawa2020}%
  \BibitemOpen
  \bibfield  {author} {\bibinfo {author} {\bibfnamefont {M.}~\bibnamefont
  {Ezawa}},\ }\href {\doibase 10.1103/PhysRevB.102.075424} {\bibfield
  {journal} {\bibinfo  {journal} {Phys. Rev. B}\ }\textbf {\bibinfo {volume}
  {102}},\ \bibinfo {pages} {075424} (\bibinfo {year} {2020})}\BibitemShut
  {NoStop}%
\bibitem [{\citenamefont {El-Ganainy}\ \emph {et~al.}(2007)\citenamefont
  {El-Ganainy}, \citenamefont {Makris}, \citenamefont {Christodoulides},\ and\
  \citenamefont {Musslimani}}]{El-Ganainy2007}%
  \BibitemOpen
  \bibfield  {author} {\bibinfo {author} {\bibfnamefont {R.}~\bibnamefont
  {El-Ganainy}}, \bibinfo {author} {\bibfnamefont {K.~G.}\ \bibnamefont
  {Makris}}, \bibinfo {author} {\bibfnamefont {D.~N.}\ \bibnamefont
  {Christodoulides}}, \ and\ \bibinfo {author} {\bibfnamefont {Z.~H.}\
  \bibnamefont {Musslimani}},\ }\href {\doibase 10.1364/OL.32.002632}
  {\bibfield  {journal} {\bibinfo  {journal} {Opt. Lett.}\ }\textbf {\bibinfo
  {volume} {32}},\ \bibinfo {pages} {2632} (\bibinfo {year}
  {2007})}\BibitemShut {NoStop}%
\bibitem [{\citenamefont {Hofmann}\ \emph {et~al.}(2016)\citenamefont
  {Hofmann}, \citenamefont {Helbig}, \citenamefont {Lee}, \citenamefont
  {Greiter},\ and\ \citenamefont {Thomale}}]{Xu2016}%
  \BibitemOpen
  \bibfield  {author} {\bibinfo {author} {\bibfnamefont {T.}~\bibnamefont
  {Hofmann}}, \bibinfo {author} {\bibfnamefont {T.}~\bibnamefont {Helbig}},
  \bibinfo {author} {\bibfnamefont {C.~H.}\ \bibnamefont {Lee}}, \bibinfo
  {author} {\bibfnamefont {M.}~\bibnamefont {Greiter}}, \ and\ \bibinfo
  {author} {\bibfnamefont {R.}~\bibnamefont {Thomale}},\ }\href {\doibase
  10.1038/nature18604} {\bibfield  {journal} {\bibinfo  {journal} {Nature}\
  }\textbf {\bibinfo {volume} {537}},\ \bibinfo {pages} {80} (\bibinfo {year}
  {2016})}\BibitemShut {NoStop}%
\bibitem [{\citenamefont {FitzHugh}(1961)}]{Fitzhugh1961}%
  \BibitemOpen
  \bibfield  {author} {\bibinfo {author} {\bibfnamefont {R.}~\bibnamefont
  {FitzHugh}},\ }\href {\doibase https://doi.org/10.1016/S0006-3495(61)86902-6}
  {\bibfield  {journal} {\bibinfo  {journal} {Biophys. J.}\ }\textbf {\bibinfo
  {volume} {1}},\ \bibinfo {pages} {445} (\bibinfo {year} {1961})}\BibitemShut
  {NoStop}%
\bibitem [{\citenamefont {Nagumo}\ \emph {et~al.}(1962)\citenamefont {Nagumo},
  \citenamefont {Arimoto},\ and\ \citenamefont {Yoshizawa}}]{Nagumo1962}%
  \BibitemOpen
  \bibfield  {author} {\bibinfo {author} {\bibfnamefont {J.}~\bibnamefont
  {Nagumo}}, \bibinfo {author} {\bibfnamefont {S.}~\bibnamefont {Arimoto}}, \
  and\ \bibinfo {author} {\bibfnamefont {S.}~\bibnamefont {Yoshizawa}},\ }\href
  {\doibase 10.1109/JRPROC.1962.288235} {\bibfield  {journal} {\bibinfo
  {journal} {Proc. IRE}\ }\textbf {\bibinfo {volume} {50}},\ \bibinfo {pages}
  {2061} (\bibinfo {year} {1962})}\BibitemShut {NoStop}%
\bibitem [{\citenamefont {Shimada}\ and\ \citenamefont
  {Nagashima}(1979)}]{Shimada1979}%
  \BibitemOpen
  \bibfield  {author} {\bibinfo {author} {\bibfnamefont {I.}~\bibnamefont
  {Shimada}}\ and\ \bibinfo {author} {\bibfnamefont {T.}~\bibnamefont
  {Nagashima}},\ }\href {\doibase 10.1143/PTP.61.1605} {\bibfield  {journal}
  {\bibinfo  {journal} {Prog. Theor. Phys.}\ }\textbf {\bibinfo {volume}
  {61}},\ \bibinfo {pages} {1605} (\bibinfo {year} {1979})}\BibitemShut
  {NoStop}%
\bibitem [{\citenamefont {Ginelli}\ \emph {et~al.}(2007)\citenamefont
  {Ginelli}, \citenamefont {Poggi}, \citenamefont {Turchi}, \citenamefont
  {Chat\'e}, \citenamefont {Livi},\ and\ \citenamefont {Politi}}]{Ginelli2007}%
  \BibitemOpen
  \bibfield  {author} {\bibinfo {author} {\bibfnamefont {F.}~\bibnamefont
  {Ginelli}}, \bibinfo {author} {\bibfnamefont {P.}~\bibnamefont {Poggi}},
  \bibinfo {author} {\bibfnamefont {A.}~\bibnamefont {Turchi}}, \bibinfo
  {author} {\bibfnamefont {H.}~\bibnamefont {Chat\'e}}, \bibinfo {author}
  {\bibfnamefont {R.}~\bibnamefont {Livi}}, \ and\ \bibinfo {author}
  {\bibfnamefont {A.}~\bibnamefont {Politi}},\ }\href {\doibase
  10.1103/PhysRevLett.99.130601} {\bibfield  {journal} {\bibinfo  {journal}
  {Phys. Rev. Lett.}\ }\textbf {\bibinfo {volume} {99}},\ \bibinfo {pages}
  {130601} (\bibinfo {year} {2007})}\BibitemShut {NoStop}%
\bibitem [{\citenamefont {Stratt}\ \emph {et~al.}(1979)\citenamefont {Stratt},
  \citenamefont {Handy},\ and\ \citenamefont {Miller}}]{Stratt1979}%
  \BibitemOpen
  \bibfield  {author} {\bibinfo {author} {\bibfnamefont {R.~M.}\ \bibnamefont
  {Stratt}}, \bibinfo {author} {\bibfnamefont {N.~C.}\ \bibnamefont {Handy}}, \
  and\ \bibinfo {author} {\bibfnamefont {W.~H.}\ \bibnamefont {Miller}},\
  }\href {\doibase 10.1063/1.438772} {\bibfield  {journal} {\bibinfo  {journal}
  {J. Chem. Phys}\ }\textbf {\bibinfo {volume} {71}},\ \bibinfo {pages} {3311}
  (\bibinfo {year} {1979})}\BibitemShut {NoStop}%
\bibitem [{\citenamefont {Ningyuan}\ \emph {et~al.}(2015)\citenamefont
  {Ningyuan}, \citenamefont {Owens}, \citenamefont {Sommer}, \citenamefont
  {Schuster},\ and\ \citenamefont {Simon}}]{Ningyuan2015}%
  \BibitemOpen
  \bibfield  {author} {\bibinfo {author} {\bibfnamefont {J.}~\bibnamefont
  {Ningyuan}}, \bibinfo {author} {\bibfnamefont {C.}~\bibnamefont {Owens}},
  \bibinfo {author} {\bibfnamefont {A.}~\bibnamefont {Sommer}}, \bibinfo
  {author} {\bibfnamefont {D.}~\bibnamefont {Schuster}}, \ and\ \bibinfo
  {author} {\bibfnamefont {J.}~\bibnamefont {Simon}},\ }\href {\doibase
  10.1103/PhysRevX.5.021031} {\bibfield  {journal} {\bibinfo  {journal} {Phys.
  Rev. X}\ }\textbf {\bibinfo {volume} {5}},\ \bibinfo {pages} {021031}
  (\bibinfo {year} {2015})}\BibitemShut {NoStop}%
\bibitem [{\citenamefont {Albert}\ \emph {et~al.}(2015)\citenamefont {Albert},
  \citenamefont {Glazman},\ and\ \citenamefont {Jiang}}]{Albert2015}%
  \BibitemOpen
  \bibfield  {author} {\bibinfo {author} {\bibfnamefont {V.~V.}\ \bibnamefont
  {Albert}}, \bibinfo {author} {\bibfnamefont {L.~I.}\ \bibnamefont {Glazman}},
  \ and\ \bibinfo {author} {\bibfnamefont {L.}~\bibnamefont {Jiang}},\ }\href
  {\doibase 10.1103/PhysRevLett.114.173902} {\bibfield  {journal} {\bibinfo
  {journal} {Phys. Rev. Lett.}\ }\textbf {\bibinfo {volume} {114}},\ \bibinfo
  {pages} {173902} (\bibinfo {year} {2015})}\BibitemShut {NoStop}%
\bibitem [{\citenamefont {Imhof}\ \emph {et~al.}(2018)\citenamefont {Imhof},
  \citenamefont {Berger}, \citenamefont {Bayer}, \citenamefont {Brehm},
  \citenamefont {Molenkamp}, \citenamefont {Kiessling}, \citenamefont
  {Schindler}, \citenamefont {Lee}, \citenamefont {Greiter}, \citenamefont
  {Neupert},\ and\ \citenamefont {Thomale}}]{Imhof2018}%
  \BibitemOpen
  \bibfield  {author} {\bibinfo {author} {\bibfnamefont {S.}~\bibnamefont
  {Imhof}}, \bibinfo {author} {\bibfnamefont {C.}~\bibnamefont {Berger}},
  \bibinfo {author} {\bibfnamefont {F.}~\bibnamefont {Bayer}}, \bibinfo
  {author} {\bibfnamefont {J.}~\bibnamefont {Brehm}}, \bibinfo {author}
  {\bibfnamefont {L.~W.}\ \bibnamefont {Molenkamp}}, \bibinfo {author}
  {\bibfnamefont {T.}~\bibnamefont {Kiessling}}, \bibinfo {author}
  {\bibfnamefont {F.}~\bibnamefont {Schindler}}, \bibinfo {author}
  {\bibfnamefont {C.~H.}\ \bibnamefont {Lee}}, \bibinfo {author} {\bibfnamefont
  {M.}~\bibnamefont {Greiter}}, \bibinfo {author} {\bibfnamefont
  {T.}~\bibnamefont {Neupert}}, \ and\ \bibinfo {author} {\bibfnamefont
  {R.}~\bibnamefont {Thomale}},\ }\href {\doibase 10.1038/s41567-018-0246-1}
  {\bibfield  {journal} {\bibinfo  {journal} {Nat. Phys.}\ }\textbf {\bibinfo
  {volume} {14}},\ \bibinfo {pages} {925} (\bibinfo {year} {2018})}\BibitemShut
  {NoStop}%
\bibitem [{\citenamefont {Ezawa}(2019)}]{Ezawa2019}%
  \BibitemOpen
  \bibfield  {author} {\bibinfo {author} {\bibfnamefont {M.}~\bibnamefont
  {Ezawa}},\ }\href {\doibase 10.1103/PhysRevB.99.121411} {\bibfield  {journal}
  {\bibinfo  {journal} {Phys. Rev. B}\ }\textbf {\bibinfo {volume} {99}},\
  \bibinfo {pages} {121411(R)} (\bibinfo {year} {2019})}\BibitemShut {NoStop}%
\end{thebibliography}%

\end{document}